\documentclass[twocolumn, astrosymb]{aastex631}
\usepackage{amsmath}
\usepackage{graphicx}

\shorttitle{Dwarf Satellite Census of NGC 2403}
\shortauthors{Carlin et al.}

\begin{document}

\title{A census of dwarf galaxy satellites around LMC-mass galaxy NGC 2403\footnote{This research is based on data collected at the Subaru Telescope, which is operated by the National Astronomical Observatory of Japan.}}

\correspondingauthor{Jeffrey L. Carlin}
\email{jcarlin@lsst.org, jeffreylcarlin@gmail.com}

\author[0000-0002-3936-9628]{Jeffrey L. Carlin}
\affiliation{AURA/Rubin Observatory, 950 North Cherry Avenue, Tucson, AZ 85719, USA}

\author[0000-0003-4102-380X]{David J. Sand}
\affiliation{Department of Astronomy/Steward Observatory, 933 North Cherry Avenue, Rm. N204, Tucson, AZ 85721-0065, USA}

\author[0000-0001-9649-4815]{Bur\c{c}in Mutlu-Pakdil}
\affiliation{Department of Physics and Astronomy, Dartmouth College, Hanover, NH 03755, USA}

\author[0000-0002-1763-4128]{Denija Crnojevi\'{c}}
\affiliation{Department of Physics \& Astronomy, University of Tampa, 401 West Kennedy Boulevard, Tampa, FL 33606, USA}

\author[0000-0001-9775-9029]{Amandine Doliva-Dolinsky}
\affiliation{Department of Physics \& Astronomy, University of Tampa, 401 West Kennedy Boulevard, Tampa, FL 33606, USA}
\affiliation{Department of Physics \& Astronomy, Dartmouth College, Hanover, NH 03755, USA}

\author[0000-0001-9061-1697]{Christopher T. Garling}
\affiliation{Department of Astronomy, University of Virginia, 530 McCormick Road, Charlottesville, VA 22904, USA}

\author[0000-0002-8040-6785]{Annika H. G. Peter}
\affiliation{CCAPP, Department of Physics, and Department of Astronomy, The Ohio State University, Columbus, OH 43210, USA}

\author[0000-0002-9658-8763]{Jean P. Brodie}
\affiliation{Centre for Astrophysics and Supercomputing, Swinburne University, John Street, Hawthorn VIC 3122, Australia}
\affiliation{Department of Astronomy \& Astrophysics, University of California Santa Cruz, 1156 High Street, Santa Cruz, CA 95064, USA}

\author[0000-0001-5590-5518]{Duncan A. Forbes}
\affiliation{Centre for Astrophysics and Supercomputing, Swinburne University, Hawthorn VIC 3122, Australia}

\author[0000-0002-8722-9806]{Jonathan R. Hargis}
\affiliation{Space Telescope Science Institute, 3700 San Martin Drive, Baltimore, MD 21218, USA}

\author[0000-0003-2473-0369]{Aaron J. Romanowsky}
\affil{University of California Observatories, 1156 High Street, Santa Cruz, CA 95064, USA}
\affil{Department of Physics \& Astronomy, San Jos\'e State University, One Washington Square, San Jose, CA 95192, USA}

\author[0000-0002-0956-7949]{Kristine Spekkens}
\affiliation{Department of Physics, Engineering Physics, and Astronomy, Queen’s University, Kingston, ON, K7L 3N6, Canada}

\author[0000-0002-1468-9668]{Jay Strader}
\affiliation{Center for Data Intensive and Time Domain Astronomy, Department of Physics and Astronomy,\\ Michigan State University, East Lansing, MI 48824, USA}

\author[0000-0003-2892-9906]{Beth Willman}
\affiliation{LSST Discovery Alliance, 933 North Cherry Avenue, Tucson, AZ 85719, USA}

\begin{abstract}

We present the first comprehensive census of the satellite population around a Large Magellanic Cloud (LMC) stellar-mass galaxy, as part of the Magellanic Analog Dwarf Companions and Stellar Halos (MADCASH) survey. We have surveyed NGC~2403 ($D=3.0$ Mpc) with the Subaru/Hyper Suprime-Cam imager out to a projected radius of 90~kpc (with partial coverage extending out to $\sim110$~kpc, or $\sim80\%$ of the virial radius of NGC~2403), resolving stars in the uppermost $\sim$2.5 mags of its red giant branch. By looking for stellar overdensities in the red giant branch spatial density map, we identify 149 satellite candidates, of which only the previously discovered MADCASH~J074238+65201-dw is a bona fide dwarf, together with the more massive and disrupting satellite DDO~44. We carefully assess the completeness of our search via injection of artificial dwarf galaxies into the images, finding that we are reliably sensitive to candidates down to $M_V\sim-7.5$ mag (and somewhat sensitive to even fainter satellites). A comparison of the satellite luminosity function of NGC 2403 down to this magnitude limit to theoretical expectations shows overall good agreement. This is the first of a full sample of 11 Magellanic Cloud-mass host galaxies we will analyze, creating a statistical sample that will provide the first quantitative constraints on hierarchical models of galaxy formation around low-mass hosts.

\end{abstract}


\section{Introduction} \label{sec:intro}

The lowest mass galaxies are powerful tools for understanding galaxy evolution, dark matter, and cosmology \citep[e.g.][]{Bullock17,Simon19,Sales22}.  In particular, the faint end of the galaxy luminosity function can probe how baryons populate the smallest dark matter halos, including how they are affected by various astrophysical processes such as reionization \citep[e.g.][]{Bullock00,Ricotti05,Applebaum21}, supernova/star formation feedback \citep[e.g.][]{maclow99,elbadry18}, and tidal/ram pressure stripping \citep[e.g.][]{Gatto13,Simpson18}. 

Ultra-faint dwarf galaxies continue to be found around the Milky Way \citep[e.g.][]{Cerny23,Simon23} and at the edge of the Local Group \citep[e.g.][]{Sand2022,Jones2023,Mcquinn23,McQuinn24}, which will remain a vital proving ground for galaxy formation and dark matter models \citep[e.g.][]{Nadler24}.  Beyond the Local Group, faint and ultra-faint dwarf galaxy populations are being identified around Milky Way-mass systems using resolved star searches \citep{Chiboucas13,Crnojevic16,Crnojevic19,Smercina18,Bennet19,Bennet20,Mutlupakdil24}, diffuse galaxy identification \citep{Bennet17,Davis21,Carlsten22,Zaritsky24} and spectroscopic surveys \citep{Geha17,Mao21,Mao24}. These programs, largely conducted on Milky Way-mass galaxies, are elucidating the typical number and scatter in satellite properties as a function of mass, environment, and accretion history \citep[e.g.][]{Bennet19,Carlsten22,Smercina22,Mutlupakdil24,Geha24}.

To complement these studies at the Milky Way-mass scale, we must also measure the low mass galaxy population in a variety of environments, from the field to galaxy cluster scale. An intriguing avenue is the dwarf satellite population of relatively massive dwarf galaxy systems at the Magellanic Cloud-mass scale, which are expected in Cold Dark Matter models \citep[e.g.][]{Dooley2017, santos-santos2022}.  At this mass scale tidal and ram pressure stripping should be weakened, allowing for a deeper understanding of these processes and how they affect the smallest galaxies.  The ultra-faint satellite population of the LMC itself is now coming into focus thanks to {\it Gaia} orbital information \citep[e.g.][]{K18,Battaglia22}. Beyond the Local Group, initial studies have yielded promising results \citep[e.g.][]{Sand2015,Carlin2016,garling2021,sand2024}, but quantitative satellite luminosity functions are largely still lacking at this mass scale.

\begin{figure}[!t]
\includegraphics[width=1.0\columnwidth, trim=0.5in 0.0in 1.0in 1.0in, clip]{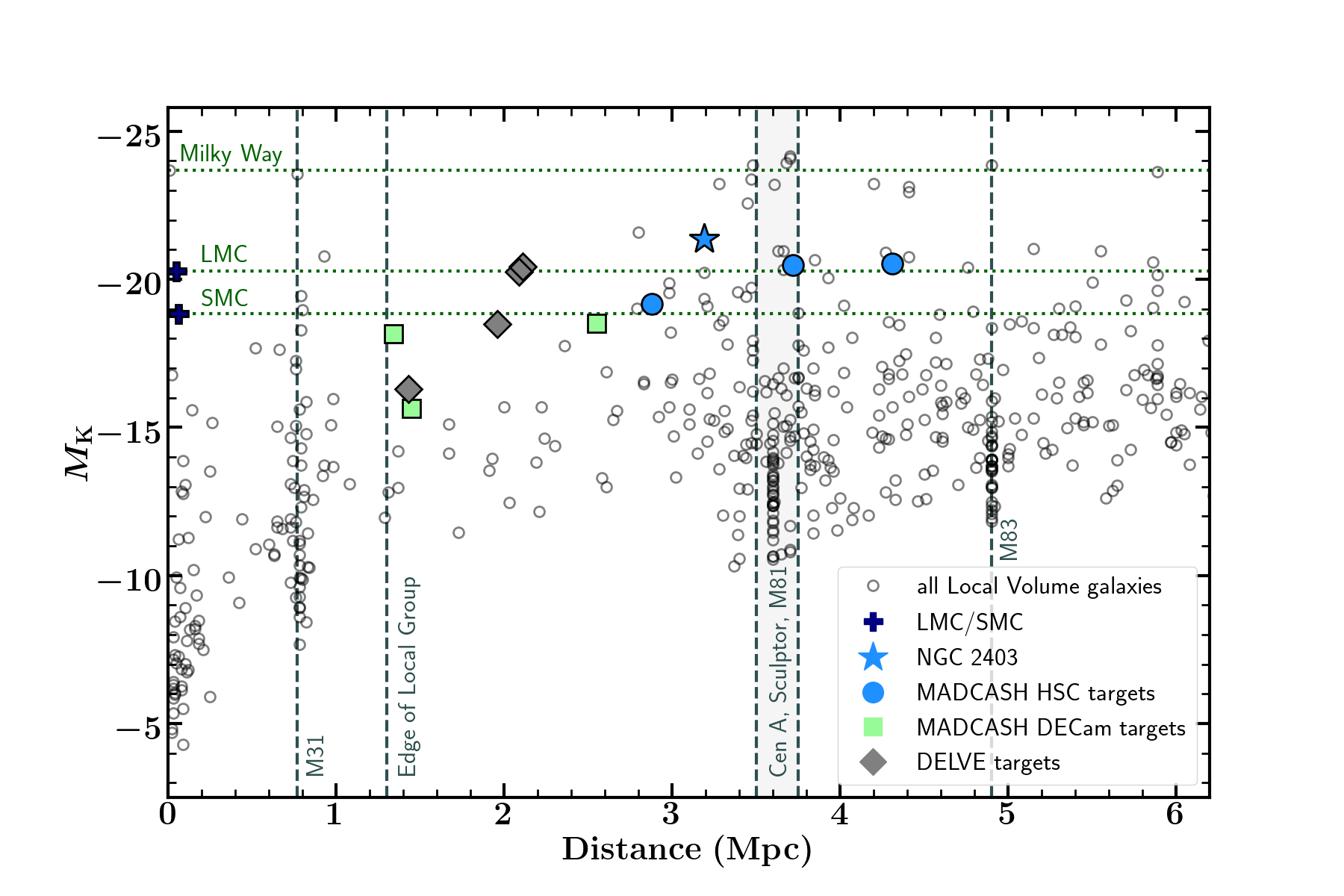}
\caption{Absolute $K$-band luminosity ($M_{\rm K}$) versus distance for Local Volume galaxies from the catalog of \citet{Karachentsev2013}. Magellanic Cloud-analog host galaxies (i.e., galaxies with stellar masses $\frac{1}{3} M_{\rm *, SMC} < M_* < 3 M_{\rm *, LMC}$) targeted by the MADCASH and DELVE-DEEP surveys are highlighted, with the subject of this work (NGC~2403) shown as a blue star.}
\label{fig:lv_gxs}
\end{figure}

\begin{table}[!t]
\label{tab:ngc2403_props}
\centering
\caption{Properties of NGC 2403}
\begin{tabular}{c c c} 
 \hline
 Parameter & Value & Reference \\ [0.5ex] 
 \hline
 RA (J2000) &  07$^{\text{h}}36^{\text{m}}51^{\text{s}}$\rlap{\hspace{-0.3em}.}4 & NED \\ 
 DEC (J2000) & $+65^\circ36\farcm09\farcs$\rlap{\hspace{-0.4em}.}2 & NED \\
 Distance (Mpc) & 3.0 & \citealt{Carlin2016} \\
 Virial Radius (kpc)\tablenotemark{a} & 140 & \citealt{mutlu-pakdil2021} \\
 Stellar Mass ($M_\odot$) & $7.2\times10^{9}$ & \citealt{Dooley2017} \\
 Halo Mass ($M_\odot$)\tablenotemark{b} & $3.4\times10^{11}$ & \citealt{mutlu-pakdil2021} \\ [0.1ex]
 \hline
\end{tabular}
\tablenotetext{a}{Calculated using the definition from \citet{bryan_norman1998}.}
\tablenotetext{b}{Inferred using the stellar mass-halo mass relation from \citet{moster2010}.}
\end{table}

\begin{deluxetable*}{lcccc}
\tablecolumns{5}
\tablewidth{0pt}
\tablecaption{Properties of the MADCASH host galaxies being observed with Subaru+HSC, plus the SMC and LMC for context. 
\label{tab:targetlist}}
\tablehead{\colhead{galaxy} & \colhead{$M_{\rm stars} (M_\odot)$\tablenotemark{a}} & \colhead{$D$ (Mpc)\tablenotemark{a}}
& \colhead{$R_{vir}$ (kpc)\tablenotemark{a}} & \colhead{$N_{\rm sat,exp}$\tablenotemark{b}} }
\startdata
***SMC & $7.0\times10^8$ & 0.06 & -- & 1-3 \\
NGC 4214 & $1.0\times10^9$ & 2.9 & 100 &  1-5 \\
***LMC & $2.6\times10^9$ & 0.05 & -- & 2-5 \\
NGC 247 & $3.2\times10^9$ & 3.7 & 120 & 2-6 \\
NGC 4244 & $3.5\times10^{9}$ & 4.3 & 120 & 2-6 \\
NGC 2403 & $7.2\times10^9$  & 3.0 & 140 & 4-8 \\
\enddata
\tablenotetext{a}{Stellar masses, distances, and virial radii are from \citet{mutlu-pakdil2021}, except the distance to NGC~2403, which is from \citet{Carlin2016}.}
\tablenotetext{b}{Predicted number of satellites with $M_{\rm stars} > 10^5~M_\odot$ from \citep{Dooley2017}, based on the \citet{garrison-kimmel2017} models. The range represents the 20th-80th percentile range of predictions for each host.}
\end{deluxetable*}

Here we present a quantitative satellite luminosity function for the nearby galaxy NGC~2403. NGC~2403 is a SAB spiral galaxy, and a stellar mass analog of the Large Magellanic Cloud ($D=3.0$ Mpc; $M_{*}$$\sim$7$\times$10$^9$ $M_{\odot}$, or 2$\times$ the stellar mass of the LMC; a summary of NGC~2403 properties adopted in this work is in Table~\ref{tab:ngc2403_props}). This galaxy is part of the ongoing MADCASH survey (see below); previous results from the survey include the initial discovery of MADCASH-1 (MADCASH J074238+652501-dw; \citealt{Carlin2016}), a faint satellite galaxy ($M_{\rm V}=-7.8$ mag) of NGC~2403 with an old and metal-poor stellar population, and no apparent H\textsc{i} gas reservoir. Follow-up imaging with the Hubble Space Telescope confirmed that MADCASH-1 contains solely ancient, metal-poor stars, and is at a distance consistent with being an NGC~2403 satellite \citep{Carlin2021}. The full red giant branch halo map of NGC~2403 revealed a tidal stream associated with the previously known massive satellite DDO~44 ($M_{\rm V}=-12.9$ mag; \citealt{Carlin2019}). The data presented here have also contributed to a study of the old star clusters in NGC~2403 and their age-metallicity relation \citep{forbes2022}. In Section~\ref{sec:madcash}, we briefly describe the MADCASH survey and its goal of probing substructure in Magellanic Cloud analog systems. In Section~\ref{sec:data} we describe the deep Subaru Hyper Suprime-Cam data of NGC~2403's halo, data reduction and the dwarf galaxy search. In Section~\ref{sec:fakes} we discuss our injection of artificial dwarf galaxies into the image-level data to quantitatively derive our completeness limits. Finally, in Section~\ref{sec:discussion} we present the satellite luminosity function of NGC~2403, placing its dwarf galaxy system in context and comparing it with expectations from cosmological simulations.

\begin{figure}[!t]
\includegraphics[width=1.0\columnwidth, trim=3.5in 0.in 0.25in 0.in, clip]{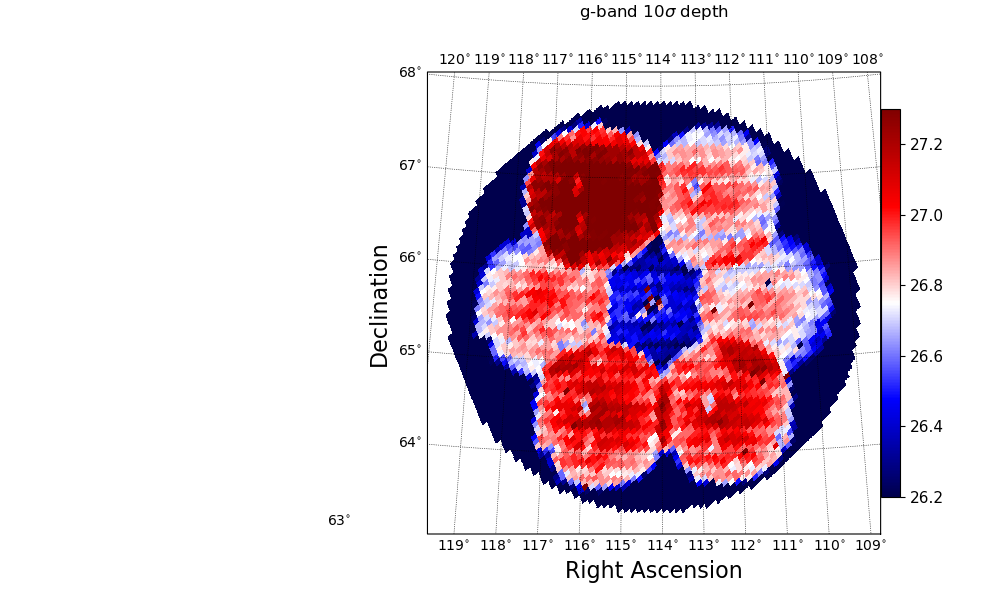}\\
\includegraphics[width=1.0\columnwidth, trim=3.5in 0.in 0.25in 0.in, clip]{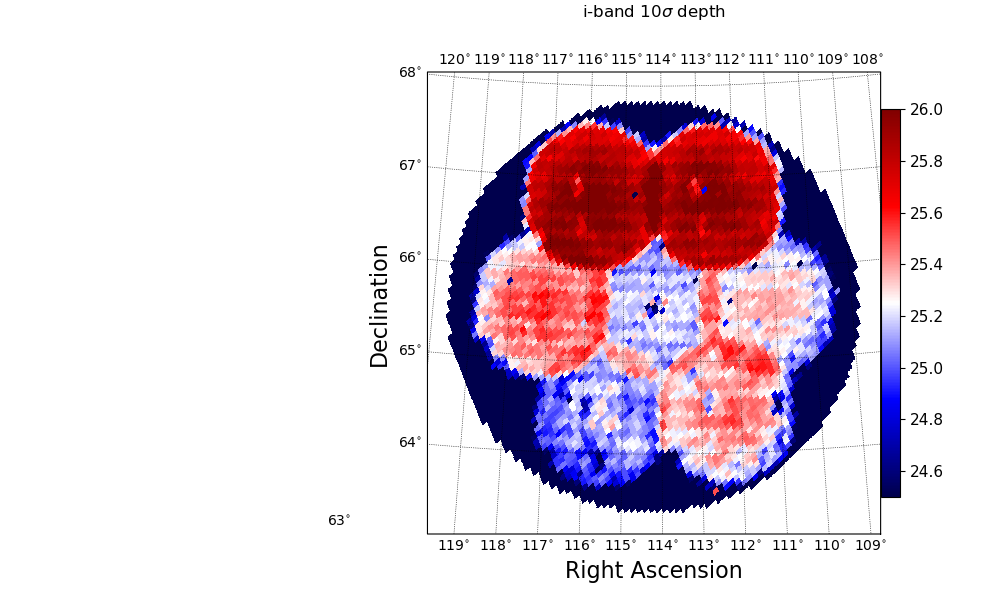}\\
\caption{The final $10$-$\sigma$ point source depths as a function of position from our Subaru/HSC dataset. The upper panel shows the $g$-band depth, and the lower panel depicts $i$-band depth. There is as much as $\sim1$~mag of variation in depth over the footprint, which is mostly due to different observing conditions (e.g., seeing; see Table~\ref{tab:obslog}) at the time the images were obtained.}
\label{fig:depth_maps}
\end{figure}

\section{The MADCASH survey} \label{sec:madcash}

In order to probe the satellite population of nearby LMC-mass galaxies, we have undertaken the Magellanic Analogs' Dwarf Companions and Stellar Halos (MADCASH) survey. MADCASH is an observational effort to obtain deep, resolved-star maps of the halos of isolated Magellanic Cloud (MC)-mass galaxies. To efficiently map large areas to depths at least 1-2 magnitudes below the RGB tip (TRGB) in nearby ($D \lesssim 4$~Mpc) MC analogs requires large field-of-view imagers on large-aperture telescopes. For MADCASH, we have used the Dark Energy Camera (DECam) on the Blanco 4-meter at CTIO, and Hyper Suprime-Cam (HSC) on the Subaru 8.2-meter in Hawaii. Figure~\ref{fig:lv_gxs} shows the absolute $K$-band luminosity versus distance for all galaxies from the Local Volume catalog of \citet{Karachentsev2013}\footnote{Available at \url{https://www.sao.ru/lv/lvgdb/}.} as small open circles, with colored, filled points representing objects targeted in our MADCASH survey (green squares and blue circles) and a sister program with DECam, the DELVE-DEEP survey \citep[][gray filled diamonds]{drlica-wagner2021}. The eleven highlighted points are the only targets with stellar masses within a factor of 3 of either the SMC or LMC, at distances where their individual RGB stars can be resolved, and accessible with either HSC or DECam.\footnote{While one can see in Figure~\ref{fig:lv_gxs} that there are more than 11 galaxies in the stellar mass and distance range of our sample selection, many of them do not make it into the final sample for a variety of reasons. Many are at declinations too far north to be observed with Subaru. In the southern sky, we limit the host sample to those within $D\lesssim2.5$~Mpc; more distant systems would require prohibitive exposure times with the 4-meter Blanco telescope. A few objects were removed because of high extinction along their line of sight, and one (NGC~404) was culled from the sample because it has a 2nd-magnitude star only a few arcminutes away.} The combined HSC and DECam datasets will map nearly the entire virial volumes of these 11 hosts. The properties of the systems we have observed with Subaru+HSC are given in Table~\ref{tab:targetlist}; DECam results will be presented in separate work. In this particular paper, we provide a systematic look at the results from a deep survey of the halo of NGC~2403 with HSC; a comprehensive census of satellites in all four systems from Table~\ref{tab:targetlist} will appear in future work.

\begin{figure}[!t]
\includegraphics[width=1.0\columnwidth, trim=0.0in 0.0in 0.25in 0.0in, clip]{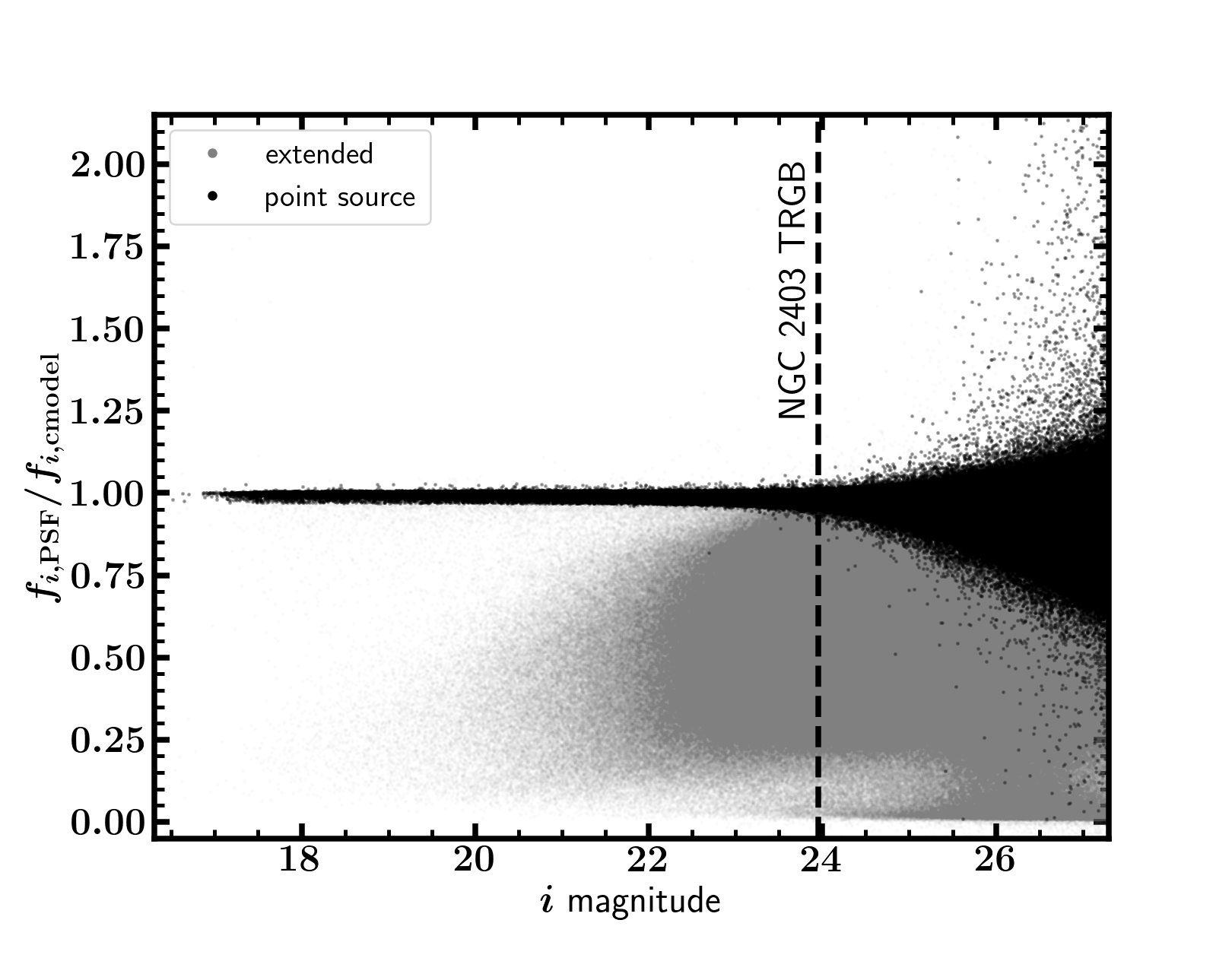}
\caption{An illustration of our method for star/galaxy separation. The ratio of point-spread-function (PSF) and \texttt{cmodel} fluxes is plotted against the $i$-band PSF magnitude. Point-like objects (i.e., ``stars'') should have $f_{\rm PSF}/f_{cmodel}\sim1$, while extended objects have more flux in the galaxy model measurement, scattering them to $f_{\rm PSF}/f_{cmodel} < 1$. Objects are classified as likely stars (highlighted in black in the diagram) if they are within $3\%$ (accounting for their measurement errors) of unity in this figure.}
\label{fig:stargalaxy}
\end{figure}

\section{Data and Analysis} \label{sec:data}

We observed seven pointings around NGC~2403 with Hyper Suprime-Cam \citep[HSC;][]{Furusawa2018, Kawanomoto2018, Komiyama2018, Miyazaki2018} on the Subaru 8.2-meter telescope at Mauna Kea (Hawaii); see Figure~\ref{fig:depth_maps} for a depiction of our data's photometric depth as a function of position. The $1.5^\circ$ diameter HSC field of view, corresponding to a projected diameter of $\sim80$~kpc at the distance of NGC~2403, enabled us to efficiently survey the surroundings of NGC~2403 out to large projected distances. The seven HSC pointings we obtained cover the entire area out to a projected radius of 90~kpc from the center of NGC~2403 (assuming $D=3.01$~Mpc as measured by \citealt{Carlin2016}), and $\sim60\%$ of the area between 90-110~kpc in projected distance. The estimated virial radius of NGC~2403 is $\sim140$~kpc \citep{mutlu-pakdil2021}; by examination of Figure~7 of \citet{Dooley2017}, we estimate that $\sim80\%$ of an LMC-mass host's satellites should be within a radius of 90~kpc.

Data were obtained on two separate observing runs -- the central, eastern, and western fields on 2016 February 9-10, and the northwest, northeast, southwest, and southeast fields on 2017 December 23-24. Skies were clear during both observing runs, with seeing between $0\farcs5-1\farcs4$ (but typically better than $0\farcs9$). To reach well below the TRGB at a distance of 3~Mpc, we observed sets of $10\times300$s exposures in $g$-band (called ``HSC-G'' at Subaru) and $10\times120$s in $i$-band (``HSC-I2''), as well as sequences of $5\times30$s exposures in each filter to prevent saturation of bright sources.  See Table~\ref{tab:obslog} for a log of our HSC observations.

The data were processed using the LSST Science Pipelines,\footnote{See \url{https://pipelines.lsst.io/index.html}.} which are being developed to process data from the Vera C. Rubin Observatory's Legacy Survey of Space and Time \citep[LSST;][]{Ivezic2019}. In particular, we used the weekly version of the Science Pipelines designated ``w\_2020\_42'' (i.e., the current version as of the 42nd week of 2020). A detailed description of the pipeline processing is presented in \citet[][see also \citealt{Bosch2019}]{Bosch2018}; processing includes instrument signature removal (e.g., bias correction, flat-fielding, etc.), a first round of source detection and matching to reference catalogs to derive photometric and astrometric calibration, point-spread function (PSF) fitting followed by refinements to the calibration, coaddition of overlapping frames, and finally detection and measurement of sources on the coadded images. The overall field covered by the data was assigned a single ``tract,'' which was subdivided into a grid of $25\times25$ ``patches'' of $4000\times4000$ pixels each. 

The full observed region is shown in Figure~\ref{fig:depth_maps}. The seven HSC pointings are clearly distinguishable in these maps, as well as their overlapping regions at the edges. In Fig.~\ref{fig:depth_maps}, the color-code represents the $10\sigma$ point source depth as a function of position (in $g$-band in the upper panel, and $i$-band in the lower panel). There is $\sim1$ magnitude of variation in the depth as a function of position, most of which can be attributed to varying conditions under which the data were obtained.

Results presented throughout this work are based on forced PSF photometry performed on the coadded images in each filter. The photometric and astrometric calibrations used PanSTARRS-1 (PS1; \citealt{Schlafly2012, Tonry2012, Magnier2013}) as a reference catalog. All data presented here have been corrected for extinction based on the \citet{Schlegel1998} dust maps, using the coefficients from \citet{Schlafly2011}.

\begin{figure}[!t]
\includegraphics[width=1.0\columnwidth, trim=0.25in 1.0in 0.25in 0.25in, clip]{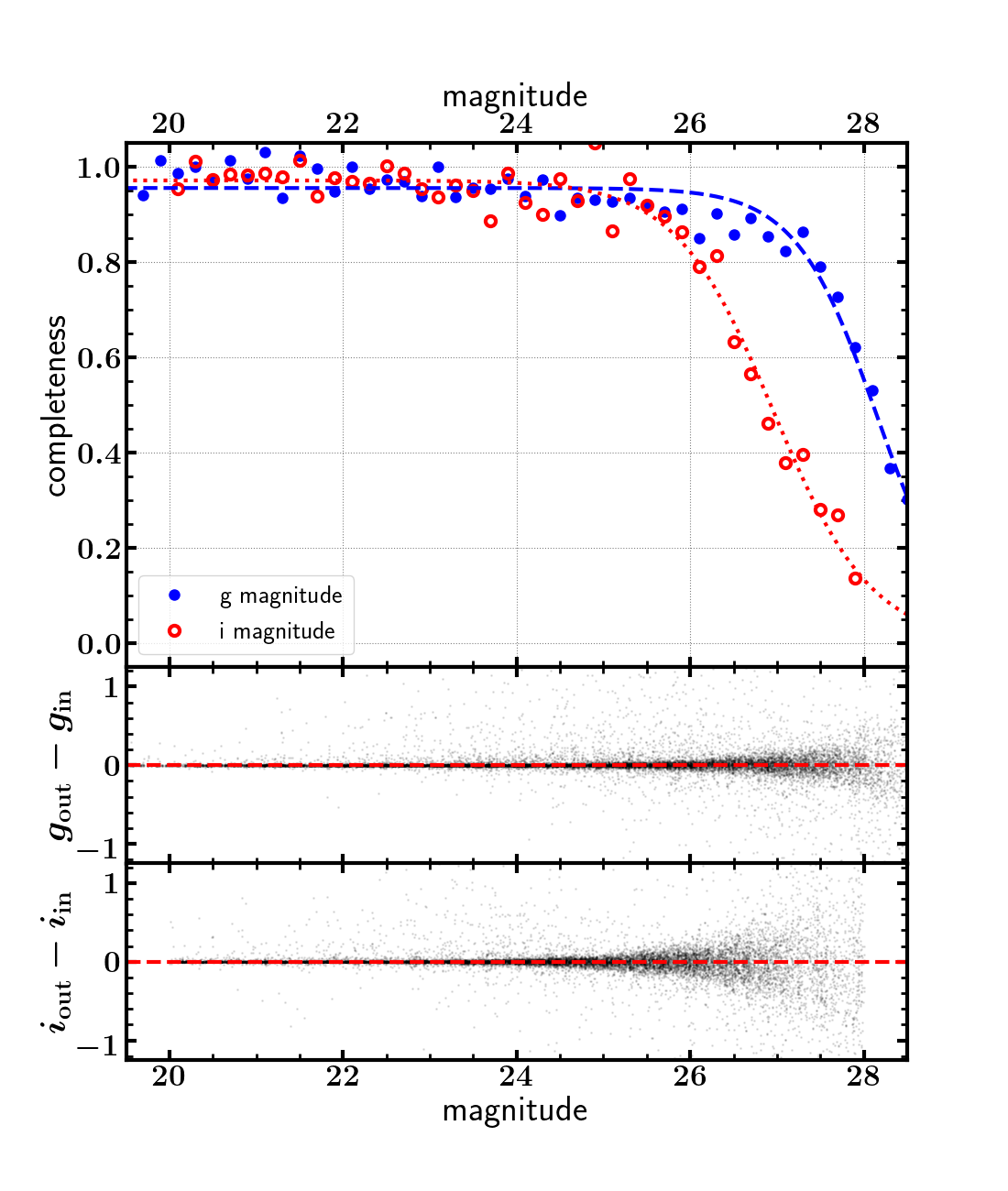}
\caption{Point-source completeness based on artificial star tests (see Section~\ref{sec:AST_completeness}) from the central 20~arcmin of the NE field. Blue and red points represent the $g$ and $i$-band completeness, respectively. The overlaid lines are the fits of Equation~\ref{eqn:compl} to the completeness values. The lower two panels compare the injected and measured magnitudes of artificial stars; there is no magnitude bias evident. The NE field is the deepest in both $g$- and $i$-band of all our 7 fields; the $50\%$ and $90\%$ completeness results from all fields are given in Table~\ref{tab:obslog} along with the $10\sigma$ point-source depths.}
\label{fig:completeness_NE}
\end{figure}

\subsection{Star-galaxy separation}

Figure~\ref{fig:stargalaxy} demonstrates our criteria for distinguishing point sources (i.e., ``stars'') from galaxies. The \texttt{cmodel} galaxy model fluxes should be equal to the PSF magnitudes for point sources, while the \texttt{cmodel} measurements capture the flux in the extended regions of resolved galaxies, whose fluxes thus diverge from PSF measurements of the same objects. We use the ratio of PSF to \texttt{cmodel} flux as a selection criterion to distinguish ``point-like'' sources from extended sources. Specifically, we select sources whose flux ratios are within $\pm0.03$ of unity (accounting for the uncertainties on their fluxes) in both the $g$ and $i$ bands. In Figure~\ref{fig:stargalaxy}, the point sources selected in this way are black points, while sources classified as extended are shown as lighter gray points.

\begin{table*}[!t]
\centering
\caption{{\it Subaru+HSC} observation log and field completeness.} \label{tab:obslog}
\begin{tabular}{lccccccc}
\tablewidth{0pt}
\hline
\hline
Field Name    & Filter &  Date & Exp\tablenotemark{a}  & seeing & 50\%\tablenotemark{b}  & 90\%\tablenotemark{c}  & 10-$\sigma$ depth \\
{}            &  {}    &  {}     & (s)  & (arcsec) & (mag) & (mag) & (mag)\\
\hline
NGC2403\_CEN	& HSC-G	& 09-Feb-2016 & 10$\times$300 & 0.9-1.4 & 26.6 & 25.1 & 26.5 \\
NGC2403\_CEN	& HSC-I2 & 09-Feb-2016 & 10$\times$120 & 0.65-0.9 & 26.0 & 24.5 & 25.2\\
NGC2403\_WEST & HSC-G & 10-Feb-2016 & 10$\times$300	& 0.6-0.75 & 27.7 & 26.4 & 26.9\\
NGC2403\_WEST & HSC-I2 & 09-Feb-2016 & 10$\times$120 & 	0.65-0.8 & 26.3 & 24.8 & 25.4\\
NGC2403\_EAST & HSC-G & 10-Feb-2016 & 10$\times$300	&	0.55-0.75 & 27.6 & 26.1 & 27.0  \\
NGC2403\_EAST & HSC-I2 & 10-Feb-2016 & 10$\times$120 & 	0.55-0.65 & 26.4 & 25.0 & 25.6  \\
NGC2403\_SE & HSC-G & 24-Dec-2017 & 10$\times$120 & 0.65-0.85 & 27.7 & 26.4 & 27.2 \\
NGC2403\_SE & HSC-I2 & 24-Dec-2017 & 10$\times$120 & 0.55-0.75 & 26.1 & 24.6 & 25.3 \\
NGC2403\_SW & HSC-G & 24-Dec-2017 & 10$\times$120 & 0.6-0.85 & 27.7 & 26.2 & 27.2 \\
NGC2403\_SW & HSC-I2 & 24-Dec-2017 & 10$\times$120 & 0.55-0.65 & 26.4 & 24.9 & 25.5 \\
NGC2403\_NE & HSC-G & 25-Dec-2017 & 10$\times$300 & 0.5-0.55 & 28.1 & 26.8 & 27.0 \\
NGC2403\_NE & HSC-I2 & 25-Dec-2017 & 10$\times$120 & 0.5-0.6 & 26.9 & 25.5 & 26.0 \\
NGC2403\_NW & HSC-G & 25-Dec-2017 & 10$\times$300 & 0.5-0.7 & 27.8 & 26.2 & 27.0 \\
NGC2403\_NW & HSC-I2 & 25-Dec-2017 & 10$\times$300 & 0.5-0.6 & 26.8 & 25.5 & 26.0 \\
\hline
\hline
\end{tabular}
\tablenotetext{a}{Total exposure time in seconds. We also took 5$\times$30s exposures in each field to increase the dynamic range at the bright end.}
\tablenotetext{b}{Magnitude at which the data are 50\% complete, based on artificial star tests.}
\tablenotetext{c}{Magnitude at which the data are 90\% complete.}
\end{table*}

\subsection{Point-source completeness via artificial star tests}\label{sec:AST_completeness}

To characterize the completeness of point-source detections in our images, we injected artificial stars into each patch of the coadded images. Artificial stars were randomly generated between magnitudes $20 < i < 28$ and with colors between $-1 < (g-i) < 2.5$, with weighting applied to generate more stars at fainter magnitudes than at the bright end. Positions were assigned randomly such that the stellar density of injected stars is on average $\sim15~{\rm arcmin}^{-2}$ throughout the entire observed field. The resulting images were then reprocessed in the same manner as the original processing, and the  measurement catalog was matched (with a 0.5~arcsec matching radius) and compared to the input list of artificial stars. Note that we did not apply star/galaxy separation criteria in this analysis, but concerned ourselves only with \textit{detection} completeness.

Figure~\ref{fig:completeness_NE} shows the results from the artificial star tests in the northeast (NE) field. The upper panel shows the completeness in $g$ (blue points) and $i$ (red points) as a function of magnitude. For similar completeness curves in each of the seven fields, we fit a function of the form:

\begin{equation}
\label{eqn:compl}
\eta(m) = \frac{A}{1 + \exp{(\frac{m-m_{50}}{\rho}})}
\end{equation}

\noindent as in \citet[][Eqn. 7]{Martin2016}, where $\eta(m)$ is the completeness as a function of magnitude, $A$ characterizes the plateau at bright magnitudes (typically near 1.0 as recovery of bright stars should be nearly complete), $\rho$ characterizes the steepness of the fall-off at the faint end, and $m_{50}$ represents magnitudes where $50\%$ of the injected stars were recovered (i.e., the $50\%$ completeness magnitude). Each 20-arcminute radius field contains $\sim18000$ artificial stars. The results from functional fits of Equation~\ref{eqn:compl} to curves in each field similar to that in  Figure~\ref{fig:completeness_NE} are summarized numerically in the $50\%$, $90\%$ completeness columns in Table~\ref{tab:obslog}, where we also provide the median $10\sigma$ limiting magnitude derived from the maps in Figure~\ref{fig:depth_maps}. There is significant variation in the depth over the observed footprint, which must be accounted for when estimating our sensitivity to the faint dwarfs we seek to detect.

\subsection{RGB candidate selection}

\begin{figure}[!t]
\includegraphics[width=1.0\columnwidth, trim=0.5in 0.0in 0.75in 0.0in, clip]{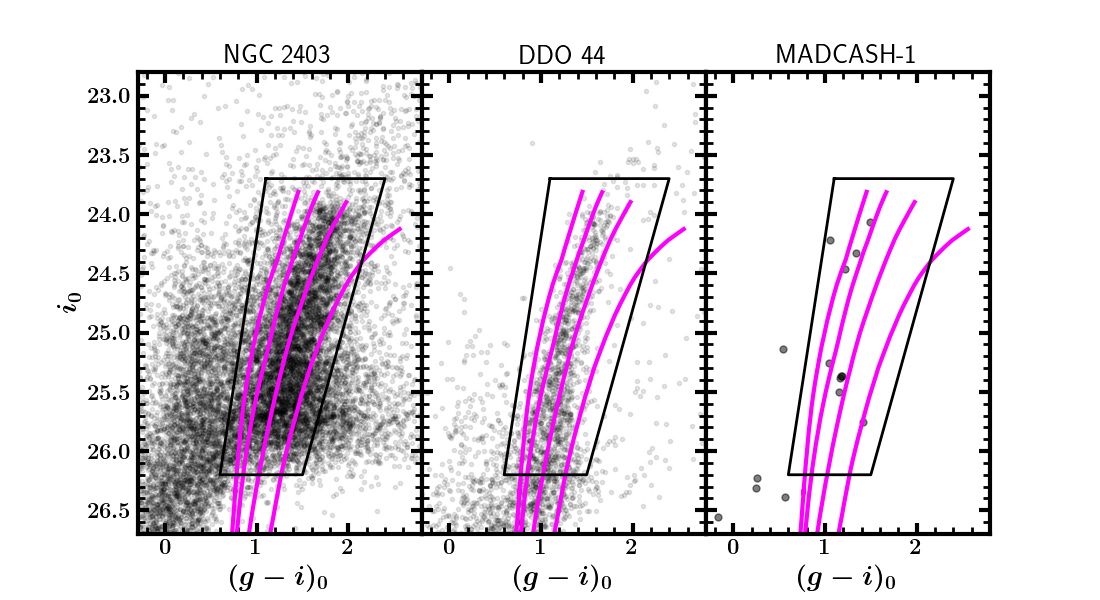}
\caption{RGB star selection. {\it Left:} Objects in an elliptical annulus between semimajor axes of $14 < a < 20$~arcmin of the center of NGC~2403, at a position angle of $124^\circ$ (Barker+2012). {\it Middle:} Between $2-4$~arcmin from the center of DDO~44. {\it Left:} Within $20$~arcsec of MADCASH-1. The black region in each panel outlines the RGB star selection. Isochrones in each panel are from \citet{bressan2012}, for metallicities of [M/H] = $-2.0, -1.5, -1.0$, and $-0.5$ and age 10~Gyr at a distance modulus of $m-M = 27.39$ \citep{Carlin2016}.
}
\label{fig:rgbsel}
\end{figure}

\begin{figure}[!t]
\includegraphics[width=1.0\columnwidth, trim=0.5in 0.0in 0.0in 0.0in, clip]{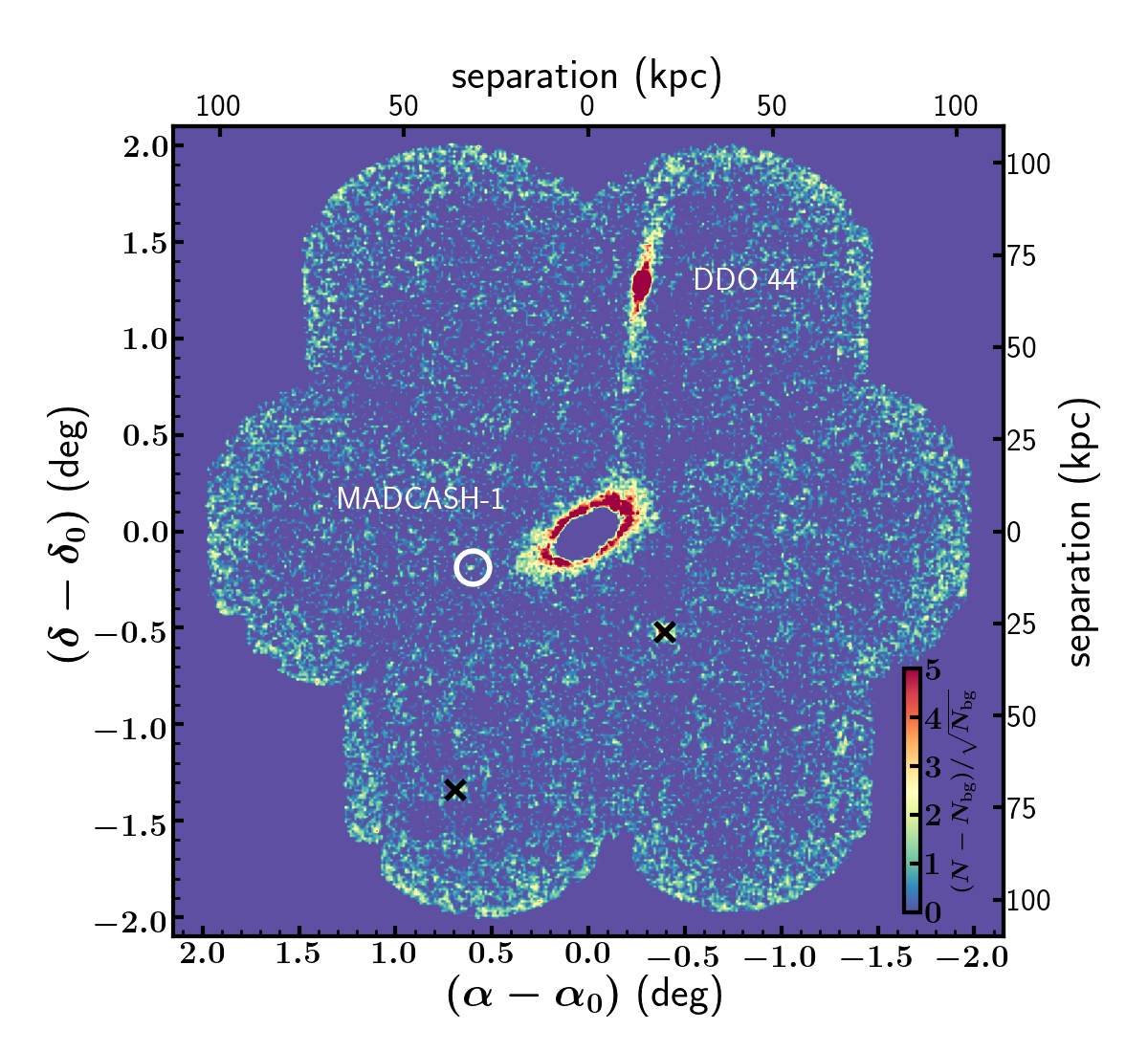}
\caption{Map of selected RGB star candidates, shown here as the RGB number density ``signal-to-noise'' above the local background, $(N-N_{\rm bg})/\sqrt{N_{\rm bg}}$. DDO~44 and its tidal stream \citep{Carlin2019} are prominent in the northern region of this panel. The faint satellite MADCASH-1 \citep{Carlin2021} is visible (and circled in white) as a small overdensity to the east (and slightly south) of the main body of NGC~2403. The black cross ($\times$) to the SE of NGC~2403 is MCG+11-10-022, and the one slightly SW is UGC-3894. The outskirts of NGC~2403 are otherwise fairly featureless, with no obvious stellar streams or tidal debris structures visible. A rough estimate based on the background density and mean flux of RGB candidates suggests that these data should be sensitive to streams with surface brightnesses as faint as $\mu_{\rm V} \sim 32$~mag~arcsec$^{-2}$.}
\label{fig:rgbmap}
\end{figure}

Figure~\ref{fig:rgbsel} demonstrates our adopted color-magnitude criteria for selecting candidate red giant branch (RGB) stars at the distance of NGC~2403. We select stars that are consistent with being metal-poor RGB stars at or near the distance of NGC~2403 (i.e., roughly consistent with the isochrones overlaid on the CMDs in Figure~\ref{fig:rgbsel}) within a box in color-magnitude space. We choose a box rather than an isochrone filter because at the distance of NGC~2403, many stars in potential dwarf galaxies may have their fluxes contaminated by unresolved flux from the dwarf galaxy's stars below the detection limit (see, e.g., \citealt{garling2021}), and thus scatter their colors and magnitudes by much more than the magnitude uncertainties estimated by automated measurement algorithms. Although the color and magnitude errors increase at the faint end, we choose to narrow the selection box to avoid as many as possible of the unresolved galaxies that dominate the number counts of objects at the faint end. These selection criteria were chosen in part by trial and error, seeking the selection criteria that maximize the signal-to-noise of known dwarfs MADCASH~1 and DDO~44 (and its tidal stream; see \citealt{Carlin2019}). 

Figure~\ref{fig:rgbmap} shows a map of the signal-to-noise (i.e., significance above the background) of surface density of RGB stars selected by our technique (a similar map that was shown in \citealt{Carlin2019} was generated from an earlier processing of the data). NGC~2403 is visible in the center of the field, with very few other obvious features. The previously discovered MADCASH dwarf \citep{Carlin2016} is barely visible to the east (and slightly south) of the NGC~2403 disk. We mark two large and bright background galaxies that erroneously contribute to our RGB sample with black ``x'' symbols. Otherwise, the only readily visible feature is the stellar tidal stream emanating from the massive satellite DDO~44, which can be traced nearly the entire $\sim70$~kpc projected distance between DDO~44 and NGC~2403. See \citet{Carlin2019} for more details about this stellar stream. From this map, we conclude that there are no other obvious stellar streams or tidal debris features in the halo of NGC~2403 to be followed up.

\subsection{Search for dwarf galaxies}

To search for candidate dwarf galaxies as overdensities of resolved stars, we adopt a simple approach. We first divide the footprint of our HSC observations into bins of 1.0~arcmin on a side, which corresponds to $\sim870$~pc at the distance of NGC~2403. For each bin, we count the number of point sources within the RGB selection box -- this constitutes the signal within each bin. The local background level is estimated by extracting the number of candidate RGB stars in bins whose centers lie in an annulus spanning a radius of 2-5 times the bin size (i.e., 2-5 arcmin) from the bin, then scaling this number to account for the different areas of the annulus and the bin. Annuli that are only partially within the observational footprint (i.e., at the outer edge of the coverage) are corrected for the fraction of the annulus that is missing. Partial annuli are also necessary near the center of NGC~2403, where the extremely high number counts of stars would create artificially high background values (we also excluded bins within 15 arcmin of the center of NGC~2403 from our search). Bins with number counts that are higher than the local background by more than 3.5~times the standard deviation of the counts in the background bins are considered candidate detections.

This search yields a total of 227 candidate overdensities. Of those, 42 are associated with the main body of NGC~2403, and are thus excluded from further consideration. Another 36 detections are associated with DDO~44 and its tidal stream; these are also removed from the sample, leaving 149 candidates for more detailed examination. We generated diagnostic plots
as seen in the Appendix in Figures~\ref{fig:ddo44_diag} through \ref{fig:fakedwarf_diag_sb29}
for all candidates, and examined each of them for evidence of a dwarf galaxy. The visual examination is critical, as any one of the panels (CMD, spatial distribution, radial density profile, image, and luminosity function) can contain features suggesting the presence of a candidate dwarf. 
Take, for example, the previously known dwarf MADCASH~1 \citep{Carlin2016}, whose diagnostic plots are seen in Figure~\ref{fig:madcash1_diag}. 
This candidate's RGB stars roughly delineate a metal-poor RGB in the CMD, the stars of which are concentrated in the spatial plot, and are visible in the image (including some unresolved, low surface brightness emission), and the luminosity function clearly differs from that of the background region.
The candidate in Figure~\ref{fig:falsedet_diag}, on the other hand, shows no clear feature in the image or the spatial distribution, and its stars do not lie along a sequence consistent with being (and having the luminosity function of) an RGB at the distance of NGC~2403. This candidate is thus an obvious false detection.

From the 149 candidates we examined, we identified only the known dwarf MADCASH~1 as a worthwhile candidate. Some candidates have enticing features in the diagnostic plots, but none make a convincing enough case to be considered bona fide dwarfs. It thus appears that the only NGC~2403 satellites detected in our dataset are MADCASH-1 and DDO~44.

\begin{figure}[!t]
\includegraphics[width=1.0\columnwidth, trim=0.0in 0.0in 0.0in 0.0in, clip]{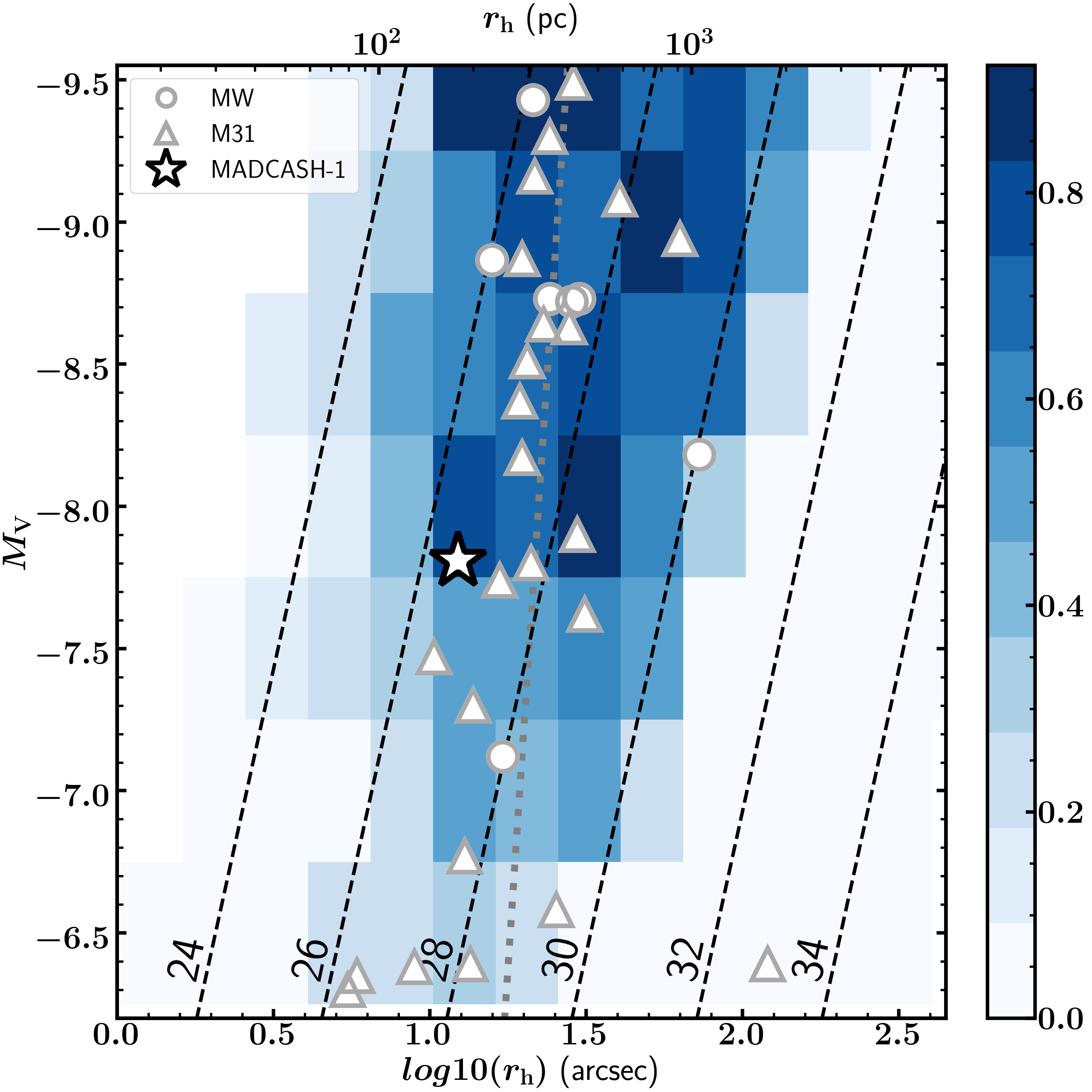}
\caption{Completeness of dwarf galaxy detection based on the injection of $>3000$ artificial dwarfs ($>40$ per bin in the figure). For comparison, we show the positions of known Milky Way (circles) and M31 (triangles) dwarfs from the Local Volume Database. NGC~2403's satellite MADCASH-1 is shown as a large star with black outline; DDO~44 is much brighter than the upper limit we explored with artificial dwarfs, and thus outside the boundaries of this figure. Diagonal lines represent constant surface brightness values, ranging from 24 to 32 mag~arcsec$^{-2}$. The gray dotted line is the interpolated $M_{\rm V}-r_{\rm half}$ relation from \citet{brasseur2011}, which we use to extract the luminosity function corrections applied to Figure~\ref{fig:lf}. Our dataset and search sensitivity should recover most typical dwarfs brighter than $M_{\rm V} < -7.0$ and $\mu_{\rm V} \lesssim 30~{\rm mag~arcsec}^{-2}$, and some fraction at even fainter absolute magnitudes. }
\label{fig:dsph_completeness}
\end{figure}

\begin{figure*}[!t]
\begin{center}
\includegraphics[width=0.32\textwidth, trim=0.5in 0.5in 0.5in 0.25in, clip]{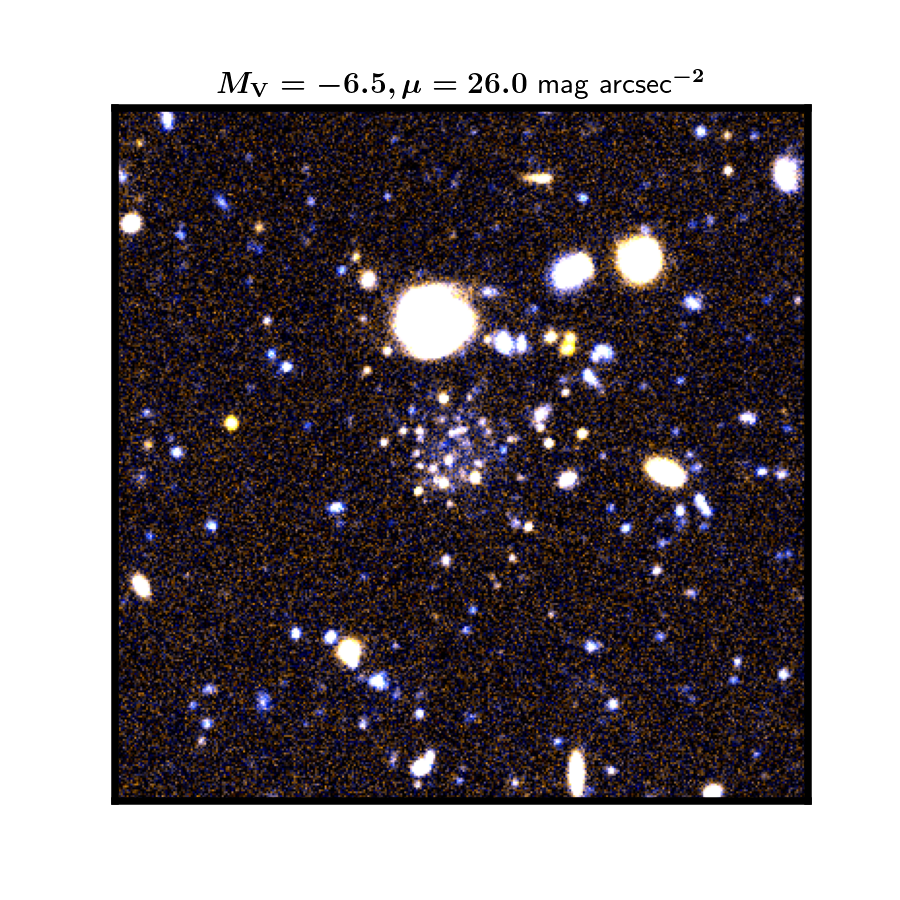}
\includegraphics[width=0.32\textwidth, trim=0.5in 0.5in 0.5in 0.25in, clip]{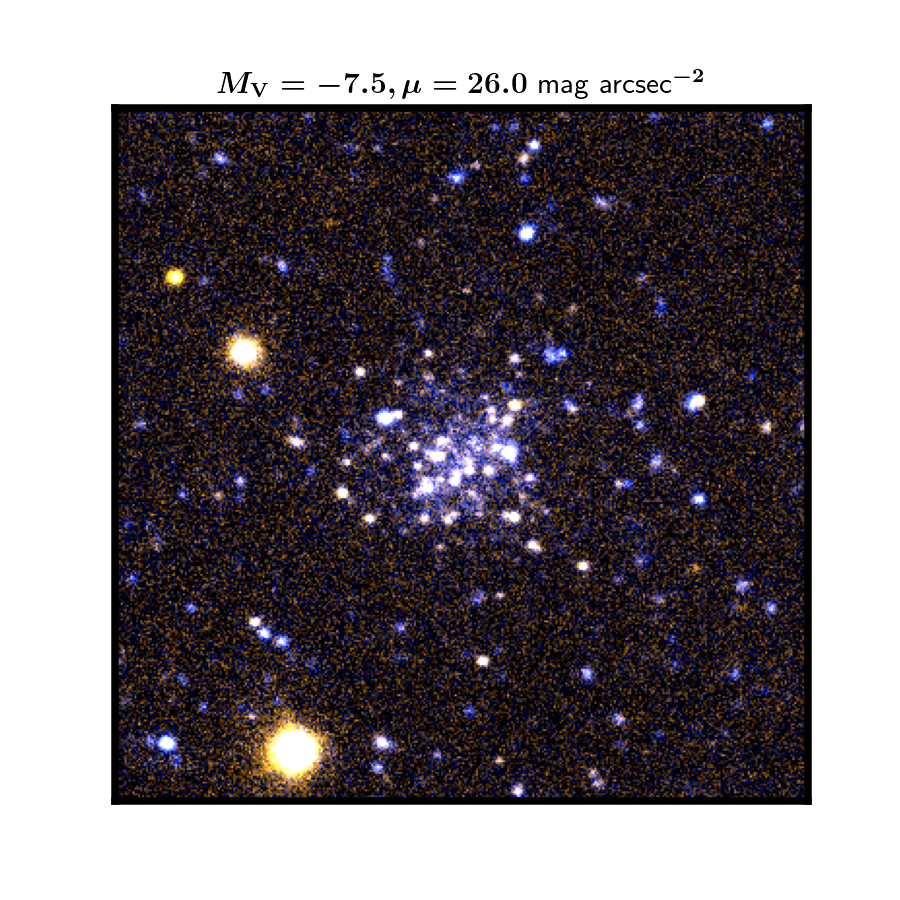}
\includegraphics[width=0.32\textwidth, trim=0.5in 0.5in 0.5in 0.25in, clip]{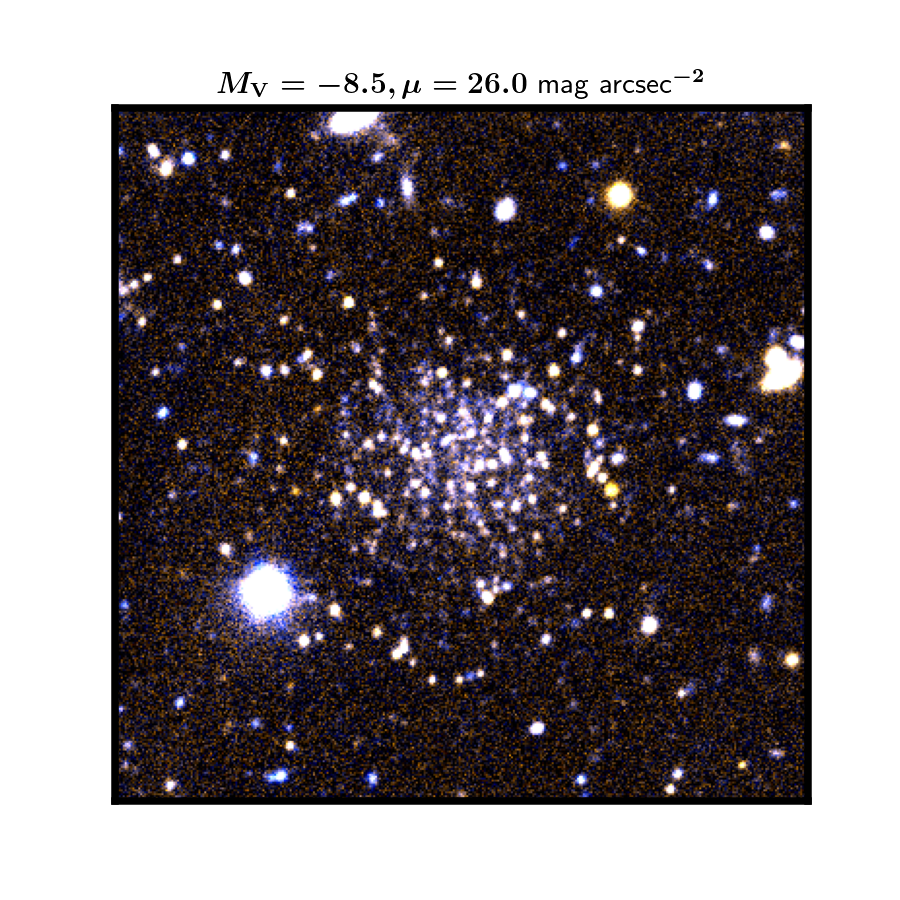}\\
\includegraphics[width=0.32\textwidth, trim=0.5in 0.5in 0.5in 0.25in, clip]{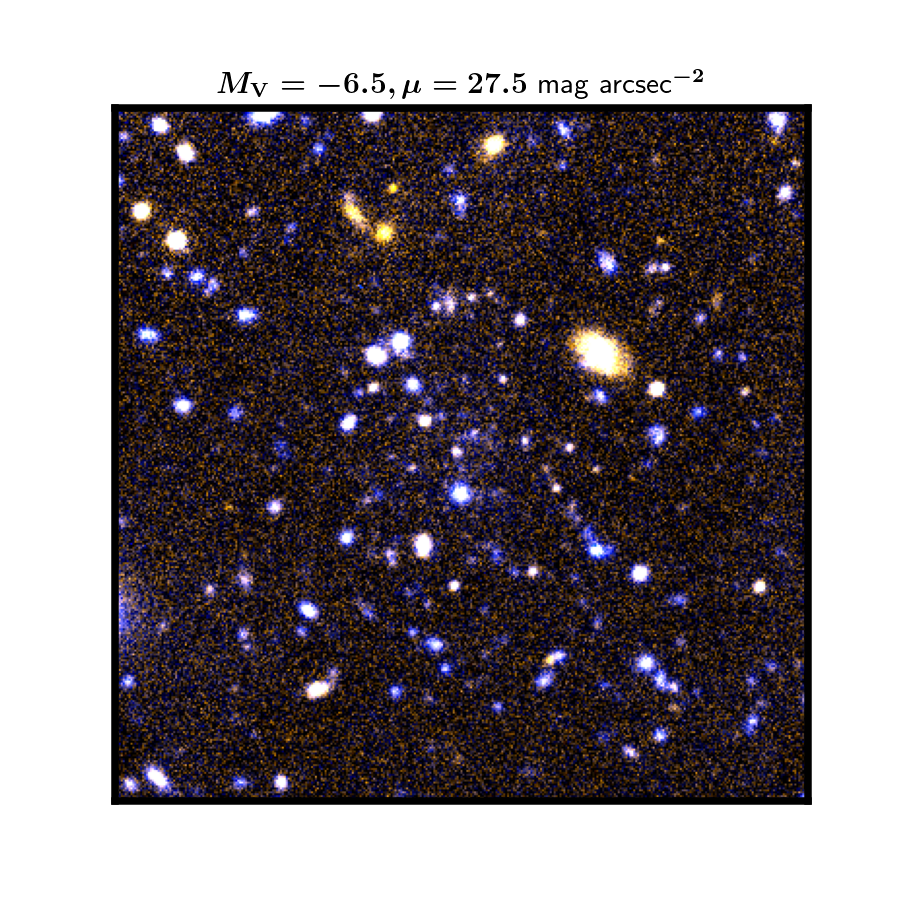}
\includegraphics[width=0.32\textwidth, trim=0.5in 0.5in 0.5in 0.25in, clip]{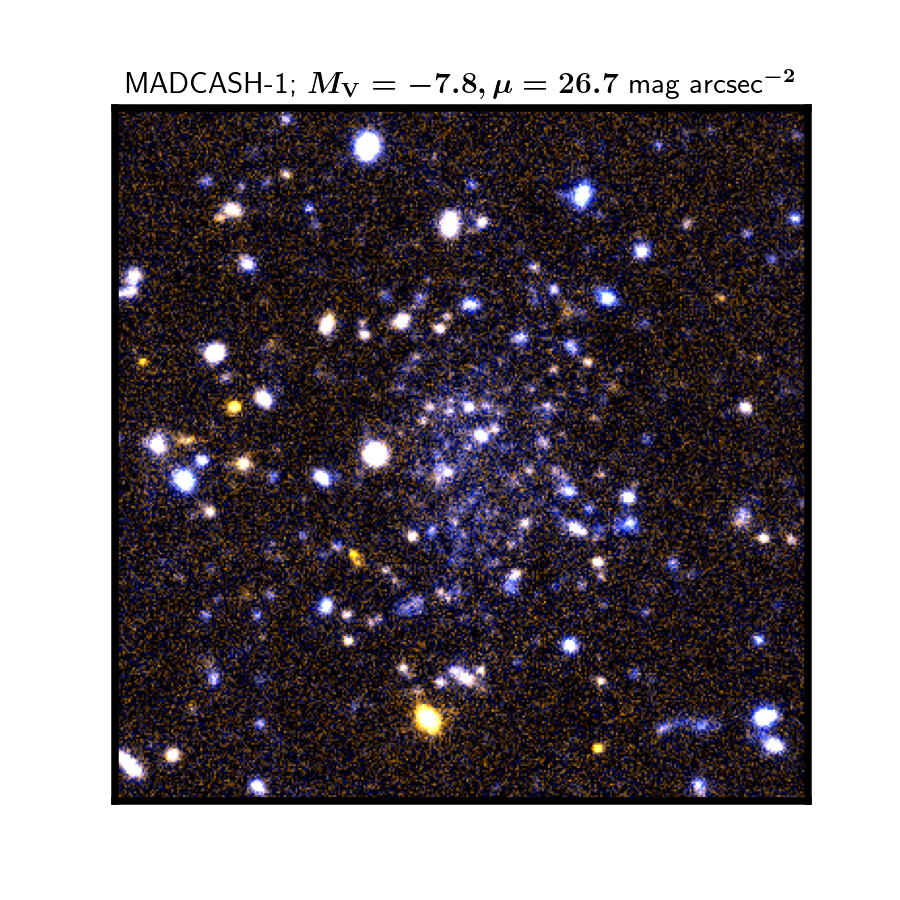}
\includegraphics[width=0.32\textwidth, trim=0.5in 0.5in 0.5in 0.25in, clip]{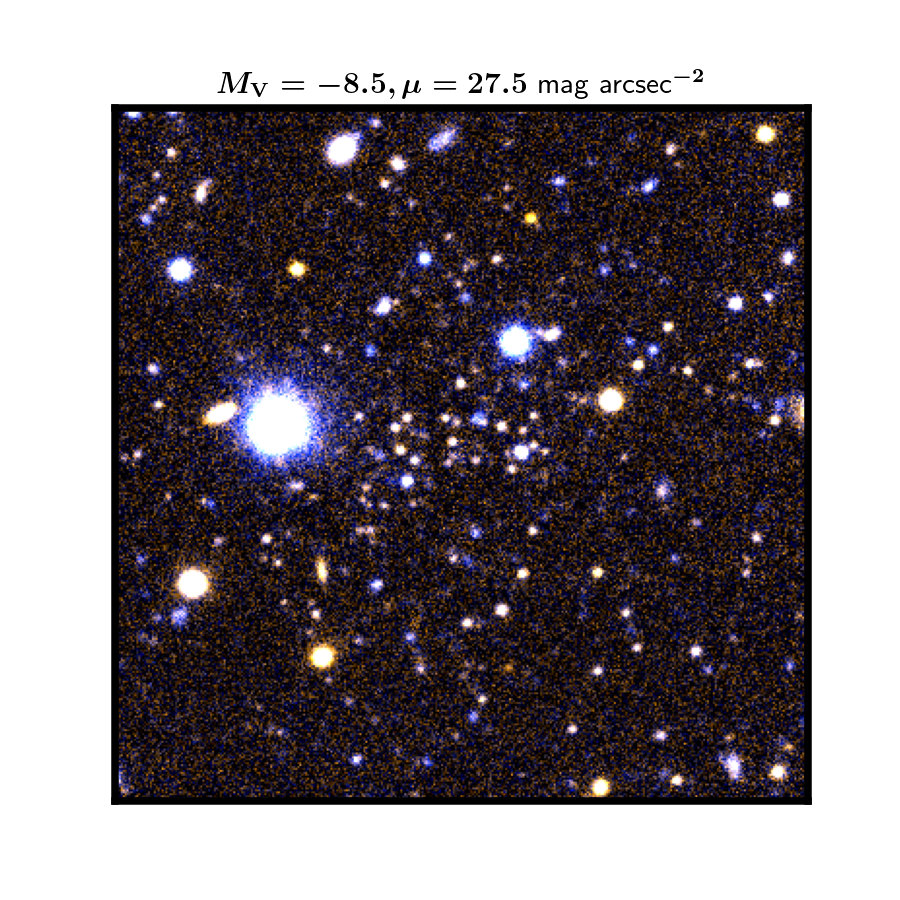}\\
\includegraphics[width=0.32\textwidth, trim=0.5in 0.5in 0.5in 0.25in, clip]{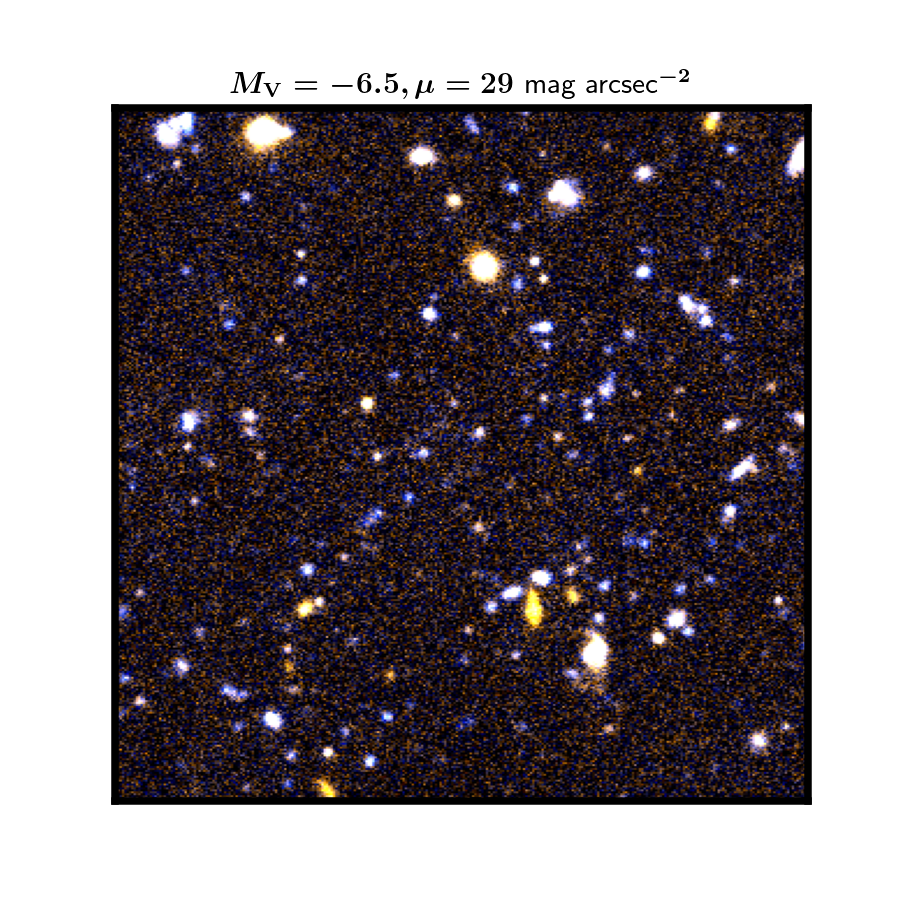}
\includegraphics[width=0.32\textwidth, trim=0.5in 0.5in 0.5in 0.25in, clip]{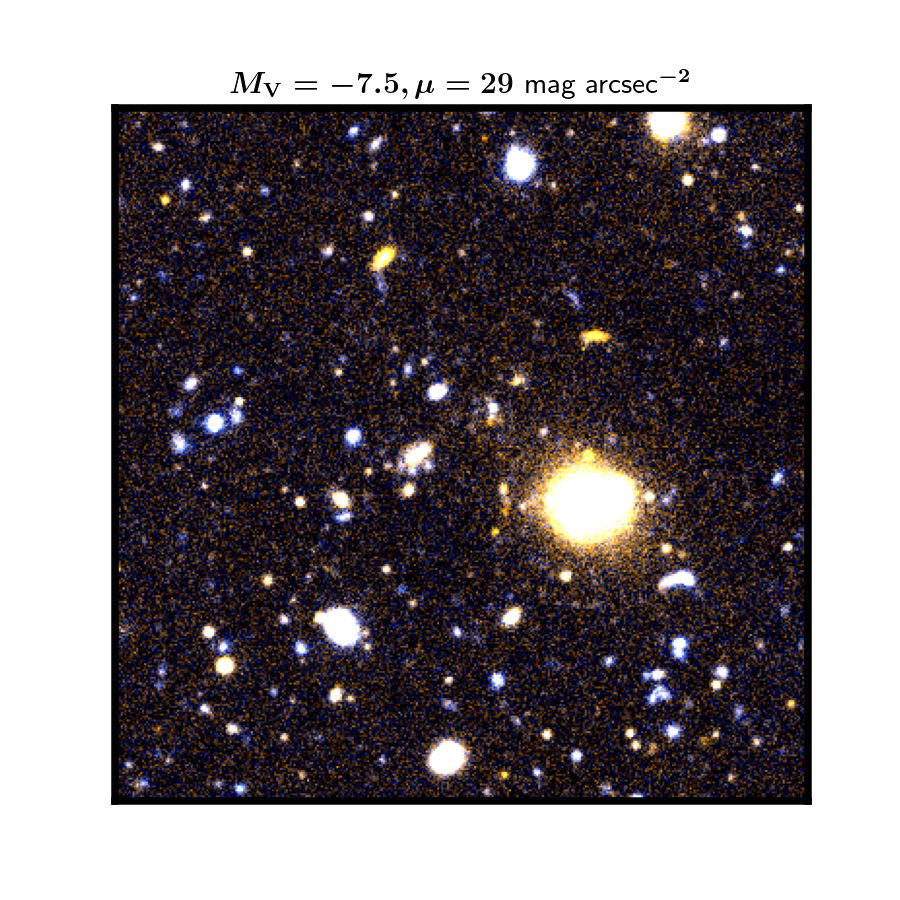}
\includegraphics[width=0.32\textwidth, trim=0.5in 0.5in 0.5in 0.25in, clip]{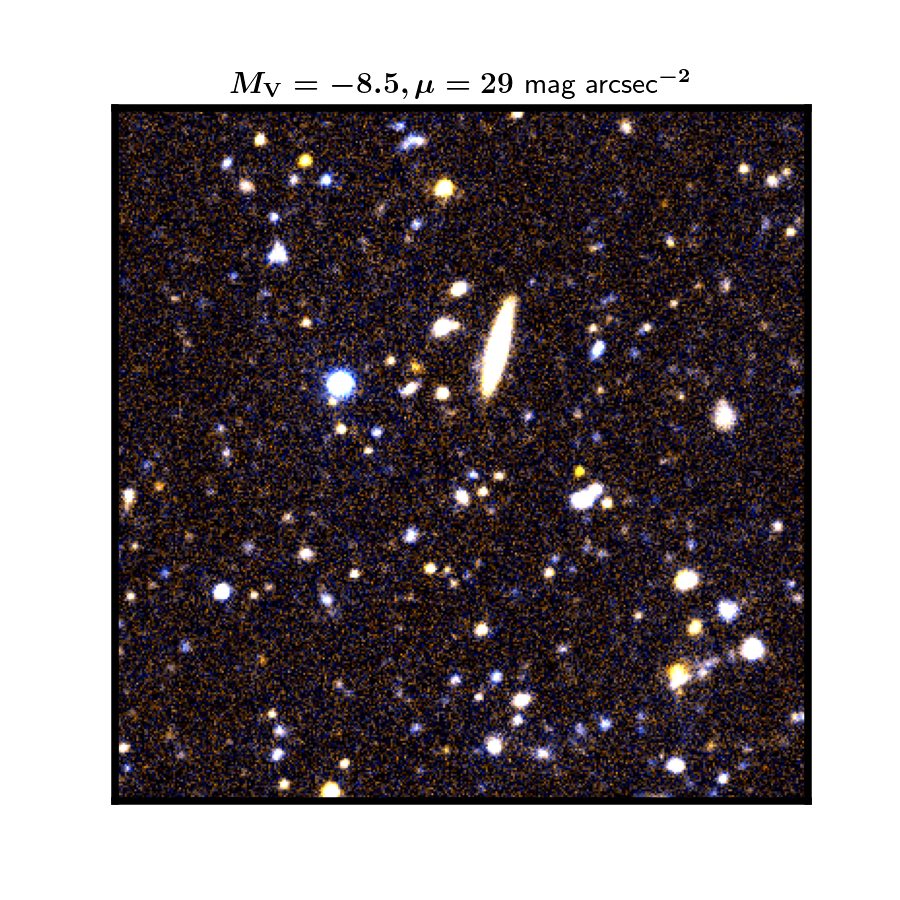}\\
\caption{Example images of synthetic dwarfs that were injected into the images to measure our recovery rate. Images are one arcminute on a side, corresponding to a projected physical width of $\sim870$~pc at a distance of 3~Mpc. Each row shows dwarfs with fixed values of surface brightness, with $\mu = 26.0, 27.5, 29.0$~mag~arcsec$^{-2}$ in the top, middle, and bottom rows. From left to right, injected dwarfs have $M_{\rm V} = -6.5, -7.5, -8.5$ in each row. However, the middle panel has been replaced with an image of the MADCASH-1 dwarf, which has $\mu = 26.7$~mag~arcsec$^{-2}$ and $M_{\rm V} = -7.8$. All of the dwarfs shown in this figure were recovered by our search. Our injected dwarf recovery fractions for similar dwarfs were 23\%, 34\%, and 63\% for the top panels, 26\% and 67\% for the left and right-hand panels in the middle row, and 26\%, 57\%, and 72\% for the dwarfs in the bottom row. The bin in Figure~\ref{fig:dsph_completeness} corresponding to MADCASH-1 had a recovery rate of 74\%.}
\label{fig:fakedwarf_images}
\end{center}
\end{figure*}

\begin{figure}[!t]
\includegraphics[width=1.0\columnwidth, trim=0.25in 0.2in 0.0in 0.0in, clip]{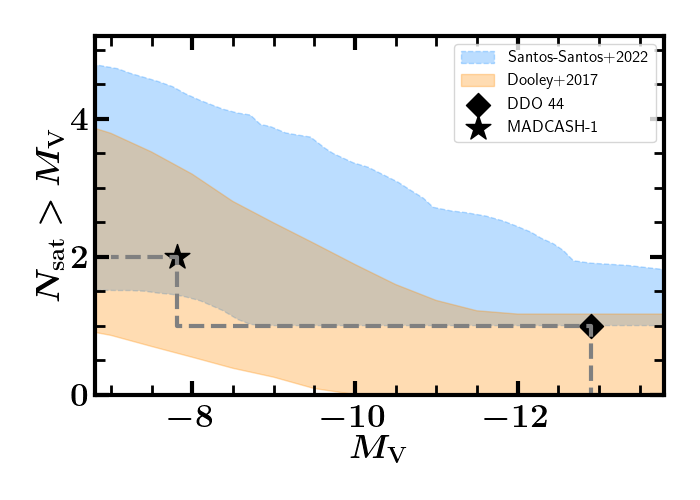}
\caption{Satellite luminosity function of NGC~2403. The filled bands represent predicted satellite populations for LMC-mass hosts from \citet[][blue]{santos-santos2022} and \citet[][orange]{Dooley2017}. The shaded regions for both comparisons represent the $1\sigma$ variation in the number of satellites over the model realizations. The predicted luminosity functions have been corrected for dwarf detection completeness as well as a volume correction to account for the observational incomplete coverage of the virial volume (see Section~\ref{sec:discussion} for details).}
\label{fig:lf}
\end{figure}

\section{Completeness via artificial dwarf galaxy tests} \label{sec:fakes}

To characterize our sensitivity to dwarf galaxies as a function of position, luminosity ($M_{\rm V}$), and size ($r_{\rm half}$; and by extension, surface brightness), we generate simulated dwarf galaxies, inject them into the images, and re-run the detection and measurement pipelines. Both our search for overdensities of RGB stars and the injected synthetic dwarfs focused on old, metal-poor stellar populations. This is a reasonable assumption given the fact that even dwarfs with recent star formation (or intact gas reservoirs) formed the majority of their stellar mass at early times (e.g., \citealt{Weisz2011}), and should thus be detectable via their old, metal-poor RGB stars.

The dwarfs were created by sampling Padova isochrones \citep{bressan2012} fixed at a 10~Gyr age and low metallicity ([Fe/H]$ = -2.0$), with a \citet{chabrier2001} log-normal luminosity function. We sampled in luminosity between $-9.5 < M_{\rm V} < -6.0$ in 0.5-mag increments, and in surface brightness (within $r_{\rm half}$) spanning $23 < \mu_{\rm V} < 34~{\rm mag~arcsec}^{-2}$ in increments of 0.5 mag~arcsec$^{-2}$ (we implicitly assume that brighter dwarfs around NGC~2403 would have been detected in prior work). Dwarfs were simulated with \citet{plummer1911} distributions, with $r_{\rm half}$ implied by the combination of luminosity and surface brightness. Dwarfs were placed at a distance of 3.0~Mpc. To reduce the computational burden, simulated stars fainter than $r = 33$~mag were removed; at 3.0~Mpc, this cut removes a few percent of the total flux of the injected dwarfs, but we confirmed via inspection of the images that the effect was not visually noticeable.
We randomly position between 2-5 dwarfs in each $4000\times4000$-pixel ``patch'' (varying the number of inserted dwarfs per patch to avoid predictable results). Dwarfs are injected as groups of individual stars using the version of \texttt{Synpipe} \citep{huang2018} that was integrated into the LSST science pipelines at the time the NGC~2403 data were processed. Example dwarf galaxy injections into the image-level data are shown in Figure~\ref{fig:fakedwarf_images}, along with the actual appearance of MADCASH-1 for comparison.

A total of 3139 artificial dwarfs were injected into the images over two separate runs; this was accomplished by randomly generating between 2-5 artificial dwarfs per 4k$\times$4k ``patch'' for each of the 394 patches in the coadded image data (the maximum of 5 dwarfs per patch was chosen to avoid overlapping of injected dwarfs). The data were fully reprocessed, and the dwarf search code run on the resulting catalogs. Dwarf detection completeness results are seen in Figure~\ref{fig:dsph_completeness}, where each bin in the luminosity-size plane had $\sim40$ artificial dwarfs injected. The completeness in each bin is simply the number detected by our algorithm divided by the total number injected in that bin. For context, we overlay known nearby dwarf galaxies from the Local Volume Database.\footnote{\url{https://github.com/apace7/local_volume_database }} Our analysis shows that we are nearly complete (i.e., $\gtrsim80\%$ detected) for dwarfs at $M_{\rm V} < -7.5$ and $\mu_{\rm V} \lesssim 30~{\rm mag~arcsec}^{-2}$, with sensitivity extending as faint as $M_{\rm V} = -6.5$. 
As shown in Figure~\ref{fig:depth_maps}, there is significant variation in depth over the observed field of view, such that Figure~\ref{fig:dsph_completeness} represents an average completeness over the full dataset. Thus a conservative estimate of the faintest dwarf to which we are reliably sensitive is $M_{\rm V} \sim -7.5$.

\begin{deluxetable*}{lcccccccccc}
\tablewidth{0pt}
\tablecaption{Properties of the satellite system of NGC 2403.}
\label{tab:n2403_sats}
\tablehead{\colhead{galaxy} & \colhead{RA} & \colhead{Dec} & \colhead{$D_{\rm TRGB}$} & \colhead{$M_{\rm V}$} & \colhead{$r_{\rm half}$} & \colhead{$r_{\rm half}$} & \colhead{$M_{\rm *}$} & \colhead{$D_{\rm proj}$} & \colhead{References} \\
\colhead{} & \colhead{(deg)} & \colhead{(deg)} & \colhead{(Mpc)} & \colhead{(mag)} & \colhead{(pc)} & \colhead{(arcsec)} & \colhead{($M_\odot$)} & \colhead{(kpc)} & \colhead{} \\
\colhead{(1)} & \colhead{(2)} & \colhead{(3)} & \colhead{(4)} & \colhead{(5)} & \colhead{(6)} & \colhead{(7)} & \colhead{(8)} & \colhead{(9)} & \colhead{(10)}}
\startdata
DDO 44 & 113.5479 & 66.8797 & $2.96\pm0.1$ & -12.9 & $740\pm20$ & $52\pm2$ & $2\times10^7$ & $\sim70$ & 1, 2, 3 \\
MADCASH-1 & 115.6642 & 65.4167 & $3.41^{+0.24}_{-0.23}$ & $-7.81\pm0.18$ & $179^{+30}_{-28}$ & $10.8\pm1.0$ & $(1.8\pm0.3)\times10^5$ & $\sim35$ & 4, 5 \\
\enddata
\tablecomments{(1) Galaxy name; (2) Right Ascension (J2000.0); (3) Declination (J2000.0); (4) TRGB distance; (5) absolute V-band magnitude; (6) physical half-light radius; (7) half-light radius on the sky; (8) stellar mass, estimated assuming $M_*/L_{\rm V} = 1.6$, a typical value for dSphs \citep{Woo2008}; (9) projected distance from NGC~2403; (10): references: 1. \citet{Whiting2007}, 2. \citet{Jerjen2001}, 3. \citet{Carlin2019}, 4. \citet{Carlin2016}, 5. \citet{Carlin2021}}
\end{deluxetable*}

\section{Discussion and Conclusions}\label{sec:discussion}

Figure~\ref{fig:lf} shows the luminosity function (LF) of NGC~2403 dwarf satellites, which consists of only two systems: DDO~44 ($M_{\rm V} = -12.9;$ \citealt{Carlin2019}) and MADCASH-1 ($M_{\rm V} = -7.8;$ \citealt{Carlin2021}); the properties of these dwarfs are summarized in Table~\ref{tab:n2403_sats}. 
For comparison we overlay the predicted satellite population for an LMC-mass host as derived by \citet{santos-santos2022} as a blue-filled region, and from \citet{Dooley2017} as an orange-shaded region. Both of these predicted satellite luminosity functions were corrected for the dwarf detection completeness of our HSC data as estimated in Section~\ref{sec:fakes} and Figure~\ref{fig:dsph_completeness}. We did not incorporate the full dependence of our dwarf sensitivity on $M_{\rm V}$ and $r_{\rm half}$, but rather interpolated along the locus occupied by MW dwarfs in the $M_{\rm V}$--$r_{\rm half}$ plane (as measured by \citealt{brasseur2011}). The predicted luminosity functions in Figure~\ref{fig:lf} were then simply scaled based on the dwarf sensitivity at each $M_{\rm V}$ value that was extracted along the \citet{brasseur2011} size-luminosity curve. Finally, we applied an additional reduction by a factor of 0.8 at all luminosities to account for our incomplete coverage of the virial radius of NGC~2403. This factor was estimated from Figure~7 of \citet{Dooley2017}, which shows the cumulative number of satellite dwarfs as a function of radius from an LMC-mass host; our complete coverage to $\sim90$~kpc should encompass $\sim80\%$ of the NGC~2403 satellites.

In Figure~\ref{fig:lf} we compare the properties of our observed NGC~2403 satellites to two $\Lambda$CDM-based predictions for satellite populations of LMC-mass hosts. The orange shaded region represents the $1\sigma$ scatter in predicted satellites from the work of \citet{Dooley2017}, which is based on the Caterpillar simulations \citep{griffen2016} and the stellar mass-halo mass relation of \citet{garrison-kimmel2017}. The blue filled region shows predicted satellites for LMC analogs from \citet{santos-santos2022}, whose work was based on the highest-resolution models from the APOSTLE simulations suite \citep{fattahi2016}. The region we include here for comparison corresponds to the ``cut-off'' model from \citet{santos-santos2022}, which predicts far fewer satellites than a standard power-law. The predictions from both models align well with the two observed satellites of NGC~2403.

\begin{figure}[!t]
\includegraphics[width=1.0\columnwidth, trim=0.25in 0.1in 0.25in 0.25in, clip]{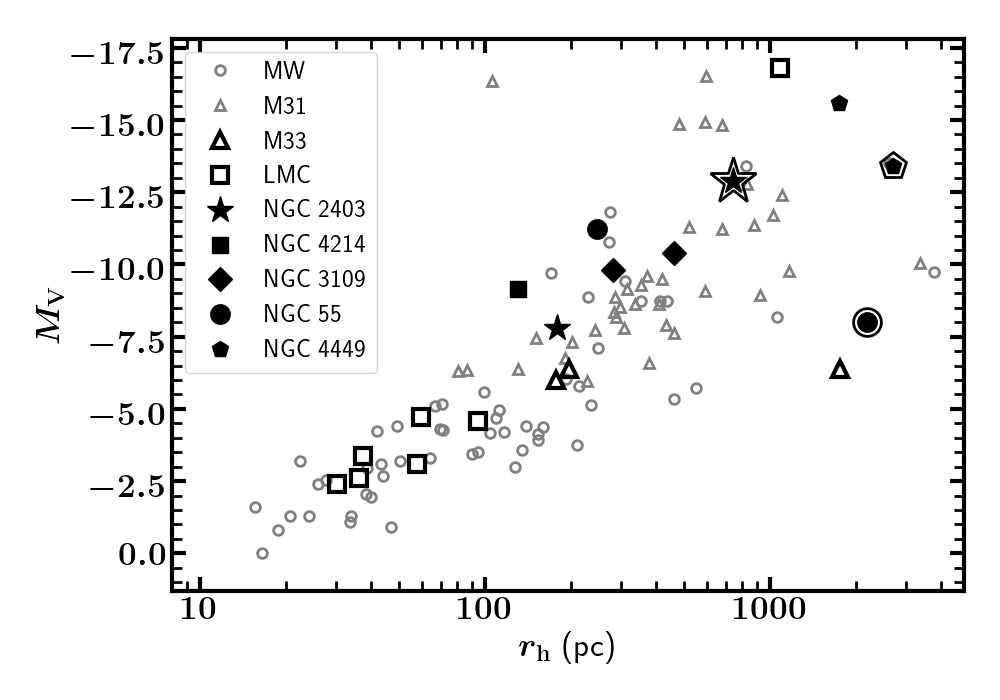}
\caption{Luminosity ($M_{\rm V}$) vs. half-light radius ($r_{\rm h}$) for known satellites of MC-mass hosts in the Local Volume. Open gray symbols show Milky Way and M31 satellites, with open black squares and triangles depicting (candidate) LMC and M33 satellites, respectively. Solid black symbols represent dwarf satellites of nearby ($1 < D < 4$~Mpc) MC-mass host galaxies; for NGC~2403 (MADCASH~1, DDO~44), NGC~4214 (MADCASH~2), NGC~3109 (Antlia, Antlia~B), NGC~55 (NGC~55-dw1, ESO~294-G010), and NGC~4449 (NGC~4449B, DDO~125). Symbols that are outlined are dwarfs that are known to be tidally disrupting.}
\label{fig:mv_rh}
\end{figure}

The LMC itself, while known to have many satellites, has only one confirmed satellite (the SMC) that is not an ultra-faint dwarf (i.e., brighter than $M_{\rm V} \sim -7.5$; see, e.g., \citealt{patel2020}), so the LMC's satellite population placed on Figure~\ref{fig:lf} would be a single line at $N=1$ (though with some uncertainty given our limited ability to kinematically confirm or refute associations between classical dwarfs and the LMC).
M33, a massive companion of the Andromeda galaxy (M31), has roughly the same stellar mass as the LMC. Its inner halo (within $r\sim50$~kpc) has been searched extensively for satellites with the PAndAS survey \citep{martin2009}, with one of the candidates (And~XXII; $M_{\rm V} = -6.5$) spectroscopically confirmed to be a likely satellite of M33 \citep{chapman2013}. Recently, Pisces~VII/Tri~III was discovered \citep{martinez-delgado2022} at a projected distance of $\sim70$~kpc from M33, and subsequently confirmed as a $M_{\rm V} = -6.0$ satellite of M33 by \citet{collins2024}. These dwarfs as well as the candidate ultra-diffuse galaxy Tri~IV \citep[$M_{\rm V} = -6.4$]{ogami2024} are shown as open black triangles in Figure~\ref{fig:mv_rh}, which places the satellites of NGC~2403 in context with the LMC and other nearby MC analogs (as well as the MW and M31). It is intriguing that all known or candidate M33 satellites thus far are ultra-faint dwarfs, though a deeper search extending to the virial radius of M33 would need to be carried out to confirm that no brighter satellites of M33 are present.

The dwarf satellite populations of Magellanic Cloud analogs \textit{beyond} the Local Group have not yet been quantitatively explored in the way we have for NGC~2403 in this work. However, a number of MC-mass systems have known dwarf satellites. We show these in context with satellites of the LMC, M33, and the MW and M31 in a luminosity--size diagram in Figure~\ref{fig:mv_rh}.\footnote{We do not include potential satellites of LMC analog NGC~300 from \citet{sand2024} in Figure~\ref{fig:mv_rh} because it is unclear which of the three dwarfs presented in that work are actually satellites of NGC~300, and which are foreground or background systems (and thus isolated, and intriguing in their own right). Follow-up studies will be necessary to assess the association of these dwarfs with NGC~300.} NGC~3109, a galaxy with roughly half the stellar mass of the SMC that resides at the outskirts of the Local Group ($D \sim 1.3$~Mpc), has two known satellites (for a comprehensive characterization of the census and its sensitivities, see Doliva-Dolinsky et al., \textit{in prep.}). Both Antlia ($M_{\rm V} = -10.4$; \citealt{sharina2008}) and Antlia~B ($M_{\rm V} = -9.7$; \citealt{Sand2015}) have H\textsc{i} gas \citep{ott2012, barnes2001} and small amounts of recent star formation \citep{McQuinn2010, Hargis2020}, but have likely been quenched via starvation due to outflows related to that star formation \citep{garling2024}. Until recently, the LMC stellar-mass galaxy NGC~55 ($D\sim2$~Mpc) had only a single known satellite, with $M_{\rm V} \approx -11$. \citet{McNanna2024} unveiled another NGC~55 satellite in a search of DES data. This dwarf, NGC~55-dw1, has $M_{\rm V} = -8.0$, but is unusually large for its luminosity. Its half-light radius of $\sim2.2$~kpc makes it the largest known dwarf galaxy fainter than $M_{\rm V} = -9$, suggesting that its size may be inflated due to a recent tidal interaction with NGC~55. Likewise, the young stellar population identified in HST imaging of MADCASH~2, an $M_{\rm V} = -9.2$ satellite of NGC~4214 \citep{Carlin2021}, a nearby ($D=2.9$~Mpc) galaxy with stellar mass roughly twice that of the SMC, could be interpreted as evidence for recent tidal interaction between MADCASH~2 and NGC~4214. Another LMC analog, NGC~4449 ($D \sim 3.8$~Mpc), has a known disrupting dwarf companion \citep{Rich2012, Martinez-Delgado2012} with stellar mass comparable to the (MW satellite) Fornax dSph. This particular satellite is the only dwarf satellite of an MC analog host with a spectroscopically measured velocity and metallicity \citep{Toloba2016}. \citet{Jahn2022} showed that tidal structures may be common around many Magellanic Cloud-mass galaxies; it seems we are seeing evidence of this in the sample that is being revealed recently.

Clearly, comprehensive searches such as the one reported here for NGC~2403 will uncover numerous dwarf satellites, as well as tidal debris and disrupting dwarfs. 
This work provides tantalizing hints at the treasures we will uncover with the Vera C. Rubin Observatory's upcoming Legacy Survey of Space and Time (LSST; \citealt{Ivezic2019, mutlu-pakdil2021}), which will reach comparable photometric depths over the entire southern sky.

\begin{acknowledgments}

We thank the referee for insightful comments that helped improve the manuscript. We acknowledge support from the following NSF grants: AST-1816196 (JLC); AST-1814208 (DC); and AST-1813628 (AHGP and CTG). 
Research by ADD is supported by NSF grant AST-1814208. DJS acknowledges support from NSF grant AST-2205863. KS acknowledges support from the Natural Sciences and Engineering Research Council of Canada (NSERC). 

This research has made use of NASA's Astrophysics Data System, and \texttt{Astropy}, a community-developed core Python package for Astronomy \citep{Price-Whelan2018b}. 

The Pan-STARRS1 Surveys have been made possible through contributions of the Institute for Astronomy, the University of Hawaii, the Pan-STARRS Project Office, the Max-Planck Society and its participating institutes, the Max Planck Institute for Astronomy, Heidelberg and the Max Planck Institute for Extraterrestrial Physics, Garching, The Johns Hopkins University, Durham University, the University of Edinburgh, Queen's University Belfast, the Harvard-Smithsonian Center for Astrophysics, the Las Cumbres Observatory Global Telescope Network Incorporated, the National Central University of Taiwan, the Space Telescope Science Institute, the National Aeronautics and Space Administration under Grant No. NNX08AR22G issued through the Planetary Science Division of the NASA Science Mission Directorate, the National Science Foundation under Grant AST-1238877, the University of Maryland, Eotvos Lorand University (ELTE), and the Los Alamos National Laboratory.

\end{acknowledgments}

\appendix

\section{Example diagnostic plots from dwarf search}

\begin{figure}[!t]
\begin{center}
\includegraphics[width=0.8\columnwidth, trim=0.25in 0.25in 0.25in 0.25in, clip]{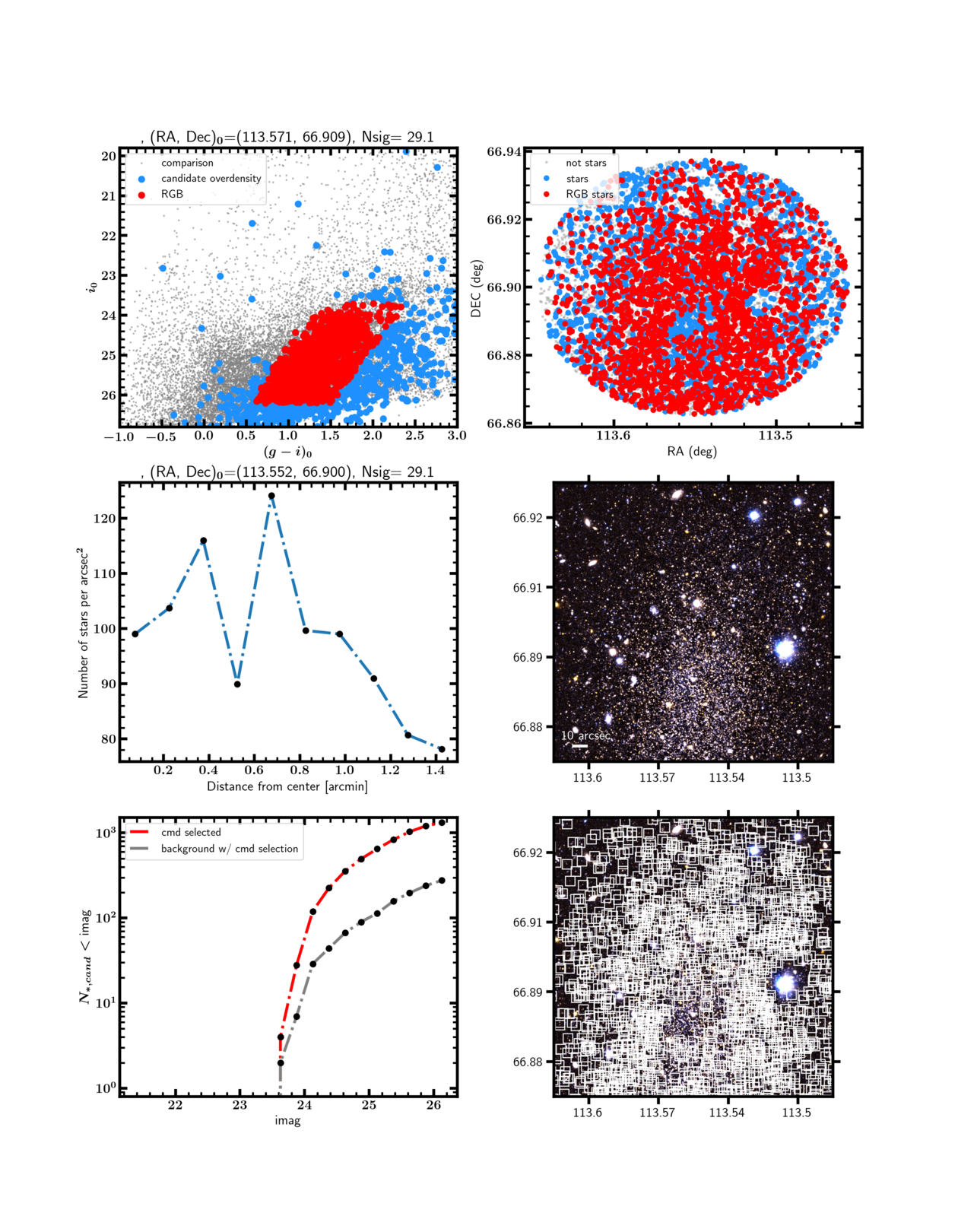}
\caption{Example diagnostic plot of a detection of (a portion of) DDO 44. The upper left panel shows a CMD, with candidate RGB stars highlighted in red, and all point sources within a 1-arcmin radius shown in blue. The upper right panel displays the spatial distribution of sources, using the same color-coding. In the middle-left plot, we show a radial surface density plot (note that the center may not be properly located on the center of the overdensity, as we have not refined the measurements at this point). The middle-right panel shows a color image of the region, and the lower right panel is the same image with RGB stars overlaid. In the lower-left plot, we show a ``luminosity function'' of candidate RGB stars; if a dwarf stellar population is present, one would expect the candidate dwarf (in red) to have a different luminosity function than the neighboring field stars (shown in gray), as it does here for DDO~44. }
\label{fig:ddo44_diag}
\end{center}
\end{figure}

\begin{figure}[!t]
\includegraphics[width=1.0\columnwidth, trim=0.25in 0.25in 0.25in 0.25in, clip]{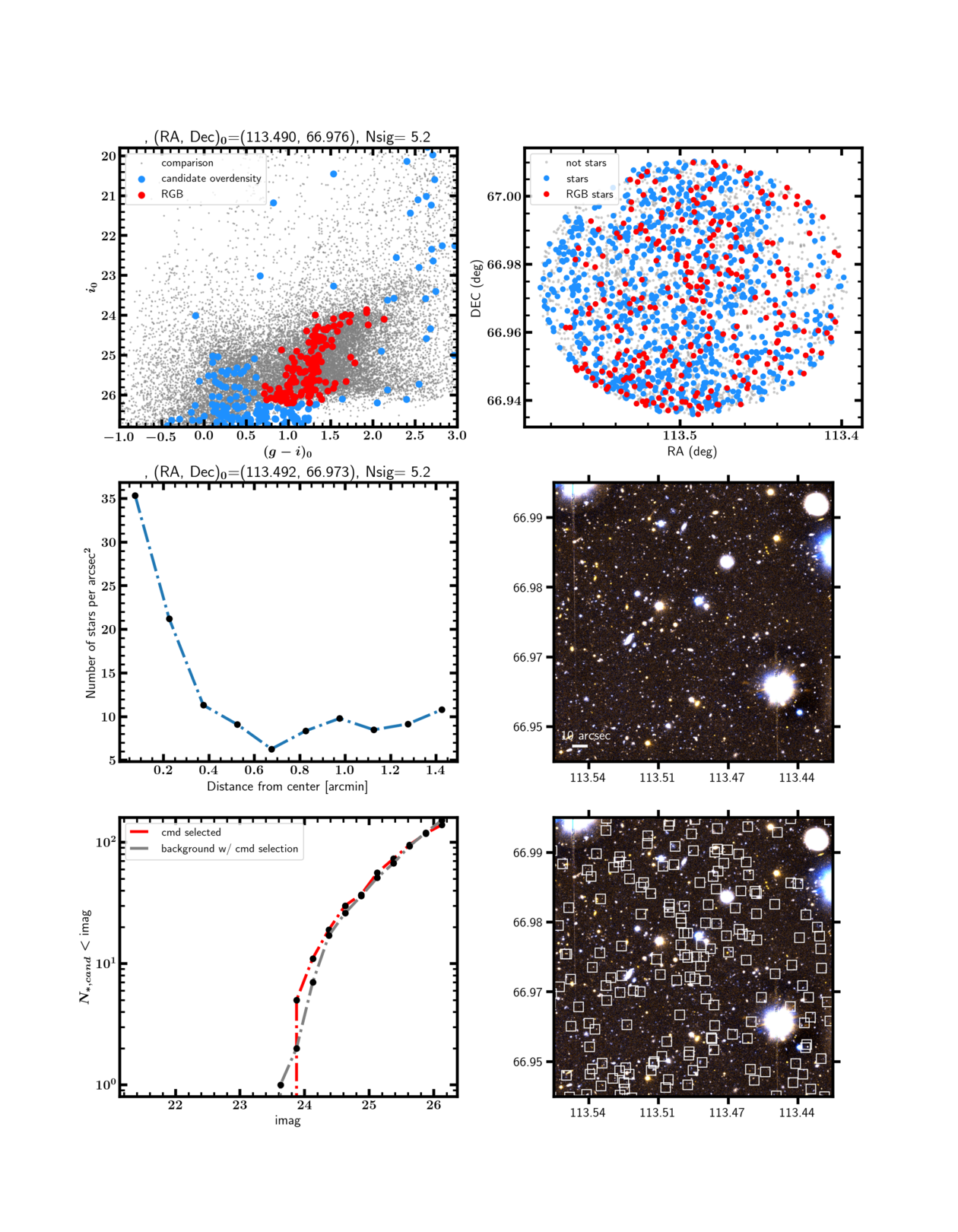}
\caption{Example diagnostic plot. This particular plot shows a portion of the DDO~44 stream, detected a few arcmin from the center of DDO~44. The RGB is clearly visible in the upper left panel (CMD).}
\label{fig:ddo44_stream_diag}
\end{figure}

\begin{figure}[!t]
\begin{center}
\includegraphics[width=0.8\columnwidth, trim=0.25in 0.25in 0.25in 0.25in, clip]{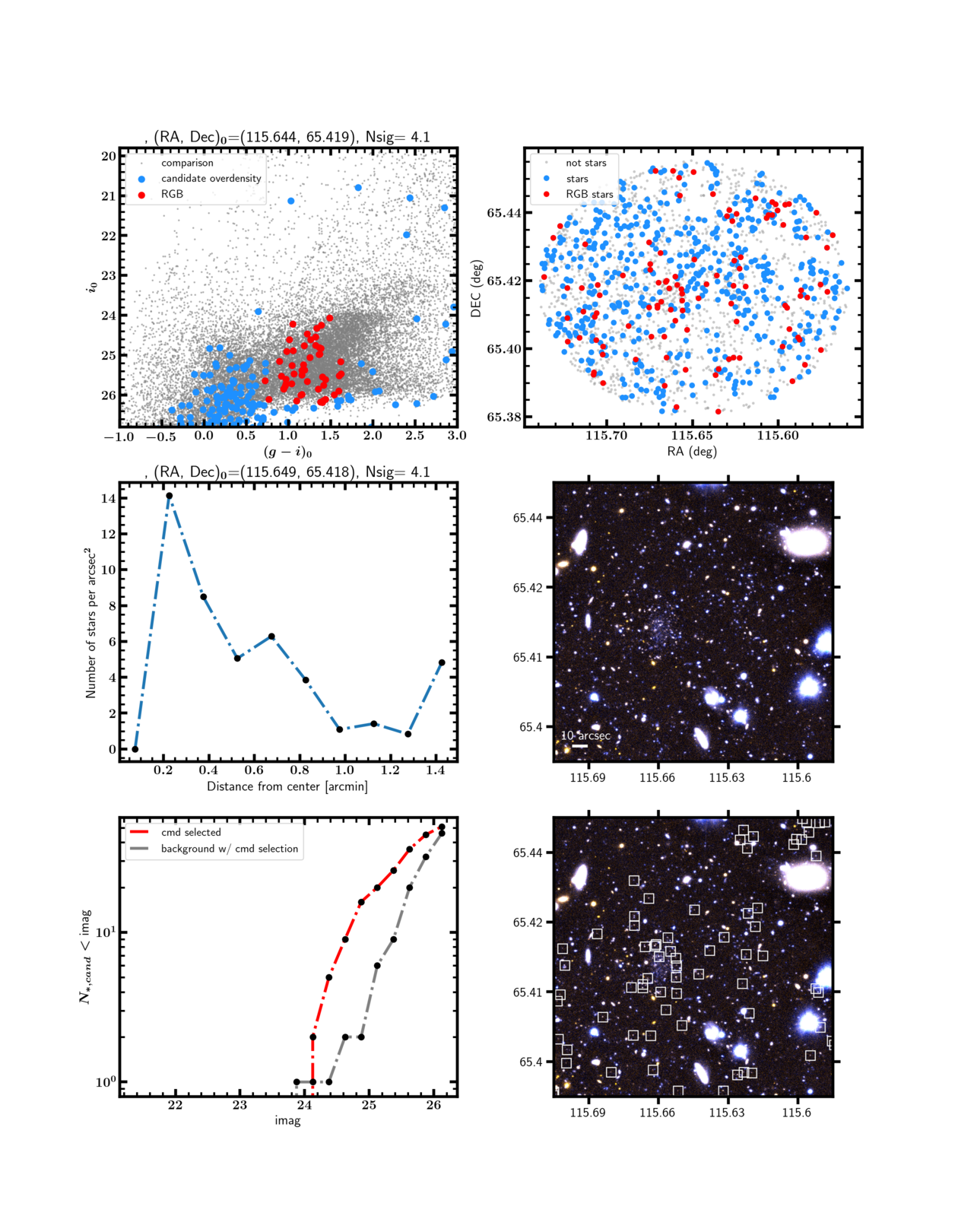}
\caption{Example diagnostic plot showing the detection of MADCASH-1, the previously known UFD near NGC~2403. Panels are as in Figure~\ref{fig:ddo44_diag}. The metal-poor RGB of MADCASH-1 is discernible in the CMD (upper left panel), but one can see in the images that the system is semi-resolved, and thus only a small number of stars are detected in the region of the semi-resolved ``fluff'' around this UFD.}
\label{fig:madcash1_diag}
\end{center}
\end{figure}

\begin{figure}[!t]
\includegraphics[width=1.0\columnwidth, trim=0.25in 0.25in 0.25in 0.25in, clip]{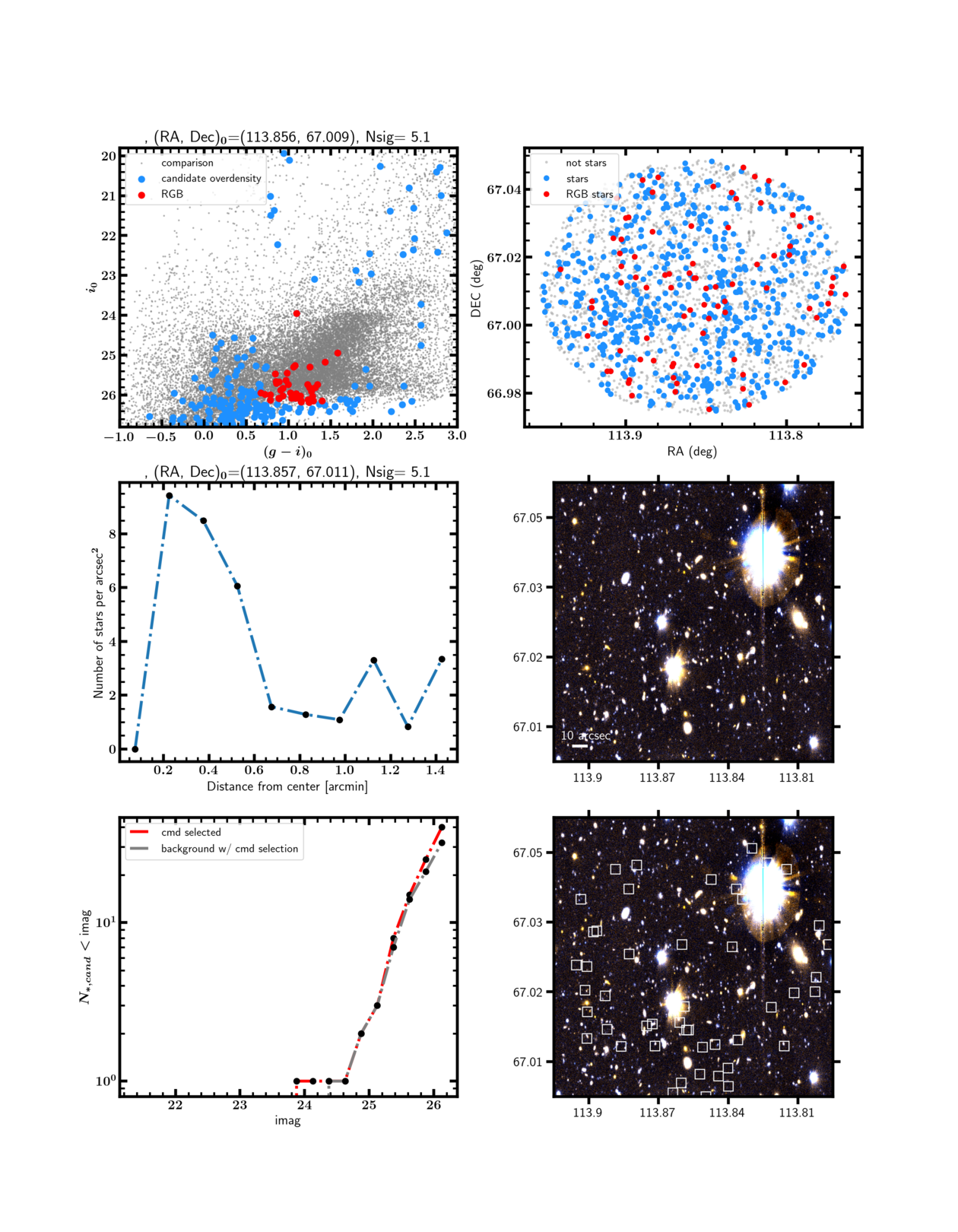}
\caption{Example diagnostic plot showing a candidate detection that is highly unlikely to be a dwarf galaxy associated with NGC~2403. Note that the luminosity function of objects detected within the RGB box does not resemble a typical LF for stars in a dwarf galaxy, but rather looks as if it was drawn from the (unresolved) background galaxy population.}
\label{fig:falsedet_diag}
\end{figure}

\begin{figure*}[!t]
\begin{center}
\includegraphics[width=0.32\textwidth, trim=0.25in 0.75in 0.25in 0.5in, clip]{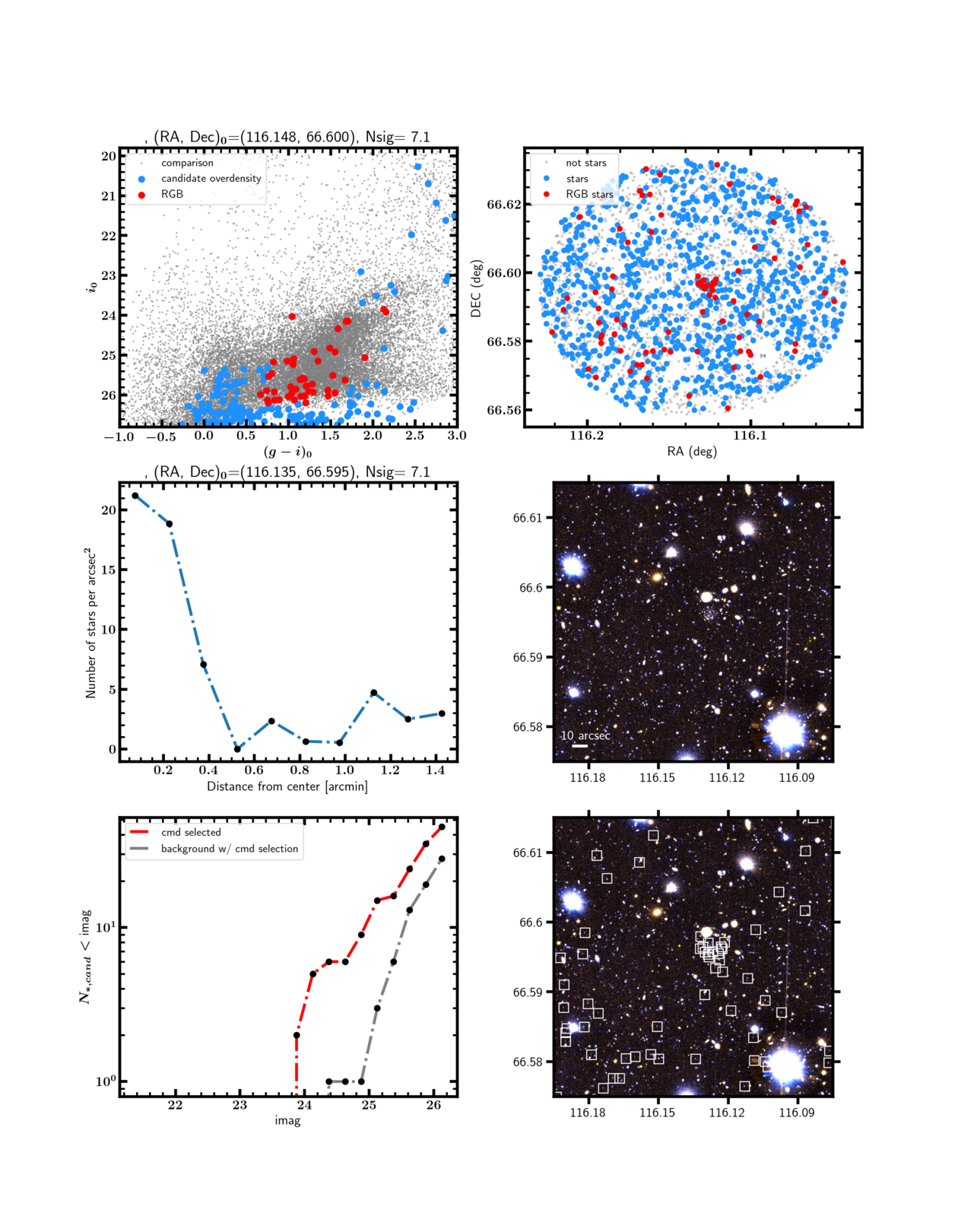}
\includegraphics[width=0.32\textwidth, trim=0.25in 0.75in 0.25in 0.5in, clip]{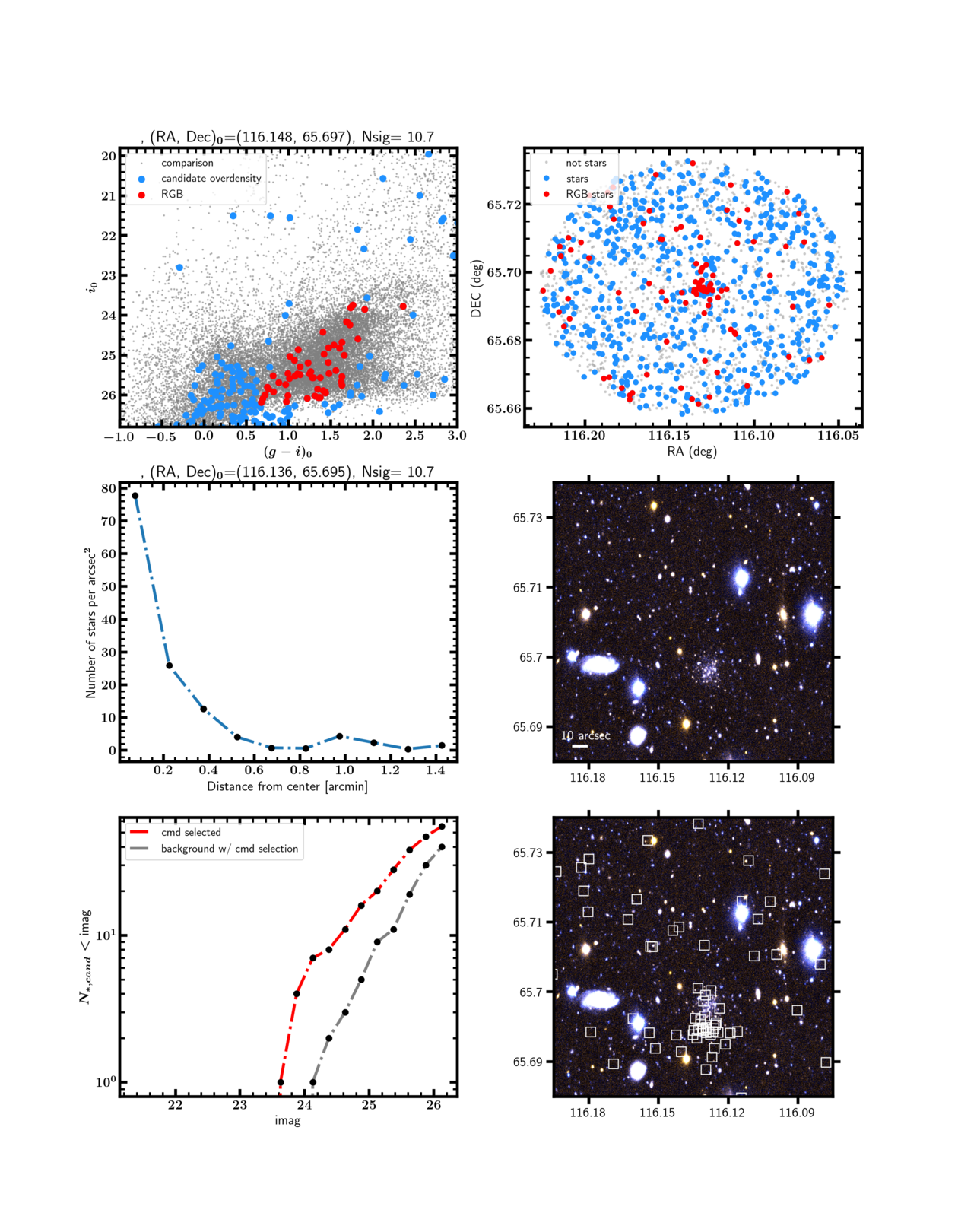}
\includegraphics[width=0.32\textwidth, trim=0.25in 0.75in 0.25in 0.5in, clip]{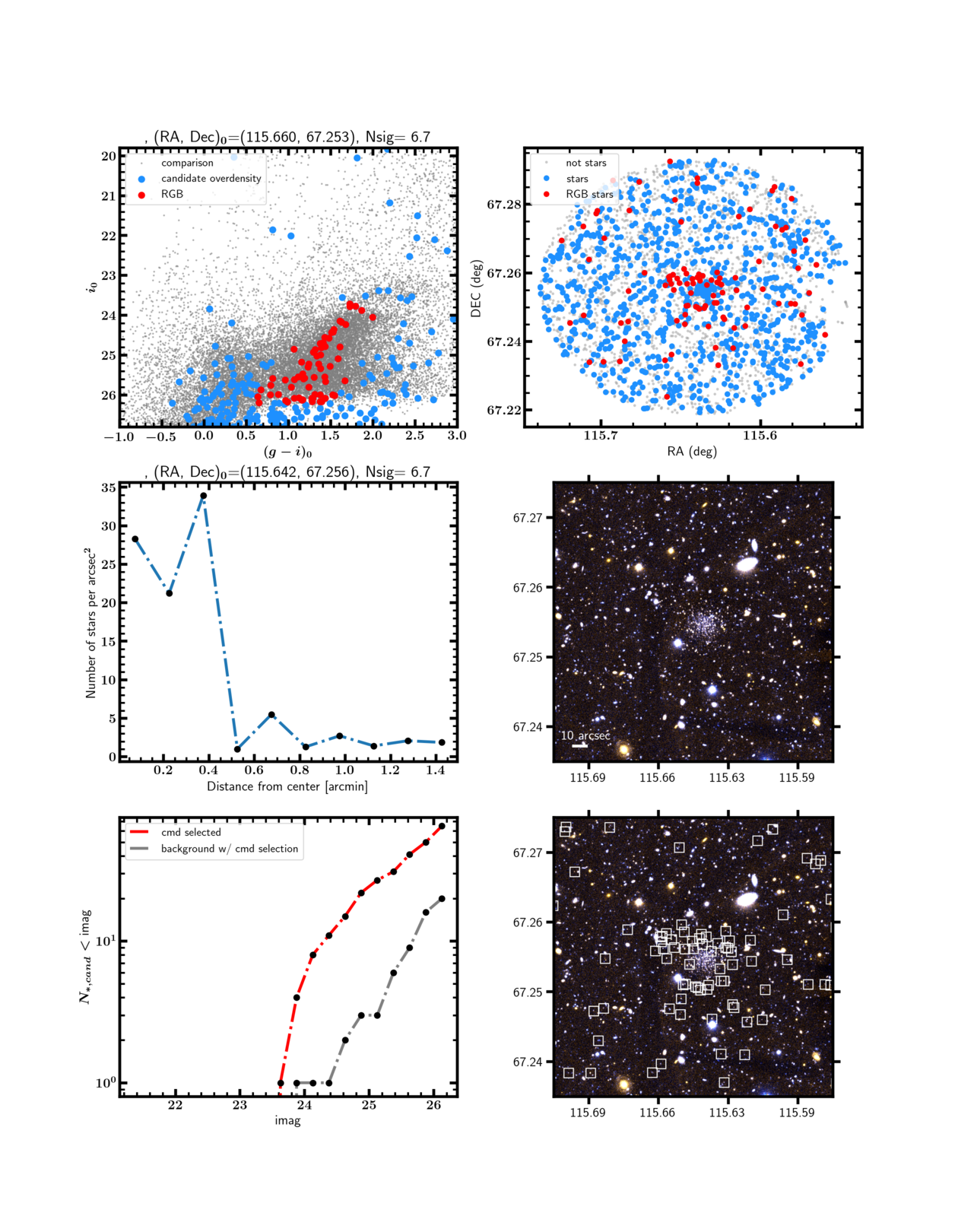}\\
\caption{Diagnostic plots for the dwarfs from the top row of Figure~\ref{fig:fakedwarf_images}, with surface brightnesses of $\sim26.0$ mag arcsec$^{-2}$, and (from left to right) $M_{\rm V} \approx -6.5, -7.5$, and $-8.5$.}
\label{fig:fakedwarf_diag_sb26}
\end{center}
\end{figure*}

\begin{figure*}[!th]
\begin{center}
\includegraphics[width=0.32\textwidth, trim=0.25in 0.75in 0.25in 0.5in, clip]{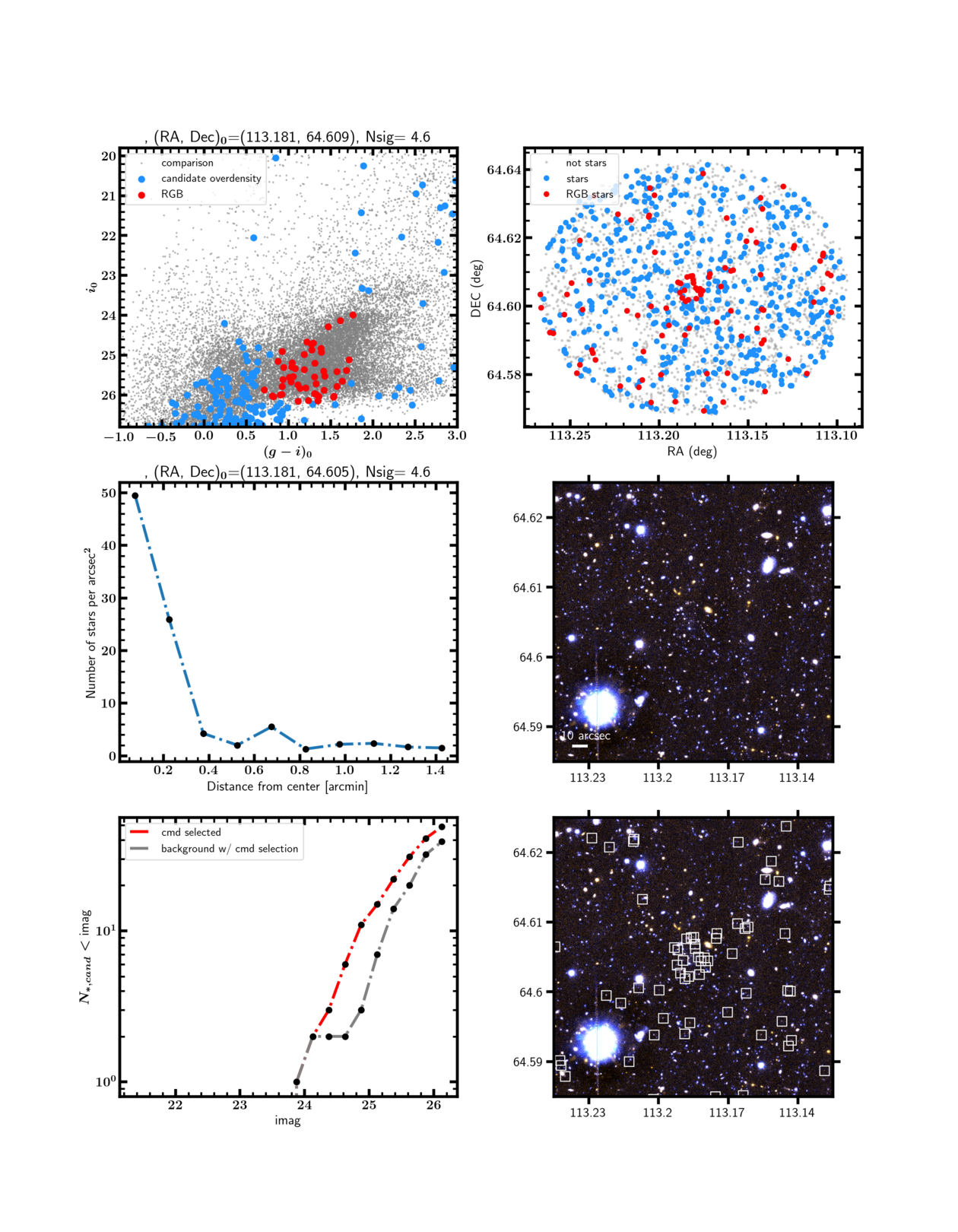}
\includegraphics[width=0.32\textwidth, trim=0.25in 0.75in 0.25in 0.5in, clip]{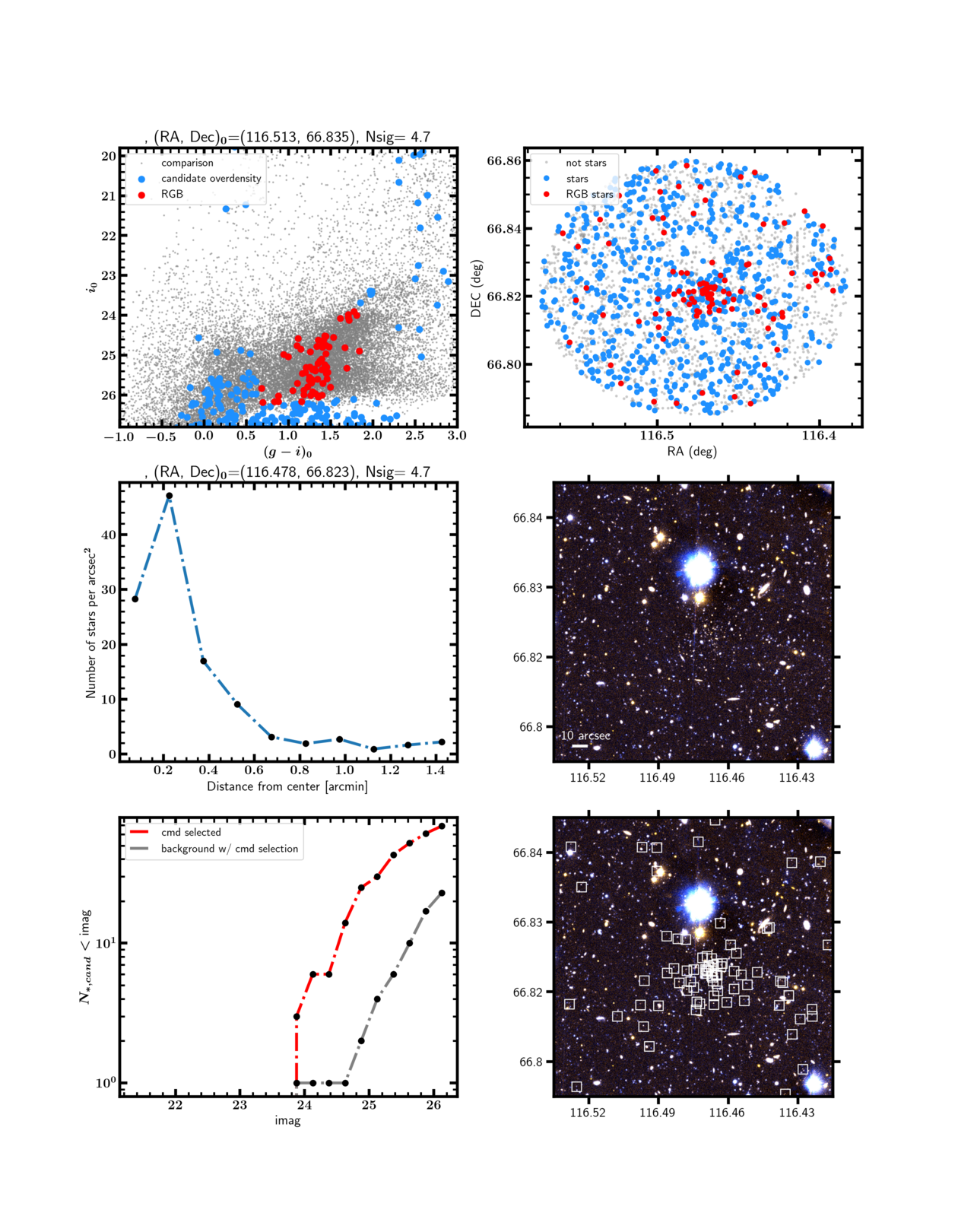}
\includegraphics[width=0.32\textwidth, trim=0.25in 0.75in 0.25in 0.5in, clip]{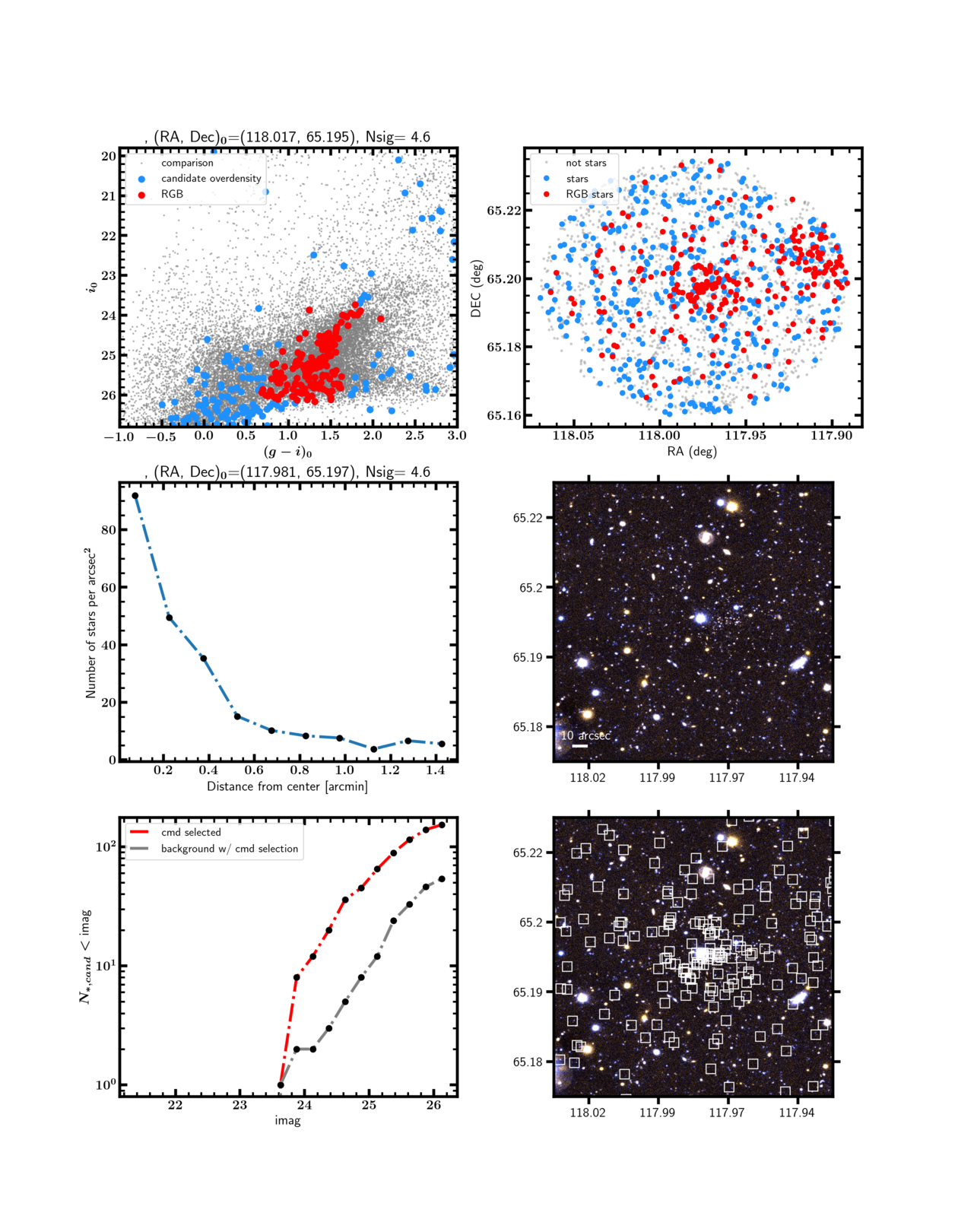}\\
\caption{Diagnostic plots for the dwarfs from the middle row of Figure~\ref{fig:fakedwarf_images}, with surface brightnesses of $\sim27.5$ mag arcsec$^{-2}$, and (from left to right) $M_{\rm V} \approx -6.5, -7.5$, and $-8.5$. Note that the middle panel in Figure~\ref{fig:fakedwarf_images} was replaced with an image of MADCASH-1; in the middle panel here, we show the diagnostic plot for a synthetic dwarf with $M_{\rm V} \approx -7.5$ and $\mu \approx 27.5$~mag~arcsec$^{-2}$.}
\label{fig:fakedwarf_diag_sb27p5}
\end{center}
\end{figure*}

\begin{figure*}[!th]
\begin{center}
\includegraphics[width=0.32\textwidth, trim=0.25in 0.75in 0.25in 0.5in, clip]{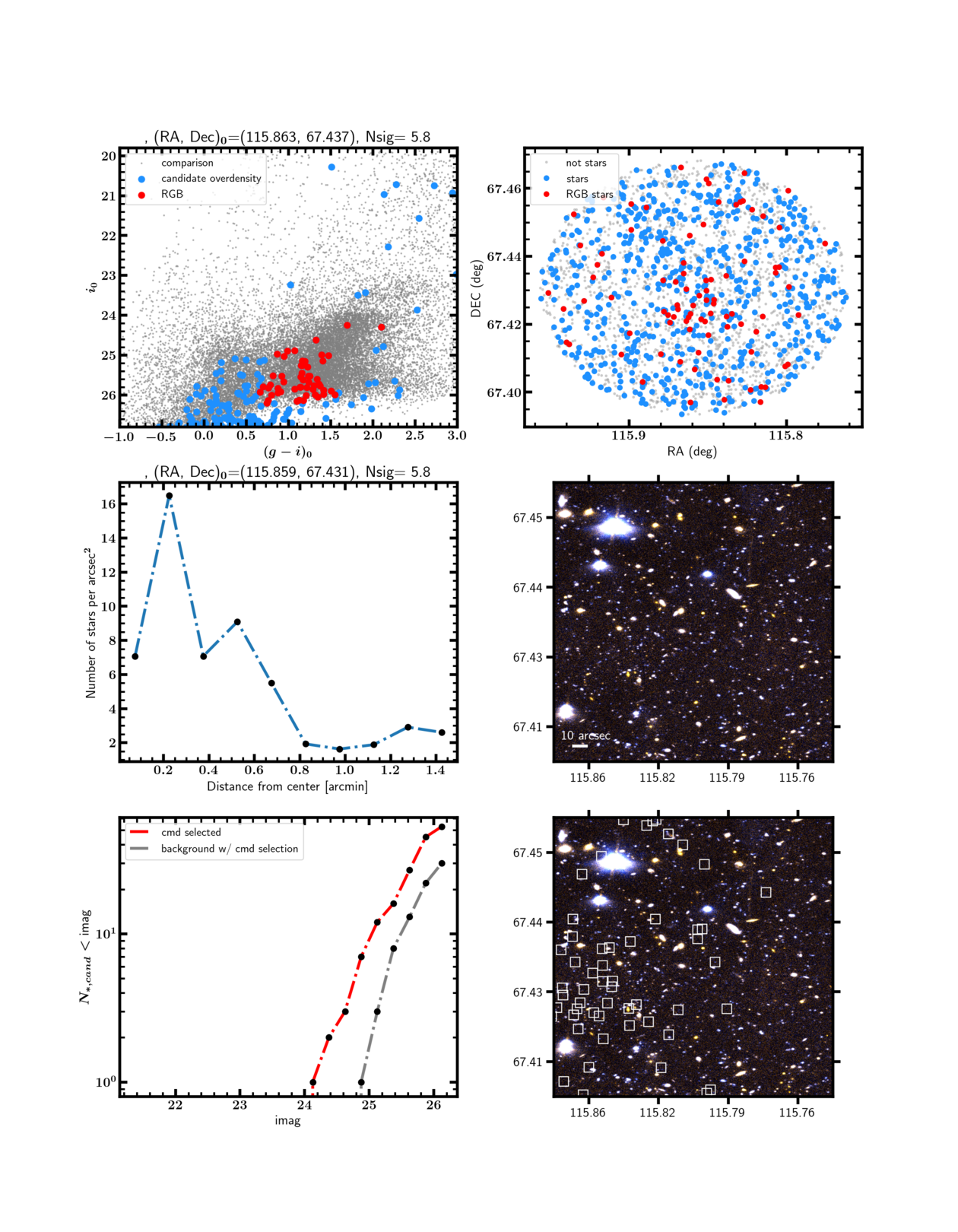}
\includegraphics[width=0.32\textwidth, trim=0.25in 0.75in 0.25in 0.5in, clip]{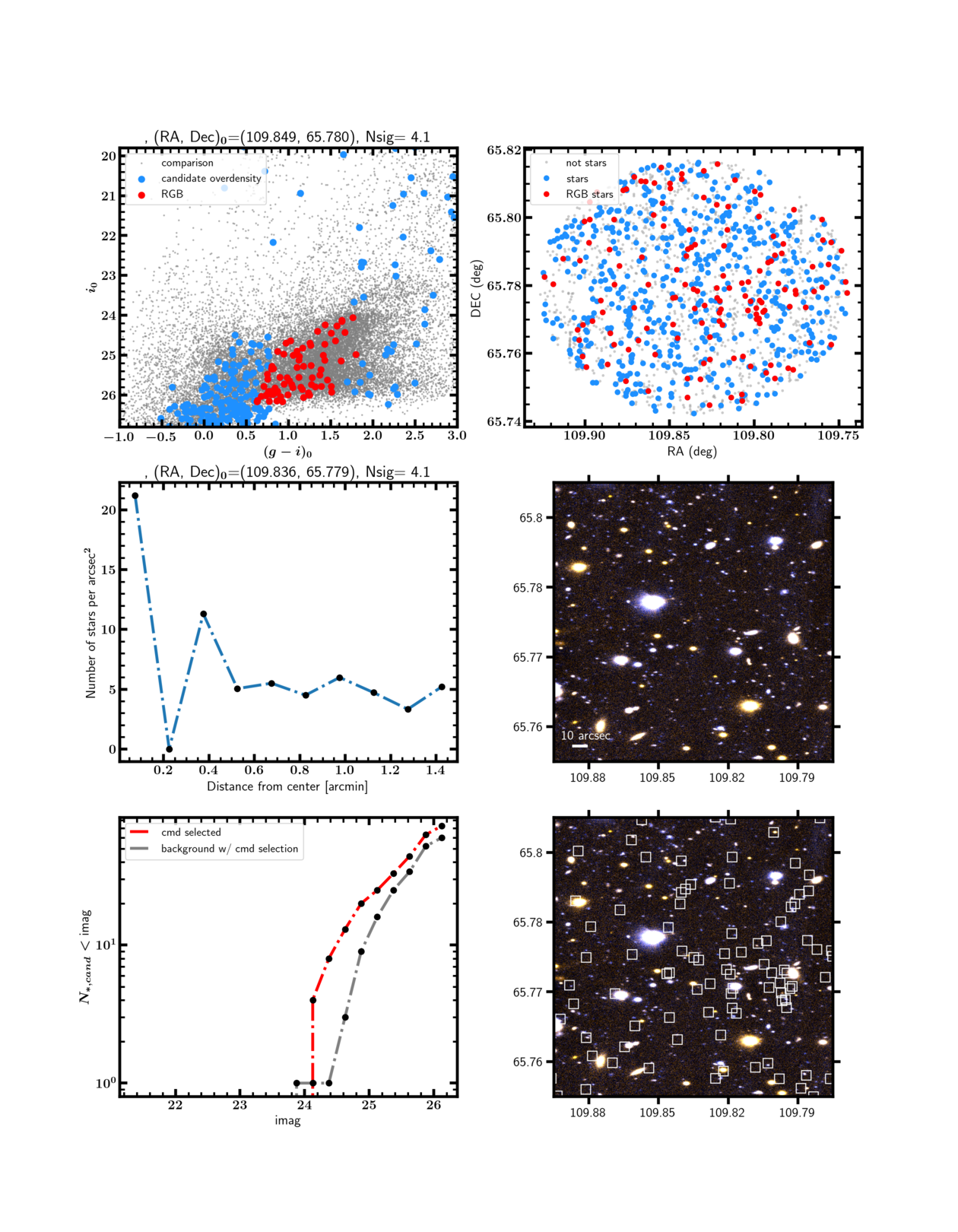}
\includegraphics[width=0.32\textwidth, trim=0.25in 0.75in 0.25in 0.5in, clip]{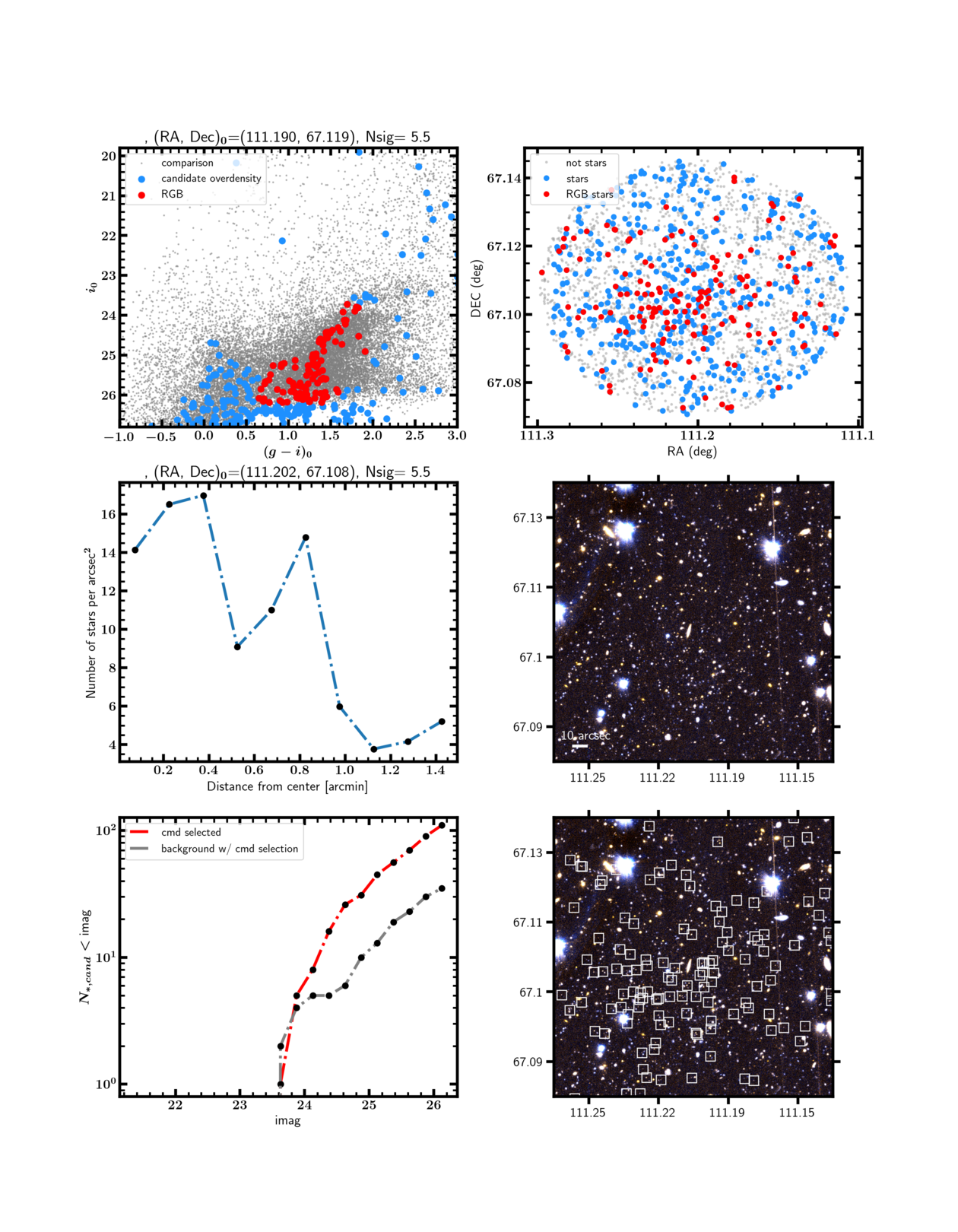}\\
\caption{Diagnostic plots for the dwarfs from the top row of Figure~\ref{fig:fakedwarf_images}, with surface brightnesses of $\sim29.0$ mag arcsec$^{-2}$, and (from left to right) $M_{\rm V} \approx -6.5, -7.5$, and $-8.5$.}
\label{fig:fakedwarf_diag_sb29}
\end{center}
\end{figure*}

Examples of the diagnostic plots we used to examine candidate dwarf detections are seen in Figures~\ref{fig:ddo44_diag} through \ref{fig:fakedwarf_diag_sb29}. Our dwarf search algorithm produced multiple detections of DDO~44 and parts of its tidal stream. An example of a DDO~44 detection is shown in Figure~\ref{fig:ddo44_diag}, and a portion of its tidal stream (identified via its coordinates that place it on the stream track) is seen in Figure~\ref{fig:ddo44_stream_diag}. While the main body of luminous dwarf DDO~44 is clearly visible in all panels of the diagnostic plot, the stream is only clearly identifiable via its signature in the CMD. The diagnostic plot for the much fainter known dwarf MADCASH-1 is seen in Figure~\ref{fig:madcash1_diag}. A metal-poor RGB sequence is visible in the CMD, and the presence of a dwarf is easily corroborated by looking at the images, where a semi-resolved dwarf is prominently seen slightly to the left of the image center.

Figure~\ref{fig:falsedet_diag} shows a fairly common type of false detection from our dwarf search. While a statistical ``excess'' over background was identified by the algorithm, it is clear from the CMD (upper left panel) that the sources contributing to the detection are predominantly faint, unresolved galaxies at the faintest end of the RGB selection box. A contributing factor to false detections like this is often the presence of a bright star or galaxy nearby. Such bright sources create ``holes'' where there are no sources detected, thus artificially decreasing the background counts and causing the appearance of an overdensity where there is none.

Finally, we include diagnostic plots for the synthetic dwarfs from Figure~\ref{fig:fakedwarf_images}. Figures~\ref{fig:fakedwarf_diag_sb26}, \ref{fig:fakedwarf_diag_sb27p5}, and \ref{fig:fakedwarf_diag_sb29} show the relevant plots for the top, middle, and bottom rows of Figure~\ref{fig:fakedwarf_images}, respectively. These figures (and the fact that all of these synthetic dwarfs were detected by our search algorithm) demonstrate that our search should have been sensitive to dwarfs as faint as $M_{\rm V} = -6.5$ to surface brightnesses as low as 29~mag~arcsec$^{-2}$. Although the $\mu \approx 29$~mag~arcsec$^{-2}$ synthetic dwarfs in Figure~\ref{fig:fakedwarf_diag_sb29} are not visible in the images, they are identifiable via their metal-poor RGB in the CMDs, and as spatial overdensities. All of the synthetic dwarfs in these figures are readily identifiable in their diagnostic plots; thus, the fact that we did not identify any similarly obvious candidates among the candidates in our Subaru/HSC data suggests that no legitimate dwarfs were present in the data.

\vspace{5mm}
\facilities{Subaru+HSC, PS1}

\software{\texttt{astropy} \citep{TheAstropyCollaboration2013,Price-Whelan2018b}, \texttt{Matplotlib} \citep{Hunter2007}, \texttt{NumPy} \citep{VanderWalt2011}, \texttt{Topcat} \citep{Taylor2005}.}

\bibliography{refs}

\begin{thebibliography}{}
\expandafter\ifx\csname natexlab\endcsname\relax\def\natexlab#1{#1}\fi
\providecommand{\url}[1]{\href{#1}{#1}}
\providecommand{\dodoi}[1]{doi:~\href{http://doi.org/#1}{\nolinkurl{#1}}}
\providecommand{\doeprint}[1]{\href{http://ascl.net/#1}{\nolinkurl{http://ascl.net/#1}}}
\providecommand{\doarXiv}[1]{\href{https://arxiv.org/abs/#1}{\nolinkurl{https://arxiv.org/abs/#1}}}

\bibitem[{{Applebaum} {et~al.}(2021){Applebaum}, {Brooks}, {Christensen},
  {Munshi}, {Quinn}, {Shen}, \& {Tremmel}}]{Applebaum21}
{Applebaum}, E., {Brooks}, A.~M., {Christensen}, C.~R., {et~al.} 2021, \apj,
  906, 96, \dodoi{10.3847/1538-4357/abcafa}

\bibitem[{{Barnes} \& {de Blok}(2001)}]{barnes2001}
{Barnes}, D.~G., \& {de Blok}, W.~J.~G. 2001, \aj, 122, 825,
  \dodoi{10.1086/321170}

\bibitem[{{Battaglia} {et~al.}(2022){Battaglia}, {Taibi}, {Thomas}, \&
  {Fritz}}]{Battaglia22}
{Battaglia}, G., {Taibi}, S., {Thomas}, G.~F., \& {Fritz}, T.~K. 2022, \aap,
  657, A54, \dodoi{10.1051/0004-6361/202141528}

\bibitem[{{Bennet} {et~al.}(2019){Bennet}, {Sand}, {Crnojevi{\'c}}, {Spekkens},
  {Karunakaran}, {Zaritsky}, \& {Mutlu-Pakdil}}]{Bennet19}
{Bennet}, P., {Sand}, D.~J., {Crnojevi{\'c}}, D., {et~al.} 2019, \apj, 885,
  153, \dodoi{10.3847/1538-4357/ab46ab}

\bibitem[{{Bennet} {et~al.}(2020){Bennet}, {Sand}, {Crnojevi{\'c}}, {Spekkens},
  {Karunakaran}, {Zaritsky}, \& {Mutlu-Pakdil}}]{Bennet20}
---. 2020, \apjl, 893, L9, \dodoi{10.3847/2041-8213/ab80c5}

\bibitem[{{Bennet} {et~al.}(2017){Bennet}, {Sand}, {Crnojevi{\'c}}, {Spekkens},
  {Zaritsky}, \& {Karunakaran}}]{Bennet17}
---. 2017, \apj, 850, 109, \dodoi{10.3847/1538-4357/aa9180}

\bibitem[{Bosch {et~al.}(2018)Bosch, Armstrong, Bickerton, Furusawa, Ikeda,
  Koike, Lupton, Mineo, Price, Takata, Tanaka, Yasuda, AlSayyad, Becker,
  Coulton, Coupon, Garmilla, Huang, Krughoff, Lang, Leauthaud, Lim, Lust,
  MacArthur, Mandelbaum, Miyatake, Miyazaki, Murata, More, Okura, Owen,
  Swinbank, Strauss, Yamada, \& Yamanoi}]{Bosch2018}
Bosch, J., Armstrong, R., Bickerton, S., {et~al.} 2018, Publications of the
  Astronomical Society of Japan, 70, \dodoi{10.1093/pasj/psx080}

\bibitem[{{Bosch} {et~al.}(2019){Bosch}, {AlSayyad}, {Armstrong}, {Bellm},
  {Chiang}, {Eggl}, {Findeisen}, {Fisher-Levine}, {Guy}, {Guyonnet},
  {Ivezi{\'c}}, {Jenness}, {Kov{\'a}cs}, {Krughoff}, {Lupton}, {Lust},
  {MacArthur}, {Meyers}, {Moolekamp}, {Morrison}, {Morton}, {O'Mullane},
  {Parejko}, {Plazas}, {Price}, {Rawls}, {Reed}, {Schellart}, {Slater},
  {Sullivan}, {Swinbank}, {Taranu}, {Waters}, \& {Wood-Vasey}}]{Bosch2019}
{Bosch}, J., {AlSayyad}, Y., {Armstrong}, R., {et~al.} 2019, in Astronomical
  Society of the Pacific Conference Series, Vol. 523, Astronomical Data
  Analysis Software and Systems XXVII, ed. P.~J. {Teuben}, M.~W. {Pound}, B.~A.
  {Thomas}, \& E.~M. {Warner}, 521, \dodoi{10.48550/arXiv.1812.03248}

\bibitem[{{Brasseur} {et~al.}(2011){Brasseur}, {Martin}, {Macci{\`o}}, {Rix},
  \& {Kang}}]{brasseur2011}
{Brasseur}, C.~M., {Martin}, N.~F., {Macci{\`o}}, A.~V., {Rix}, H.-W., \&
  {Kang}, X. 2011, \apj, 743, 179, \dodoi{10.1088/0004-637X/743/2/179}

\bibitem[{{Bressan} {et~al.}(2012){Bressan}, {Marigo}, {Girardi}, {Salasnich},
  {Dal Cero}, {Rubele}, \& {Nanni}}]{bressan2012}
{Bressan}, A., {Marigo}, P., {Girardi}, L., {et~al.} 2012, \mnras, 427, 127,
  \dodoi{10.1111/j.1365-2966.2012.21948.x}

\bibitem[{{Bryan} \& {Norman}(1998)}]{bryan_norman1998}
{Bryan}, G.~L., \& {Norman}, M.~L. 1998, \apj, 495, 80, \dodoi{10.1086/305262}

\bibitem[{{Bullock} \& {Boylan-Kolchin}(2017)}]{Bullock17}
{Bullock}, J.~S., \& {Boylan-Kolchin}, M. 2017, \araa, 55, 343,
  \dodoi{10.1146/annurev-astro-091916-055313}

\bibitem[{{Bullock} {et~al.}(2000){Bullock}, {Kravtsov}, \&
  {Weinberg}}]{Bullock00}
{Bullock}, J.~S., {Kravtsov}, A.~V., \& {Weinberg}, D.~H. 2000, \apj, 539, 517,
  \dodoi{10.1086/309279}

\bibitem[{{Carlin} {et~al.}(2016){Carlin}, {Sand}, {Price}, {Willman},
  {Karunakaran}, {Spekkens}, {Bell}, {Brodie}, {Crnojevi{\'c}}, {Forbes},
  {Hargis}, {Kirby}, {Lupton}, {Peter}, {Romanowsky}, \&
  {Strader}}]{Carlin2016}
{Carlin}, J.~L., {Sand}, D.~J., {Price}, P., {et~al.} 2016, \apjl, 828, L5,
  \dodoi{10.3847/2041-8205/828/1/L5}

\bibitem[{{Carlin} {et~al.}(2019){Carlin}, {Garling}, {Peter}, {Crnojevi{\'c}},
  {Forbes}, {Hargis}, {Mutlu-Pakdil}, {Pucha}, {Romanowsky}, {Sand},
  {Spekkens}, {Strader}, \& {Willman}}]{Carlin2019}
{Carlin}, J.~L., {Garling}, C.~T., {Peter}, A. H.~G., {et~al.} 2019, \apj, 886,
  109, \dodoi{10.3847/1538-4357/ab4c32}

\bibitem[{{Carlin} {et~al.}(2021){Carlin}, {Mutlu-Pakdil}, {Crnojevi{\'c}},
  {Garling}, {Karunakaran}, {Peter}, {Tollerud}, {Forbes}, {Hargis}, {Lim},
  {Romanowsky}, {Sand}, {Spekkens}, \& {Strader}}]{Carlin2021}
{Carlin}, J.~L., {Mutlu-Pakdil}, B., {Crnojevi{\'c}}, D., {et~al.} 2021, \apj,
  909, 211, \dodoi{10.3847/1538-4357/abe040}

\bibitem[{{Carlsten} {et~al.}(2022){Carlsten}, {Greene}, {Beaton}, {Danieli},
  \& {Greco}}]{Carlsten22}
{Carlsten}, S.~G., {Greene}, J.~E., {Beaton}, R.~L., {Danieli}, S., \& {Greco},
  J.~P. 2022, \apj, 933, 47, \dodoi{10.3847/1538-4357/ac6fd7}

\bibitem[{{Cerny} {et~al.}(2023){Cerny}, {Mart{\'\i}nez-V{\'a}zquez},
  {Drlica-Wagner}, {Pace}, {Mutlu-Pakdil}, {Li}, {Riley}, {Crnojevi{\'c}},
  {Bom}, {Carballo-Bello}, {Carlin}, {Chiti}, {Choi}, {Collins},
  {Darragh-Ford}, {Ferguson}, {Geha}, {Mart{\'\i}nez-Delgado}, {Massana},
  {Mau}, {Medina}, {Mu{\~n}oz}, {Nadler}, {No{\"e}l}, {Olsen}, {Pieres},
  {Sakowska}, {Simon}, {Stringfellow}, {Tollerud}, {Vivas}, {Walker},
  {Wechsler}, \& {Delve Collaboration}}]{Cerny23}
{Cerny}, W., {Mart{\'\i}nez-V{\'a}zquez}, C.~E., {Drlica-Wagner}, A., {et~al.}
  2023, \apj, 953, 1, \dodoi{10.3847/1538-4357/acdd78}

\bibitem[{{Chabrier}(2001)}]{chabrier2001}
{Chabrier}, G. 2001, \apj, 554, 1274, \dodoi{10.1086/321401}

\bibitem[{{Chapman} {et~al.}(2013){Chapman}, {Widrow}, {Collins}, {Dubinski},
  {Ibata}, {Rich}, {Ferguson}, {Irwin}, {Lewis}, {Martin}, {McConnachie},
  {Pe{\~n}arrubia}, \& {Tanvir}}]{chapman2013}
{Chapman}, S.~C., {Widrow}, L., {Collins}, M.~L.~M., {et~al.} 2013, \mnras,
  430, 37, \dodoi{10.1093/mnras/sts392}

\bibitem[{{Chiboucas} {et~al.}(2013){Chiboucas}, {Jacobs}, {Tully}, \&
  {Karachentsev}}]{Chiboucas13}
{Chiboucas}, K., {Jacobs}, B.~A., {Tully}, R.~B., \& {Karachentsev}, I.~D.
  2013, \aj, 146, 126, \dodoi{10.1088/0004-6256/146/5/126}

\bibitem[{{Collins} {et~al.}(2024){Collins}, {Karim}, {Martinez-Delgado},
  {Monelli}, {Tollerud}, {Donatiello}, {Navabi}, {Charles}, \&
  {Boschin}}]{collins2024}
{Collins}, M. L.~M., {Karim}, N., {Martinez-Delgado}, D., {et~al.} 2024,
  \mnras, 528, 2614, \dodoi{10.1093/mnras/stae199}

\bibitem[{{Crnojevi{\'c}} {et~al.}(2016){Crnojevi{\'c}}, {Sand}, {Spekkens},
  {Caldwell}, {Guhathakurta}, {McLeod}, {Seth}, {Simon}, {Strader}, \&
  {Toloba}}]{Crnojevic16}
{Crnojevi{\'c}}, D., {Sand}, D.~J., {Spekkens}, K., {et~al.} 2016, \apj, 823,
  19, \dodoi{10.3847/0004-637X/823/1/19}

\bibitem[{{Crnojevi{\'c}} {et~al.}(2019){Crnojevi{\'c}}, {Sand}, {Bennet},
  {Pasetto}, {Spekkens}, {Caldwell}, {Guhathakurta}, {McLeod}, {Seth}, {Simon},
  {Strader}, \& {Toloba}}]{Crnojevic19}
{Crnojevi{\'c}}, D., {Sand}, D.~J., {Bennet}, P., {et~al.} 2019, \apj, 872, 80,
  \dodoi{10.3847/1538-4357/aafbe7}

\bibitem[{{Davis} {et~al.}(2021){Davis}, {Nierenberg}, {Peter}, {Garling},
  {Greco}, {Kochanek}, {Utomo}, {Casey}, {Pogge}, {Roberts}, {Sand}, \&
  {Sardone}}]{Davis21}
{Davis}, A.~B., {Nierenberg}, A.~M., {Peter}, A. H.~G., {et~al.} 2021, \mnras,
  500, 3854, \dodoi{10.1093/mnras/staa3246}

\bibitem[{{Dooley} {et~al.}(2017){Dooley}, {Peter}, {Carlin}, {Frebel},
  {Bechtol}, \& {Willman}}]{Dooley2017}
{Dooley}, G.~A., {Peter}, A. H.~G., {Carlin}, J.~L., {et~al.} 2017, \mnras,
  472, 1060, \dodoi{10.1093/mnras/stx2001}

\bibitem[{{Drlica-Wagner} {et~al.}(2021){Drlica-Wagner}, {Carlin}, {Nidever},
  {Ferguson}, {Kuropatkin}, {Adam{\'o}w}, {Cerny}, {Choi}, {Esteves},
  {Mart{\'\i}nez-V{\'a}zquez}, {Mau}, {Miller}, {Mutlu-Pakdil}, {Neilsen},
  {Olsen}, {Pace}, {Riley}, {Sakowska}, {Sand}, {Santana-Silva}, {Tollerud},
  {Tucker}, {Vivas}, {Zaborowski}, {Zenteno}, {Abbott}, {Allam}, {Bechtol},
  {Bell}, {Bell}, {Bilaji}, {Bom}, {Carballo-Bello}, {Crnojevi{\'c}}, {Cioni},
  {Diaz-Ocampo}, {de Boer}, {Erkal}, {Gruendl}, {Hernandez-Lang}, {Hughes},
  {James}, {Johnson}, {Li}, {Mao}, {Mart{\'\i}nez-Delgado}, {Massana},
  {McNanna}, {Morgan}, {Nadler}, {No{\"e}l}, {Palmese}, {Peter}, {Rykoff},
  {S{\'a}nchez}, {Shipp}, {Simon}, {Smercina}, {Soares-Santos}, {Stringfellow},
  {Tavangar}, {van der Marel}, {Walker}, {Wechsler}, {Wu}, {Yanny},
  {Fitzpatrick}, {Huang}, {Jacques}, {Nikutta}, {Scott}, \& {Astro Data
  Lab}}]{drlica-wagner2021}
{Drlica-Wagner}, A., {Carlin}, J.~L., {Nidever}, D.~L., {et~al.} 2021, \apjs,
  256, 2, \dodoi{10.3847/1538-4365/ac079d}

\bibitem[{{El-Badry} {et~al.}(2018){El-Badry}, {Quataert}, {Wetzel}, {Hopkins},
  {Weisz}, {Chan}, {Fitts}, {Boylan-Kolchin}, {Kere{\v{s}}},
  {Faucher-Gigu{\`e}re}, \& {Garrison-Kimmel}}]{elbadry18}
{El-Badry}, K., {Quataert}, E., {Wetzel}, A., {et~al.} 2018, \mnras, 473, 1930,
  \dodoi{10.1093/mnras/stx2482}

\bibitem[{{Fattahi} {et~al.}(2016){Fattahi}, {Navarro}, {Sawala}, {Frenk},
  {Oman}, {Crain}, {Furlong}, {Schaller}, {Schaye}, {Theuns}, \&
  {Jenkins}}]{fattahi2016}
{Fattahi}, A., {Navarro}, J.~F., {Sawala}, T., {et~al.} 2016, \mnras, 457, 844,
  \dodoi{10.1093/mnras/stv2970}

\bibitem[{{Forbes} {et~al.}(2022){Forbes}, {Ferr{\'e}-Mateu}, {Gannon},
  {Romanowsky}, {Carlin}, {Brodie}, \& {Day}}]{forbes2022}
{Forbes}, D.~A., {Ferr{\'e}-Mateu}, A., {Gannon}, J.~S., {et~al.} 2022, \mnras,
  512, 802, \dodoi{10.1093/mnras/stac503}

\bibitem[{{Furusawa} {et~al.}(2018){Furusawa}, {Koike}, {Takata}, {Okura},
  {Miyatake}, {Lupton}, {Bickerton}, {Price}, {Bosch}, {Yasuda}, {Mineo},
  {Yamada}, {Miyazaki}, {Nakata}, {Koshida}, {Komiyama}, {Utsumi},
  {Kawanomoto}, {Jeschke}, {Noumaru}, {Schubert}, {Iwata}, {Finet},
  {Fujiyoshi}, {Tajitsu}, {Terai}, \& {Lee}}]{Furusawa2018}
{Furusawa}, H., {Koike}, M., {Takata}, T., {et~al.} 2018, \pasj, 70, S3,
  \dodoi{10.1093/pasj/psx079}

\bibitem[{{Garling} {et~al.}(2021){Garling}, {Peter}, {Kochanek}, {Sand}, \&
  {Crnojevi{\'c}}}]{garling2021}
{Garling}, C.~T., {Peter}, A. H.~G., {Kochanek}, C.~S., {Sand}, D.~J., \&
  {Crnojevi{\'c}}, D. 2021, \mnras, 507, 4764, \dodoi{10.1093/mnras/stab2447}

\bibitem[{{Garling} {et~al.}(2024){Garling}, {Peter}, {Spekkens}, {Sand},
  {Hargis}, {Crnojevi{\'c}}, \& {Carlin}}]{garling2024}
{Garling}, C.~T., {Peter}, A. H.~G., {Spekkens}, K., {et~al.} 2024, \mnras,
  528, 365, \dodoi{10.1093/mnras/stae014}

\bibitem[{{Garrison-Kimmel} {et~al.}(2017){Garrison-Kimmel}, {Bullock},
  {Boylan-Kolchin}, \& {Bardwell}}]{garrison-kimmel2017}
{Garrison-Kimmel}, S., {Bullock}, J.~S., {Boylan-Kolchin}, M., \& {Bardwell},
  E. 2017, \mnras, 464, 3108, \dodoi{10.1093/mnras/stw2564}

\bibitem[{{Gatto} {et~al.}(2013){Gatto}, {Fraternali}, {Read}, {Marinacci},
  {Lux}, \& {Walch}}]{Gatto13}
{Gatto}, A., {Fraternali}, F., {Read}, J.~I., {et~al.} 2013, \mnras, 433, 2749,
  \dodoi{10.1093/mnras/stt896}

\bibitem[{{Geha} {et~al.}(2017){Geha}, {Wechsler}, {Mao}, {Tollerud}, {Weiner},
  {Bernstein}, {Hoyle}, {Marchi}, {Marshall}, {Mu{\~n}oz}, \& {Lu}}]{Geha17}
{Geha}, M., {Wechsler}, R.~H., {Mao}, Y.-Y., {et~al.} 2017, \apj, 847, 4,
  \dodoi{10.3847/1538-4357/aa8626}

\bibitem[{{Geha} {et~al.}(2024){Geha}, {Mao}, {Wechsler}, {Asali}, {Kado-Fong},
  {Kallivayalil}, {Nadler}, {Tollerud}, {Weiner}, {de los Reyes}, {Wang}, \&
  {Wu}}]{Geha24}
{Geha}, M., {Mao}, Y.-Y., {Wechsler}, R.~H., {et~al.} 2024, arXiv e-prints,
  arXiv:2404.14499, \dodoi{10.48550/arXiv.2404.14499}

\bibitem[{{Griffen} {et~al.}(2016){Griffen}, {Ji}, {Dooley}, {G{\'o}mez},
  {Vogelsberger}, {O'Shea}, \& {Frebel}}]{griffen2016}
{Griffen}, B.~F., {Ji}, A.~P., {Dooley}, G.~A., {et~al.} 2016, \apj, 818, 10,
  \dodoi{10.3847/0004-637X/818/1/10}

\bibitem[{{Hargis} {et~al.}(2020){Hargis}, {Albers}, {Crnojevi{\'c}}, {Sand},
  {Weisz}, {Carlin}, {Spekkens}, {Willman}, {Peter}, {Grillmair}, \&
  {Dolphin}}]{Hargis2020}
{Hargis}, J.~R., {Albers}, S., {Crnojevi{\'c}}, D., {et~al.} 2020, \apj, 888,
  31, \dodoi{10.3847/1538-4357/ab58d2}

\bibitem[{{Huang} {et~al.}(2018){Huang}, {Leauthaud}, {Murata}, {Bosch},
  {Price}, {Lupton}, {Mandelbaum}, {Lackner}, {Bickerton}, {Miyazaki},
  {Coupon}, \& {Tanaka}}]{huang2018}
{Huang}, S., {Leauthaud}, A., {Murata}, R., {et~al.} 2018, \pasj, 70, S6,
  \dodoi{10.1093/pasj/psx126}

\bibitem[{Hunter(2007)}]{Hunter2007}
Hunter, J.~D. 2007, Computing in Science \& Engineering, 9, 90,
  \dodoi{10.1109/MCSE.2007.55}

\bibitem[{Ivezi{\'{c}} {et~al.}(2019)Ivezi{\'{c}}, Kahn, Tyson, Abel, Acosta,
  Allsman, Alonso, AlSayyad, Anderson, Andrew, {P. Angel}, Angeli, Ansari,
  Antilogus, Araujo, Armstrong, Arndt, Astier, Aubourg, Auza, Axelrod, Bard,
  Barr, Barrau, Bartlett, Bauer, Bauman, Baumont, Bechtol, Bechtol, Becker,
  Becla, Beldica, Bellavia, Bianco, Biswas, Blanc, Blazek, Blandford, Bloom,
  Bogart, Bond, Booth, Borgland, Borne, Bosch, Boutigny, Brackett, Bradshaw,
  Brandt, Brown, Bullock, Burchat, Burke, Cagnoli, Calabrese, Callahan, Callen,
  Carlin, Carlson, Chandrasekharan, Charles-Emerson, Chesley, Cheu, Chiang,
  Chiang, Chirino, Chow, Ciardi, Claver, Cohen-Tanugi, Cockrum, Coles,
  Connolly, Cook, Cooray, Covey, Cribbs, Cui, Cutri, Daly, Daniel, Daruich,
  Daubard, Daues, Dawson, Delgado, Dellapenna, de~Peyster, de~Val-Borro, Digel,
  Doherty, Dubois, Dubois-Felsmann, Durech, Economou, Eifler, Eracleous,
  Emmons, Neto, Ferguson, Figueroa, Fisher-Levine, Focke, Foss, Frank, Freemon,
  Gangler, Gawiser, Geary, Gee, Geha, Gessner, Gibson, Gilmore, Glanzman,
  Glick, Goldina, Goldstein, Goodenow, Graham, Gressler, Gris, Guy, Guyonnet,
  Haller, Harris, Hascall, Haupt, Hernandez, Herrmann, Hileman, Hoblitt,
  Hodgson, Hogan, Howard, Huang, Huffer, Ingraham, Innes, Jacoby, Jain, Jammes,
  Jee, Jenness, Jernigan, Jevremovi{\'{c}}, Johns, Johnson, Johnson, Jones,
  Juramy-Gilles, Juri{\'{c}}, Kalirai, Kallivayalil, Kalmbach, Kantor, Karst,
  Kasliwal, Kelly, Kessler, Kinnison, Kirkby, Knox, Kotov, Krabbendam,
  Krughoff, Kub{\'{a}}nek, Kuczewski, Kulkarni, Ku, Kurita, Lage, Lambert,
  Lange, Langton, Guillou, Levine, Liang, Lim, Lintott, Long, Lopez, Lotz,
  Lupton, Lust, MacArthur, Mahabal, Mandelbaum, Markiewicz, Marsh, Marshall,
  Marshall, May, McKercher, McQueen, Meyers, Migliore, Miller, Mills, Miraval,
  Moeyens, Moolekamp, Monet, Moniez, Monkewitz, Montgomery, Morrison, Mueller,
  Muller, Arancibia, Neill, Newbry, Nief, Nomerotski, Nordby, O'Connor, Oliver,
  Olivier, Olsen, O'Mullane, Ortiz, Osier, Owen, Pain, Palecek, Parejko,
  Parsons, Pease, Peterson, Peterson, Petravick, Petrick, Petry, Pierfederici,
  Pietrowicz, Pike, Pinto, Plante, Plate, Plutchak, Price, Prouza, Radeka,
  Rajagopal, Rasmussen, Regnault, Reil, Reiss, Reuter, Ridgway, Riot, Ritz,
  Robinson, Roby, Roodman, Rosing, Roucelle, Rumore, Russo, Saha, Sassolas,
  Schalk, Schellart, Schindler, Schmidt, Schneider, Schneider, Schoening,
  Schumacher, Schwamb, Sebag, Selvy, Sembroski, Seppala, Serio, Serrano, Shaw,
  Shipsey, Sick, Silvestri, Slater, Smith, Smith, Sobhani, Soldahl,
  Storrie-Lombardi, Stover, Strauss, Street, Stubbs, Sullivan, Sweeney,
  Swinbank, Szalay, Takacs, Tether, Thaler, Thayer, Thomas, Thornton, Thukral,
  Tice, Trilling, Turri, Berg, Berk, Vetter, Virieux, Vucina, Wahl, Walkowicz,
  Walsh, Walter, Wang, Wang, Warner, Wiecha, Willman, Winters, Wittman, Wolff,
  Wood-Vasey, Wu, Xin, Yoachim, \& Zhan}]{Ivezic2019}
Ivezi{\'{c}}, {\v{Z}}., Kahn, S.~M., Tyson, J.~A., {et~al.} 2019, The
  Astrophysical Journal, 873, 111, \dodoi{10.3847/1538-4357/ab042c}

\bibitem[{{Jahn} {et~al.}(2022){Jahn}, {Sales}, {Wetzel}, {Samuel}, {El-Badry},
  {Boylan-Kolchin}, \& {Bullock}}]{Jahn2022}
{Jahn}, E.~D., {Sales}, L.~V., {Wetzel}, A., {et~al.} 2022, \mnras, 513, 2673,
  \dodoi{10.1093/mnras/stac811}

\bibitem[{{Jerjen} {et~al.}(2001){Jerjen}, {Rekola}, {Takalo}, {Coleman}, \&
  {Valtonen}}]{Jerjen2001}
{Jerjen}, H., {Rekola}, R., {Takalo}, L., {Coleman}, M., \& {Valtonen}, M.
  2001, \aap, 380, 90, \dodoi{10.1051/0004-6361:20011408}

\bibitem[{{Jones} {et~al.}(2023){Jones}, {Mutlu-Pakdil}, {Sand}, {Donnerstein},
  {Crnojevi{\'c}}, {Bennet}, {Fielder}, {Karunakaran}, {Spekkens}, {Strader},
  {Urquhart}, \& {Zaritsky}}]{Jones2023}
{Jones}, M.~G., {Mutlu-Pakdil}, B., {Sand}, D.~J., {et~al.} 2023, \apjl, 957,
  L5, \dodoi{10.3847/2041-8213/ad0130}

\bibitem[{{Kallivayalil} {et~al.}(2018){Kallivayalil}, {Sales}, {Zivick},
  {Fritz}, {Del Pino}, {Sohn}, {Besla}, {van der Marel}, {Navarro}, \&
  {Sacchi}}]{K18}
{Kallivayalil}, N., {Sales}, L.~V., {Zivick}, P., {et~al.} 2018, \apj, 867, 19,
  \dodoi{10.3847/1538-4357/aadfee}

\bibitem[{{Karachentsev} {et~al.}(2013){Karachentsev}, {Makarov}, \&
  {Kaisina}}]{Karachentsev2013}
{Karachentsev}, I.~D., {Makarov}, D.~I., \& {Kaisina}, E.~I. 2013, \aj, 145,
  101, \dodoi{10.1088/0004-6256/145/4/101}

\bibitem[{{Kawanomoto} {et~al.}(2018){Kawanomoto}, {Uraguchi}, {Komiyama},
  {Miyazaki}, {Furusawa}, {Finet}, {Hattori}, {Wang}, {Yasuda}, \&
  {Suzuki}}]{Kawanomoto2018}
{Kawanomoto}, S., {Uraguchi}, F., {Komiyama}, Y., {et~al.} 2018, \pasj, 70, 66,
  \dodoi{10.1093/pasj/psy056}

\bibitem[{{Komiyama} {et~al.}(2018){Komiyama}, {Obuchi}, {Nakaya}, {Kamata},
  {Kawanomoto}, {Utsumi}, {Miyazaki}, {Uraguchi}, {Furusawa}, {Morokuma},
  {Uchida}, {Miyatake}, {Mineo}, {Fujimori}, {Aihara}, {Karoji}, {Gunn}, \&
  {Wang}}]{Komiyama2018}
{Komiyama}, Y., {Obuchi}, Y., {Nakaya}, H., {et~al.} 2018, \pasj, 70, S2,
  \dodoi{10.1093/pasj/psx069}

\bibitem[{{Mac Low} \& {Ferrara}(1999)}]{maclow99}
{Mac Low}, M.-M., \& {Ferrara}, A. 1999, \apj, 513, 142, \dodoi{10.1086/306832}

\bibitem[{Magnier {et~al.}(2013)Magnier, Schlafly, Finkbeiner, Juric, Tonry,
  Burgett, Chambers, Flewelling, Kaiser, Kudritzki, Morgan, Price, Sweeney, \&
  Stubbs}]{Magnier2013}
Magnier, E.~A., Schlafly, E., Finkbeiner, D., {et~al.} 2013, The Astrophysical
  Journal Supplement Series, 205, 20, \dodoi{10.1088/0067-0049/205/2/20}

\bibitem[{{Mao} {et~al.}(2021){Mao}, {Geha}, {Wechsler}, {Weiner}, {Tollerud},
  {Nadler}, \& {Kallivayalil}}]{Mao21}
{Mao}, Y.-Y., {Geha}, M., {Wechsler}, R.~H., {et~al.} 2021, \apj, 907, 85,
  \dodoi{10.3847/1538-4357/abce58}

\bibitem[{{Mao} {et~al.}(2024){Mao}, {Geha}, {Wechsler}, {Asali}, {Wang},
  {Kado-Fong}, {Kallivayalil}, {Nadler}, {Tollerud}, {Weiner}, {de los Reyes},
  \& {Wu}}]{Mao24}
---. 2024, arXiv e-prints, arXiv:2404.14498, \dodoi{10.48550/arXiv.2404.14498}

\bibitem[{{Martin} {et~al.}(2009){Martin}, {McConnachie}, {Irwin}, {Widrow},
  {Ferguson}, {Ibata}, {Dubinski}, {Babul}, {Chapman}, {Fardal}, {Lewis},
  {Navarro}, \& {Rich}}]{martin2009}
{Martin}, N.~F., {McConnachie}, A.~W., {Irwin}, M., {et~al.} 2009, \apj, 705,
  758, \dodoi{10.1088/0004-637X/705/1/758}

\bibitem[{{Martin} {et~al.}(2016){Martin}, {Ibata}, {Lewis}, {McConnachie},
  {Babul}, {Bate}, {Bernard}, {Chapman}, {Collins}, {Conn}, {Crnojevi{\'c}},
  {Fardal}, {Ferguson}, {Irwin}, {Mackey}, {McMonigal}, {Navarro}, \&
  {Rich}}]{Martin2016}
{Martin}, N.~F., {Ibata}, R.~A., {Lewis}, G.~F., {et~al.} 2016, \apj, 833, 167,
  \dodoi{10.3847/1538-4357/833/2/167}

\bibitem[{{Mart{\'\i}nez-Delgado} {et~al.}(2022){Mart{\'\i}nez-Delgado},
  {Karim}, {Charles}, {Boschin}, {Monelli}, {Collins}, {Donatiello}, \&
  {Alfaro}}]{martinez-delgado2022}
{Mart{\'\i}nez-Delgado}, D., {Karim}, N., {Charles}, E. J.~E., {et~al.} 2022,
  \mnras, 509, 16, \dodoi{10.1093/mnras/stab2797}

\bibitem[{{Mart{\'\i}nez-Delgado} {et~al.}(2012){Mart{\'\i}nez-Delgado},
  {Romanowsky}, {Gabany}, {Annibali}, {Arnold}, {Fliri}, {Zibetti}, {van der
  Marel}, {Rix}, {Chonis}, {Carballo-Bello}, {Aloisi}, {Macci{\`o}},
  {Gallego-Laborda}, {Brodie}, \& {Merrifield}}]{Martinez-Delgado2012}
{Mart{\'\i}nez-Delgado}, D., {Romanowsky}, A.~J., {Gabany}, R.~J., {et~al.}
  2012, \apjl, 748, L24, \dodoi{10.1088/2041-8205/748/2/L24}

\bibitem[{{McNanna} {et~al.}(2024){McNanna}, {Bechtol}, {Mau}, {Nadler},
  {Medoff}, {Drlica-Wagner}, {Cerny}, {Crnojevi{\'c}}, {Mutlu-Pakd{\i}l},
  {Vivas}, {Pace}, {Carlin}, {Collins}, {Ferguson}, {Mart{\'\i}nez-Delgado},
  {Mart{\'\i}nez-V{\'a}zquez}, {Noel}, {Riley}, {Sand}, {Smercina}, {Tollerud},
  {Wechsler}, {Abbott}, {Aguena}, {Alves}, {Bacon}, {Bom}, {Brooks}, {Burke},
  {Carballo-Bello}, {Carnero Rosell}, {Carretero}, {da Costa}, {Davis}, {de
  Vicente}, {Diehl}, {Doel}, {Ferrero}, {Frieman}, {Giannini}, {Gruen},
  {Gutierrez}, {Gruendl}, {Hinton}, {Hollowood}, {Honscheid}, {James}, {Kuehn},
  {Marshall}, {Mena-Fern{\'a}ndez}, {Miquel}, {Pereira}, {Pieres},
  {Malag{\'o}n}, {Sakowska}, {Sanchez}, {Sanchez Cid}, {Santiago},
  {Sevilla-Noarbe}, {Smith}, {Stringfellow}, {Suchyta}, {Swanson}, {Tarle},
  {Weaverdyck}, {Wiseman}, {DES Collaboration}, \& {DELVE
  Collaboration}}]{McNanna2024}
{McNanna}, M., {Bechtol}, K., {Mau}, S., {et~al.} 2024, \apj, 961, 126,
  \dodoi{10.3847/1538-4357/ad07d0}

\bibitem[{{McQuinn} {et~al.}(2023){McQuinn}, {Mao}, {Buckley}, {Shih}, {Cohen},
  \& {Dolphin}}]{Mcquinn23}
{McQuinn}, K. B.~W., {Mao}, Y.-Y., {Buckley}, M.~R., {et~al.} 2023, \apj, 944,
  14, \dodoi{10.3847/1538-4357/acaec9}

\bibitem[{{McQuinn} {et~al.}(2024){McQuinn}, {Mao}, {Tollerud}, {Cohen},
  {Shih}, {Buckley}, \& {Dolphin}}]{McQuinn24}
{McQuinn}, K. B.~W., {Mao}, Y.-Y., {Tollerud}, E.~J., {et~al.} 2024, \apj, 967,
  161, \dodoi{10.3847/1538-4357/ad429b}

\bibitem[{{McQuinn} {et~al.}(2010){McQuinn}, {Skillman}, {Cannon}, {Dalcanton},
  {Dolphin}, {Hidalgo-Rodr{\'\i}guez}, {Holtzman}, {Stark}, {Weisz}, \&
  {Williams}}]{McQuinn2010}
{McQuinn}, K. B.~W., {Skillman}, E.~D., {Cannon}, J.~M., {et~al.} 2010, \apj,
  721, 297, \dodoi{10.1088/0004-637X/721/1/297}

\bibitem[{{Miyazaki} {et~al.}(2018){Miyazaki}, {Komiyama}, {Kawanomoto}, {Doi},
  {Furusawa}, {Hamana}, {Hayashi}, {Ikeda}, {Kamata}, {Karoji}, {Koike},
  {Kurakami}, {Miyama}, {Morokuma}, {Nakata}, {Namikawa}, {Nakaya}, {Nariai},
  {Obuchi}, {Oishi}, {Okada}, {Okura}, {Tait}, {Takata}, {Tanaka}, {Tanaka},
  {Terai}, {Tomono}, {Uraguchi}, {Usuda}, {Utsumi}, {Yamada}, {Yamanoi},
  {Aihara}, {Fujimori}, {Mineo}, {Miyatake}, {Oguri}, {Uchida}, {Tanaka},
  {Yasuda}, {Takada}, {Murayama}, {Nishizawa}, {Sugiyama}, {Chiba}, {Futamase},
  {Wang}, {Chen}, {Ho}, {Liaw}, {Chiu}, {Ho}, {Lai}, {Lee}, {Jeng}, {Iwamura},
  {Armstrong}, {Bickerton}, {Bosch}, {Gunn}, {Lupton}, {Loomis}, {Price},
  {Smith}, {Strauss}, {Turner}, {Suzuki}, {Miyazaki}, {Muramatsu}, {Yamamoto},
  {Endo}, {Ezaki}, {Ito}, {Kawaguchi}, {Sofuku}, {Taniike}, {Akutsu}, {Dojo},
  {Kasumi}, {Matsuda}, {Imoto}, {Miwa}, {Suzuki}, {Takeshi}, \&
  {Yokota}}]{Miyazaki2018}
{Miyazaki}, S., {Komiyama}, Y., {Kawanomoto}, S., {et~al.} 2018, \pasj, 70, S1,
  \dodoi{10.1093/pasj/psx063}

\bibitem[{{Moster} {et~al.}(2010){Moster}, {Somerville}, {Maulbetsch}, {van den
  Bosch}, {Macci{\`o}}, {Naab}, \& {Oser}}]{moster2010}
{Moster}, B.~P., {Somerville}, R.~S., {Maulbetsch}, C., {et~al.} 2010, \apj,
  710, 903, \dodoi{10.1088/0004-637X/710/2/903}

\bibitem[{{Mutlu-Pakdil} {et~al.}(2021){Mutlu-Pakdil}, {Sand}, {Crnojevi{\'c}},
  {Drlica-Wagner}, {Caldwell}, {Guhathakurta}, {Seth}, {Simon}, {Strader}, \&
  {Toloba}}]{mutlu-pakdil2021}
{Mutlu-Pakdil}, B., {Sand}, D.~J., {Crnojevi{\'c}}, D., {et~al.} 2021, \apj,
  918, 88, \dodoi{10.3847/1538-4357/ac0db8}

\bibitem[{{Mutlu-Pakdil} {et~al.}(2024){Mutlu-Pakdil}, {Sand}, {Crnojevi{\'c}},
  {Bennet}, {Jones}, {Spekkens}, {Karunakaran}, {Zaritsky}, {Caldwell},
  {Fielder}, {Guhathakurta}, {Seth}, {Simon}, {Strader}, \&
  {Toloba}}]{Mutlupakdil24}
---. 2024, \apj, 966, 188, \dodoi{10.3847/1538-4357/ad36c4}

\bibitem[{{Nadler} {et~al.}(2024){Nadler}, {Gluscevic}, {Driskell}, {Wechsler},
  {Moustakas}, {Benson}, \& {Mao}}]{Nadler24}
{Nadler}, E.~O., {Gluscevic}, V., {Driskell}, T., {et~al.} 2024, \apj, 967, 61,
  \dodoi{10.3847/1538-4357/ad3bb1}

\bibitem[{{Ogami} {et~al.}(2024){Ogami}, {Komiyama}, {Chiba}, {Tanaka},
  {Guhathakurta}, {Kirby}, {Wyse}, {Filion}, {Kirihara}, {Ishigaki}, \&
  {Hayashi}}]{ogami2024}
{Ogami}, I., {Komiyama}, Y., {Chiba}, M., {et~al.} 2024, arXiv e-prints,
  arXiv:2407.07481, \dodoi{10.48550/arXiv.2407.07481}

\bibitem[{{Ott} {et~al.}(2012){Ott}, {Stilp}, {Warren}, {Skillman},
  {Dalcanton}, {Walter}, {de Blok}, {Koribalski}, \& {West}}]{ott2012}
{Ott}, J., {Stilp}, A.~M., {Warren}, S.~R., {et~al.} 2012, \aj, 144, 123,
  \dodoi{10.1088/0004-6256/144/4/123}

\bibitem[{{Patel} {et~al.}(2020){Patel}, {Kallivayalil}, {Garavito-Camargo},
  {Besla}, {Weisz}, {van der Marel}, {Boylan-Kolchin}, {Pawlowski}, \&
  {G{\'o}mez}}]{patel2020}
{Patel}, E., {Kallivayalil}, N., {Garavito-Camargo}, N., {et~al.} 2020, \apj,
  893, 121, \dodoi{10.3847/1538-4357/ab7b75}

\bibitem[{{Plummer}(1911)}]{plummer1911}
{Plummer}, H.~C. 1911, \mnras, 71, 460, \dodoi{10.1093/mnras/71.5.460}

\bibitem[{Price-Whelan {et~al.}(2018)Price-Whelan, Sipőcz, G{\"{u}}nther, Lim,
  Crawford, Conseil, Shupe, Craig, Dencheva, Ginsburg, VanderPlas, Bradley,
  P{\'{e}}rez-Su{\'{a}}rez, de~Val-Borro, Aldcroft, Cruz, Robitaille, Tollerud,
  Ardelean, Babej, Bach, Bachetti, Bakanov, Bamford, Barentsen, Barmby,
  Baumbach, Berry, Biscani, Boquien, Bostroem, Bouma, Brammer, Bray,
  Breytenbach, Buddelmeijer, Burke, Calderone, Rodr{\'{i}}guez, Cara, Cardoso,
  Cheedella, Copin, Corrales, Crichton, D'Avella, Deil, Depagne, Dietrich,
  Donath, Droettboom, Earl, Erben, Fabbro, Ferreira, Finethy, Fox, Garrison,
  Gibbons, Goldstein, Gommers, Greco, Greenfield, Groener, Grollier, Hagen,
  Hirst, Homeier, Horton, Hosseinzadeh, Hu, Hunkeler, Ivezi{\'{c}}, Jain,
  Jenness, Kanarek, Kendrew, Kern, Kerzendorf, Khvalko, King, Kirkby, Kulkarni,
  Kumar, Lee, Lenz, Littlefair, Ma, Macleod, Mastropietro, McCully, Montagnac,
  Morris, Mueller, Mumford, Muna, Murphy, Nelson, Nguyen, Ninan, N{\"{o}}the,
  Ogaz, Oh, Parejko, Parley, Pascual, Patil, Patil, Plunkett, Prochaska,
  Rastogi, Janga, Sabater, Sakurikar, Seifert, Sherbert, Sherwood-Taylor, Shih,
  Sick, Silbiger, Singanamalla, Singer, Sladen, Sooley, Sornarajah, Streicher,
  Teuben, Thomas, Tremblay, Turner, Terr{\'{o}}n, van Kerkwijk, de~la Vega,
  Watkins, Weaver, Whitmore, Woillez, Zabalza, Zabalza, \&
  Contributors}]{Price-Whelan2018b}
Price-Whelan, A.~M., Sipőcz, B.~M., G{\"{u}}nther, H.~M., {et~al.} 2018, The
  Astronomical Journal, 156, 123, \dodoi{10.3847/1538-3881/aabc4f}

\bibitem[{{Rich} {et~al.}(2012){Rich}, {Collins}, {Black}, {Longstaff}, {Koch},
  {Benson}, \& {Reitzel}}]{Rich2012}
{Rich}, R.~M., {Collins}, M.~L.~M., {Black}, C.~M., {et~al.} 2012, \nat, 482,
  192, \dodoi{10.1038/nature10837}

\bibitem[{{Ricotti} \& {Gnedin}(2005)}]{Ricotti05}
{Ricotti}, M., \& {Gnedin}, N.~Y. 2005, \apj, 629, 259, \dodoi{10.1086/431415}

\bibitem[{Robitaille {et~al.}(2013)Robitaille, Tollerud, Greenfield,
  Droettboom, Bray, Aldcroft, Davis, Ginsburg, Price-Whelan, Kerzendorf,
  Conley, Crighton, Barbary, Muna, Ferguson, Grollier, Parikh, Nair,
  G{\"{u}}nther, Deil, Woillez, Conseil, Kramer, Turner, Singer, Fox, Weaver,
  Zabalza, Edwards, {Azalee Bostroem}, Burke, Casey, Crawford, Dencheva, Ely,
  Jenness, Labrie, Lim, Pierfederici, Pontzen, Ptak, Refsdal, Servillat, \&
  Streicher}]{TheAstropyCollaboration2013}
Robitaille, T.~P., Tollerud, E.~J., Greenfield, P., {et~al.} 2013, Astronomy \&
  Astrophysics, 558, A33, \dodoi{10.1051/0004-6361/201322068}

\bibitem[{{Sales} {et~al.}(2022){Sales}, {Wetzel}, \& {Fattahi}}]{Sales22}
{Sales}, L.~V., {Wetzel}, A., \& {Fattahi}, A. 2022, Nature Astronomy, 6, 897,
  \dodoi{10.1038/s41550-022-01689-w}

\bibitem[{{Sand} {et~al.}(2015){Sand}, {Spekkens}, {Crnojevi{\'c}}, {Hargis},
  {Willman}, {Strader}, \& {Grillmair}}]{Sand2015}
{Sand}, D.~J., {Spekkens}, K., {Crnojevi{\'c}}, D., {et~al.} 2015, \apjl, 812,
  L13, \dodoi{10.1088/2041-8205/812/1/L13}

\bibitem[{{Sand} {et~al.}(2022){Sand}, {Mutlu-Pakdil}, {Jones}, {Karunakaran},
  {Wang}, {Yang}, {Chiti}, {Bennet}, {Crnojevi{\'c}}, \& {Spekkens}}]{Sand2022}
{Sand}, D.~J., {Mutlu-Pakdil}, B., {Jones}, M.~G., {et~al.} 2022, \apjl, 935,
  L17, \dodoi{10.3847/2041-8213/ac85ee}

\bibitem[{{Sand} {et~al.}(2024){Sand}, {Mutlu-Pakdil}, {Jones}, {Karunakaran},
  {Andrews}, {Bennet}, {Crnojevic}, {Donatiello}, {Drlica-Wagner}, {Fielder},
  {Martinez-Delgado}, {Martinez-Vazquez}, {Spekkens}, {Doliva-Dolinsky},
  {Hunger}, {Carlin}, {Cerny}, {Hai}, {McQuinn}, {Pace}, \&
  {Smercina}}]{sand2024}
---. 2024, arXiv e-prints, arXiv:2409.16345, \dodoi{10.48550/arXiv.2409.16345}

\bibitem[{{Santos-Santos} {et~al.}(2022){Santos-Santos}, {Sales}, {Fattahi}, \&
  {Navarro}}]{santos-santos2022}
{Santos-Santos}, I. M.~E., {Sales}, L.~V., {Fattahi}, A., \& {Navarro}, J.~F.
  2022, \mnras, 515, 3685, \dodoi{10.1093/mnras/stac2057}

\bibitem[{Schlafly \& Finkbeiner(2011)}]{Schlafly2011}
Schlafly, E.~F., \& Finkbeiner, D.~P. 2011, The Astrophysical Journal, 737,
  103, \dodoi{10.1088/0004-637X/737/2/103}

\bibitem[{Schlafly {et~al.}(2012)Schlafly, Finkbeiner, Juri{\'{c}}, Magnier,
  Burgett, Chambers, Grav, Hodapp, Kaiser, Kudritzki, Martin, Morgan, Price,
  Rix, Stubbs, Tonry, \& Wainscoat}]{Schlafly2012}
Schlafly, E.~F., Finkbeiner, D.~P., Juri{\'{c}}, M., {et~al.} 2012, The
  Astrophysical Journal, 756, 158, \dodoi{10.1088/0004-637X/756/2/158}

\bibitem[{{Schlegel} {et~al.}(1998){Schlegel}, {Finkbeiner}, \&
  {Davis}}]{Schlegel1998}
{Schlegel}, D.~J., {Finkbeiner}, D.~P., \& {Davis}, M. 1998, \apj, 500, 525,
  \dodoi{10.1086/305772}

\bibitem[{{Sharina} {et~al.}(2008){Sharina}, {Karachentsev}, {Dolphin},
  {Karachentseva}, {Tully}, {Karataeva}, {Makarov}, {Makarova}, {Sakai},
  {Shaya}, {Nikolaev}, \& {Kuznetsov}}]{sharina2008}
{Sharina}, M.~E., {Karachentsev}, I.~D., {Dolphin}, A.~E., {et~al.} 2008,
  \mnras, 384, 1544, \dodoi{10.1111/j.1365-2966.2007.12814.x}

\bibitem[{{Simon}(2019)}]{Simon19}
{Simon}, J.~D. 2019, \araa, 57, 375,
  \dodoi{10.1146/annurev-astro-091918-104453}

\bibitem[{{Simpson} {et~al.}(2018){Simpson}, {Grand}, {G{\'o}mez}, {Marinacci},
  {Pakmor}, {Springel}, {Campbell}, \& {Frenk}}]{Simpson18}
{Simpson}, C.~M., {Grand}, R. J.~J., {G{\'o}mez}, F.~A., {et~al.} 2018, \mnras,
  478, 548, \dodoi{10.1093/mnras/sty774}

\bibitem[{{Smercina} {et~al.}(2018){Smercina}, {Bell}, {Price}, {D'Souza},
  {Slater}, {Bailin}, {Monachesi}, \& {Nidever}}]{Smercina18}
{Smercina}, A., {Bell}, E.~F., {Price}, P.~A., {et~al.} 2018, \apj, 863, 152,
  \dodoi{10.3847/1538-4357/aad2d6}

\bibitem[{{Smercina} {et~al.}(2022){Smercina}, {Bell}, {Samuel}, \&
  {D'Souza}}]{Smercina22}
{Smercina}, A., {Bell}, E.~F., {Samuel}, J., \& {D'Souza}, R. 2022, \apj, 930,
  69, \dodoi{10.3847/1538-4357/ac5d56}

\bibitem[{{Smith} {et~al.}(2023){Smith}, {Jensen}, {Roediger}, {Sestito},
  {Hayes}, {McConnachie}, {Cuillandre}, {Gwyn}, {Magnier}, {Chambers},
  {Hammer}, {Hudson}, {Martin}, {Navarro}, \& {Scott}}]{Simon23}
{Smith}, S. E.~T., {Jensen}, J., {Roediger}, J., {et~al.} 2023, \aj, 166, 76,
  \dodoi{10.3847/1538-3881/acdd77}

\bibitem[{Taylor(2005)}]{Taylor2005}
Taylor, M. 2005, in Astronomical Society of the Pacific Conference Series, Vol.
  347, Astronomical Data Analysis Software and Systems XIV, ed. P.~Shopbell,
  M.~Britton, \& R.~Ebert, 29

\bibitem[{{Toloba} {et~al.}(2016){Toloba}, {Guhathakurta}, {Romanowsky},
  {Brodie}, {Mart{\'\i}nez-Delgado}, {Arnold}, {Ramachandran}, \&
  {Theakanath}}]{Toloba2016}
{Toloba}, E., {Guhathakurta}, P., {Romanowsky}, A.~J., {et~al.} 2016, \apj,
  824, 35, \dodoi{10.3847/0004-637X/824/1/35}

\bibitem[{Tonry {et~al.}(2012)Tonry, Stubbs, Lykke, Doherty, Shivvers, Burgett,
  Chambers, Hodapp, Kaiser, Kudritzki, Magnier, Morgan, Price, \&
  Wainscoat}]{Tonry2012}
Tonry, J.~L., Stubbs, C.~W., Lykke, K.~R., {et~al.} 2012, The Astrophysical
  Journal, 750, 99, \dodoi{10.1088/0004-637X/750/2/99}

\bibitem[{van~der Walt {et~al.}(2011)van~der Walt, Colbert, \&
  Varoquaux}]{VanderWalt2011}
van~der Walt, S., Colbert, S.~C., \& Varoquaux, G. 2011, Computing in Science
  \& Engineering, 13, 22, \dodoi{10.1109/MCSE.2011.37}

\bibitem[{{Weisz} {et~al.}(2011){Weisz}, {Dalcanton}, {Williams}, {Gilbert},
  {Skillman}, {Seth}, {Dolphin}, {McQuinn}, {Gogarten}, {Holtzman}, {Rosema},
  {Cole}, {Karachentsev}, \& {Zaritsky}}]{Weisz2011}
{Weisz}, D.~R., {Dalcanton}, J.~J., {Williams}, B.~F., {et~al.} 2011, \apj,
  739, 5, \dodoi{10.1088/0004-637X/739/1/5}

\bibitem[{{Whiting} {et~al.}(2007){Whiting}, {Hau}, {Irwin}, \&
  {Verdugo}}]{Whiting2007}
{Whiting}, A.~B., {Hau}, G. K.~T., {Irwin}, M., \& {Verdugo}, M. 2007, \aj,
  133, 715, \dodoi{10.1086/510309}

\bibitem[{{Woo} {et~al.}(2008){Woo}, {Courteau}, \& {Dekel}}]{Woo2008}
{Woo}, J., {Courteau}, S., \& {Dekel}, A. 2008, \mnras, 390, 1453,
  \dodoi{10.1111/j.1365-2966.2008.13770.x}

\bibitem[{{Zaritsky} {et~al.}(2024){Zaritsky}, {Golini}, {Donnerstein},
  {Trujillo}, {Akhlaghi}, {Chamba}, {D'Onofrio}, {Eskandarlou}, {Zahra
  Hosseini-ShahiSavandi}, {Infante-Sainz}, {Martin}, {Montes}, {Rom{\'a}n},
  {Sedighi}, \& {Sharbaf}}]{Zaritsky24}
{Zaritsky}, D., {Golini}, G., {Donnerstein}, R., {et~al.} 2024, arXiv e-prints,
  arXiv:2406.01912, \dodoi{10.48550/arXiv.2406.01912}

\end{thebibliography}
\bibliographystyle{aasjournal}

\end{document}